\documentclass[pdflatex, sn-mathphys-num]{sn-jnl}

\usepackage{graphicx}%
\usepackage{multirow}%
\usepackage{amsmath,amssymb,amsfonts}%
\usepackage{amsthm}%
\usepackage{mathrsfs}%
\usepackage[title]{appendix}%
\usepackage{xcolor}%
\usepackage{textcomp}%
\usepackage{manyfoot}%
\usepackage{booktabs}%
\usepackage{algorithm}%
\usepackage{algorithmicx}%
\usepackage{algpseudocode}%
\usepackage{listings}%
\usepackage{makecell}%

\begin{document}

\title[Simulation of proton-induced activation]{Simulation of proton-induced activation for low-Earth orbit high energy astrophysics missions}


\author*[1,2]{\fnm{Riccardo} \sur{Campana}}\email{riccardo.campana@inaf.it}

\affil*[1]{\orgdiv{INAF}, \orgname{OAS}, \orgaddress{\street{Via Gobetti 101}, \city{Bologna}, \postcode{40129}, \country{Italy}}}

\affil[2]{\orgdiv{INFN}, \orgname{Sezione di Bologna}, \orgaddress{\street{Viale Berti Pichat 6/7}, \city{Bologna}, \postcode{40127},  \country{Italy}}}


\abstract{
Proton-induced activation represents a major source of instrumental background for high-energy astrophysics missions in low-Earth orbit, where trapped protons, particularly during transits within the South Atlantic Anomaly region, irradiate spacecraft materials and generate radioactive isotopes. Direct Monte Carlo simulations of activation and of the ensuing decays are computationally inefficient, due to the low probability of nuclide production and the large number of decay events required for sufficient statistical accuracy. 

In this paper we provide a new implementation of an efficient three-step algorithm that decouples isotope production, radioactive-decay evolution, and background synthesis, enabling rapid reconstruction of activation-induced background for arbitrary irradiation histories. The method combines Geant4-based identification of all radioisotopes produced by monochromatic proton irradiations, numerical solutions of the Bateman equations for linearized decay chains, and simulation of the detector response to each isotope decay emissions. The approach greatly reduces the computational cost while maintaining accuracy, as demonstrated through validation against direct simulations, which show excellent agreement over many orders of magnitude in activity and time.

This method is applied to two representative case studies: HERMES and eXTP/LAD and WFM, covering different detector technologies and orbital configurations. The presented framework enables fast exploration of design and operational scenarios (e.g., orbit selection, radiation models, or duty cycles) and is well suited for background budgeting and optimization of future high-energy space missions.
}



\maketitle

\section{Introduction}\label{sec1}
The Earth atmosphere is opaque to X-rays and $\gamma$-rays; as a result, high-energy astrophysics observation can be conduced only from space, with instrumentation typically flown aboard satellites in low-altitude orbits (500–600 km Low Earth Orbits, or LEOs). This orbital environment induces a significant instrumental background, primarily due to interactions between cosmic rays and other energetic particles with the detector and the surrounding spacecraft structures \cite{campana22}.

The overall background for a space-borne high-energy instrument can be divided into two main components. The first is the \emph{prompt} background, caused by direct energy deposits from photons and charged particles in the detector. The second component arises from \emph{activation} processes, where high-energy particles produce nuclear reactions in the spacecraft and detector materials, transmuting them into unstable isotopes. These isotopes subsequently decay over varying timescales, emitting $\alpha$, $\beta$, or $\gamma$ radiation, which contributes to the \emph{delayed} background component.

The characteristics of activation background depend strongly on the spacecraft and detector geometry, the materials used, and the specific irradiation profile. For missions in LEOs, much of this contribution arises from geomagnetically trapped protons in the Earth magnetosphere, such as the South Atlantic Anomaly (SAA) region. The activation background typically comprises two components: a short-lived contribution from isotopes with half-lives ranging from seconds to days, which usually diminishes rapidly (depending on the frequency of the transits through the trapped proton regions), and a long-lived component from isotopes with half-lives on the scale of days to years, which usually accumulates over the mission duration.

Estimating the onboard background levels for an X/$\gamma$-ray space experiment is an essential part of mission design and planning. Based on the anticipated environment, various strategies can be implemented to meet scientific requirements and reduce background interference, such as hardware shielding, onboard software filtering, or mission-wide measures like optimized pointing plans.

A common approach involves Monte Carlo simulations using radiation-transport codes to model how photons and particles interact with the spacecraft and detector components, defined by their geometry and materials. Geant4 \cite{agostinelli2003} is the standard framework for these high-energy physics simulations. It provides a set of C++ libraries to construct a virtual model of the instrument, simulate interactions with environmental particles, and extract relevant results such as energy deposits in the detector.

The computational effort needed for simulations depends on the desired statistical accuracy and the level of detail in the instrument mass model. To balance precision and efficiency, less critical parts, such as spacecraft structures, are typically represented with simplified geometries and approximate compositions. In contrast, critical components (e.g., the detector itself) are modeled with high detail, often derived from CAD designs. 

In general, a simulation for the activation of a satellite or detector can be performed by irradiating a mass model with high-energy protons and examining the delayed emissions over suitable timescales, e.g. ranging from $10^{-3}$~s to $10^6$~s. However, this \emph{direct} approach is highly inefficient. Radio-activation is a comparatively rare process, competing with numerous other physical interaction mechanisms, resulting in a low probability of generating unstable nuclei. For example, when a 1~cm thick Cu target is irradiated with 20~MeV monochromatic protons, on average, only one activation occurs for every $\sim$370 primary protons simulated. Moreover, achieving sufficient statistical accuracy requires generating a large number of decay events for each desired post-irradiation time bin, significantly increasing the computational cost.

To address these challenges, a three-step efficient algorithm has been proposed by several authors. Initially introduced in the MGGPOD software framework \cite{weidenspointner2005}, this approach was subsequently implemented in MEGAlib \cite{zoglauer2006} and adapted in various works for specific experiments \cite{odaka2018, galgoczi2020}. The core concept involves decoupling the generation of radioactive nuclides in a detector or spacecraft mass model from the simulation of the background induced by their decay products. Additionally, the activity of each radioactive species within a given time frame is determined through a numerical solution of the radioactive decay equations.

In this paper, we present a variation of this approach that enables the construction of a database containing all relevant information required to reconstruct the expected activation-induced background for \emph{arbitrary} irradiation profiles (both in terms of time and energy).

As case studies, this approach is applied in two scenarios: the \emph{High Energy Modular Ensemble of Satellites} (HERMES) cubesat constellation \cite{fiore21}, and two experiments onboard the \emph{enhanced X-ray Timing Polarimeter} (eXTP) mission, i.e., the \emph{Large Area Detector} (LAD) and the \emph{Wide Field Monitor} (WFM) \cite{feroci24,hernanz24}.

This paper is structured as follows. In Section~\ref{s:2} a general introduction to the activation of materials after a proton irradiation is provided, followed in Section~\ref{s:3} by an overview of the trapped high energy proton environment encountered in a typical low-Earth orbit. Section~\ref{s:4} discusses the activation simulation algorithm, and its validation. The following Sections \ref{s:hermes}, and \ref{s:extp} present our case studies.

\section{Activation of nuclides from proton irradiation}\label{s:2}
Nuclear isotope activation through proton irradiation involves bombarding a stable target nucleus with high-energy protons, causing nuclear reactions that result in the formation of radionuclides. The interaction of protons with nuclei can occur through a variety of mechanisms, including elastic and inelastic scattering. These interactions are governed by their cross-sections, which vary with the energy of the incoming protons and the properties of the target nucleus.

\begin{figure}
    \centering
    \includegraphics[width=0.7\textwidth]{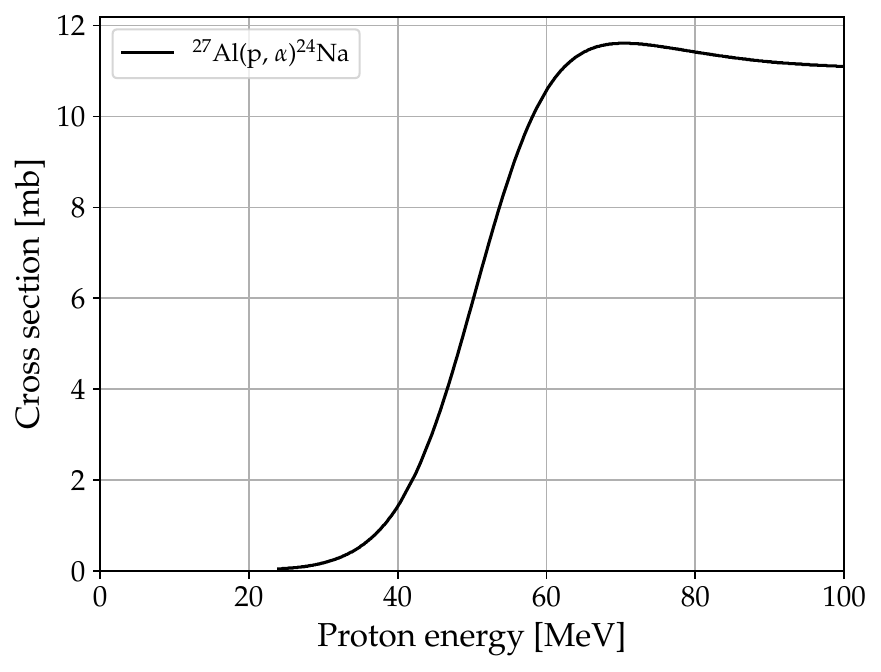}
    \caption{Cross section (in millibarns, 1 mb = 10$^{-27}$ cm$^2$) for the $^{27}$Al(p, $\alpha$)$^{24}$Na reaction. Data from the IAEA database, \url{https://www-nds.iaea.org}.}
    \label{fig:cross-sections}
\end{figure}

Proton irradiation induces nuclear reactions such as (p, n), (p, $\alpha$), and (p, $\gamma$), where a proton collides with a nucleus, resulting in the ejection of one or more particles (e.g., neutrons or $\alpha$) and the formation of a different nuclide. For instance, the irradiation of a nickel target with protons can produce $^{64}$Cu through the $^{64}$Ni(p, n)$^{64}$Cu reaction\footnote{We employ here the standard notation for nuclear reactions, in which $X(a, b)Y$ means $X + a \rightarrow Y + b$.}. 
Another example is the activation of aluminum: proton irradiation of naturally occurring $^{27}$Al can lead to the formation of $^{24}$Na  via the  $^{27}$Al(p, $\alpha$)$^{24}$Na reaction.  An example of the cross-sections for the latter reaction is shown in Figure~\ref{fig:cross-sections}, and is typical of these processes: a minimum proton energy of a few MeVs (from 3--5 up to $\sim$20~MeV, depending on the target atomic number $Z$) is required to penetrate the nuclear Coulomb barrier and induce a reaction, with the cross-section rapidly rising with energy, peaking and then reaching a sort of plateau.

\section{The trapped proton environment in LEO}\label{s:3}
Most of the high energy astrophysics missions are flown on Low Earth Orbits (LEOs), spanning altitudes from approximately 400 to 600 km above the surface. Although they have several advantages with respect to other types of orbits (e.g., launcher performance, ground communications, general radiation levels \citep{campana22}), these represent a complex radiation environment dominated by trapped energetic particles within Earth's magnetosphere. Among these particles, trapped protons, besides being of the most significant radiation hazards for spacecraft, astronauts, and sensitive electronic systems operating in this orbital environment, are the main source of activation for scientific instruments.

Trapped protons in LEO originate primarily from the cosmic ray albedo neutron decay and solar energetic particle (SEP) events \cite{campana22}. The first process occurs when high-energy cosmic rays interact with the atmosphere, producing neutrons that subsequently decay into protons and electrons while moving through the magnetosphere. These newly created protons, along with solar protons that penetrate the magnetosphere during geomagnetic disturbances, become trapped by the geomagnetic field lines in quasi-stable orbits around the planet.
The motion of the trapped particles creates the characteristic toroidal  structure of the Van Allen radiation belts \cite{vanallen58}, with particles bouncing between magnetic mirror points in the conjugate hemispheres.

The trapped proton environment exhibits significant spatial variability, strongly influenced by the fact that the geomagnetic dipole field is tilted and offset with respect to the Earth center and rotation axis. The South Atlantic Anomaly (SAA) represents the most prominent feature, where the inner radiation belt dips closest to Earth surface. Within the SAA region, proton fluxes can exceed those found elsewhere in LEO by several orders of magnitude, with energies ranging from hundreds of keV to several hundred MeV.
Outside the SAA, trapped proton populations in LEO are generally much less intense but still present significant challenges. The energy spectrum typically follows a power law distribution, with lower energy protons (1--10 MeV) being more abundant than higher energy particles ($>$50 MeV). However, the higher energy protons pose greater penetration concerns for spacecraft shielding and can cause more severe single event effects in electronic devices.
The trapped proton environment in LEO exhibits variability across multiple timescales, due, e.g., to geomagnetic storms (hours-days scale) and/or to solar cycle variations (11-year scale).

The characterization of trapped particle environments has evolved significantly through decades of satellite-based measurements and modeling efforts. The foundational NASA models AP8 \cite{sawyer76} and AE8 \cite{vette91}, developed from 1960s--1970s satellite data, established the aerospace industry standard for predicting omnidirectional proton and electron fluxes across energy ranges of 0.1--400~MeV and 0.04--7~MeV, respectively. While these models distinguish between solar minimum and maximum conditions, they lack dynamic geomagnetic field modeling and exhibit substantial uncertainties in regions with steep spatial gradients, particularly at the SAA boundaries.
Subsequent modeling efforts have addressed many of these limitations through enhanced datasets and methodological improvements. An ESA-sponsored low-altitude proton model utilized SAMPEX satellite measurements to provide more accurate flux predictions below 600~km altitude for solar minimum conditions \cite{heynderickx99}. Most significantly, the comprehensive AE9/AP9/SPM model suite \cite{ginet13}, developed collaboratively by the National Reconnaissance Office and Air Force Research Laboratory, incorporated over 30 satellite datasets spanning 1976--2011. This advanced framework introduced higher spatial resolution, quantified uncertainties through Monte Carlo analysis, and provided statistical flux thresholds at various percentile levels while incorporating space weather variability and instrumental uncertainties.
Despite these advances, modeling challenges persist, particularly for low-inclination, low-altitude orbital regimes where observational data remains sparse. 

The Space Environment Information System (SPENVIS) platform\footnote{\url{https://www.spenvis.oma.be}} has emerged as a comprehensive web-based interface, enabling mission planners to access multiple environmental models and compute orbit-specific radiation exposures. This integrated approach represents the current state-of-the-art for trapped particle environment assessment, though continued model refinement remains essential as new satellite missions provide additional observational constraints and as space weather understanding continues to advance.

\section{The simulation framework}\label{s:4}

\subsection{The three-step algorithm}

 This method leverages a set of large-scale initial simulations and involves the following steps:
\begin{enumerate}
    \item \textbf{Simulation of nuclide generation}: Identify the radioactive nuclides produced in the mass model upon irradiation by protons.
    \item \textbf{Numerical calculation of decay chains and activities}: Solve the radioactive decay equations to determine the \emph{activity} of each species over time, in every volume of the simulated mass model.
    \item \textbf{Simulation of background contributions}: Determine the background induced by the decay of each nuclide within a specified volume.
\end{enumerate}

This approach significantly improves computational efficiency while maintaining accuracy, making it particularly suited for activation studies in high-energy astrophysics and related fields.

\subsubsection{Step 1}\label{s:step1}

In the first step, the aim is to determine the number, type and location of each radioisotope generated by proton irradiation. The Monte Carlo simulation for each primary particle generated in the run proceeds until a radioisotope is generated within any volume, recording all the relevant information such as the nuclide atomic number $Z$, mass number $A$, excitation level $E$, and location (i.e., in which volume of the mass model the isotope has been generated). It is essential that the specific track of the event is then stopped after the generation of the isotope, to avoid to accumulate also its daughters.

In order to separate ``prompt'' radiation to ``delayed'' decays, a suitable threshold on half-life can be implemented: if a given isotope has an half-life shorter than a characteristic time (which could be, for example, the detector dead time or linked to the timescales that one wants to investigate) the simulation is not stopped and the interaction of the decay products are treated as an instantaneous contribution to the instrumental background count rate.

In Geant4, this step is specifically implemented in an instantiation of the \texttt{G4SteppingAction} class. This is the class which allows the user to interact with the simulation at the ``step'' level\footnote{A Geant4 simulation consists, for each \emph{event} in a given \emph{run}, of several \emph{tracks}, corresponding to the primary particle tracked in its interaction and of all its generated secondary particles. Each track is then a collection of \emph{steps}, each representing one physical interaction with the mass model.}.

To ensure a full flexibility in the following steps, the simulation can be performed once by simulating several monochromatic proton energies, for example in the range from 4--5 up to 700~MeV (protons with lower energies are not able to induce activation, and the usual fluxes at higher energies are generally negligible). Any specific irradiation spectral profile can be then derived by a suitable weighting of each simulation results, as will be discussed in the following.

As a result of the first step, we obtain a list of all the generated isotopes (which can be conveniently normalized per unit of incident proton fluence, protons/cm$^{2}$) in each volume of the mass model, with their characteristics $(Z, A, E)$.

\subsubsection{Step 2}\label{s:step2}

In order to evaluate the activity for each isotope in each volume at an arbitrary time $t$ after irradiation, we need to solve the equations of the radioactive decay. This requires, in the general case of a complex decay chain, to numerically solve the Bateman system of differential equations, which is summarized in Appendix~\ref{a:bateman}.

\begin{figure}
    \centering
    \includegraphics[width=0.7\textwidth]{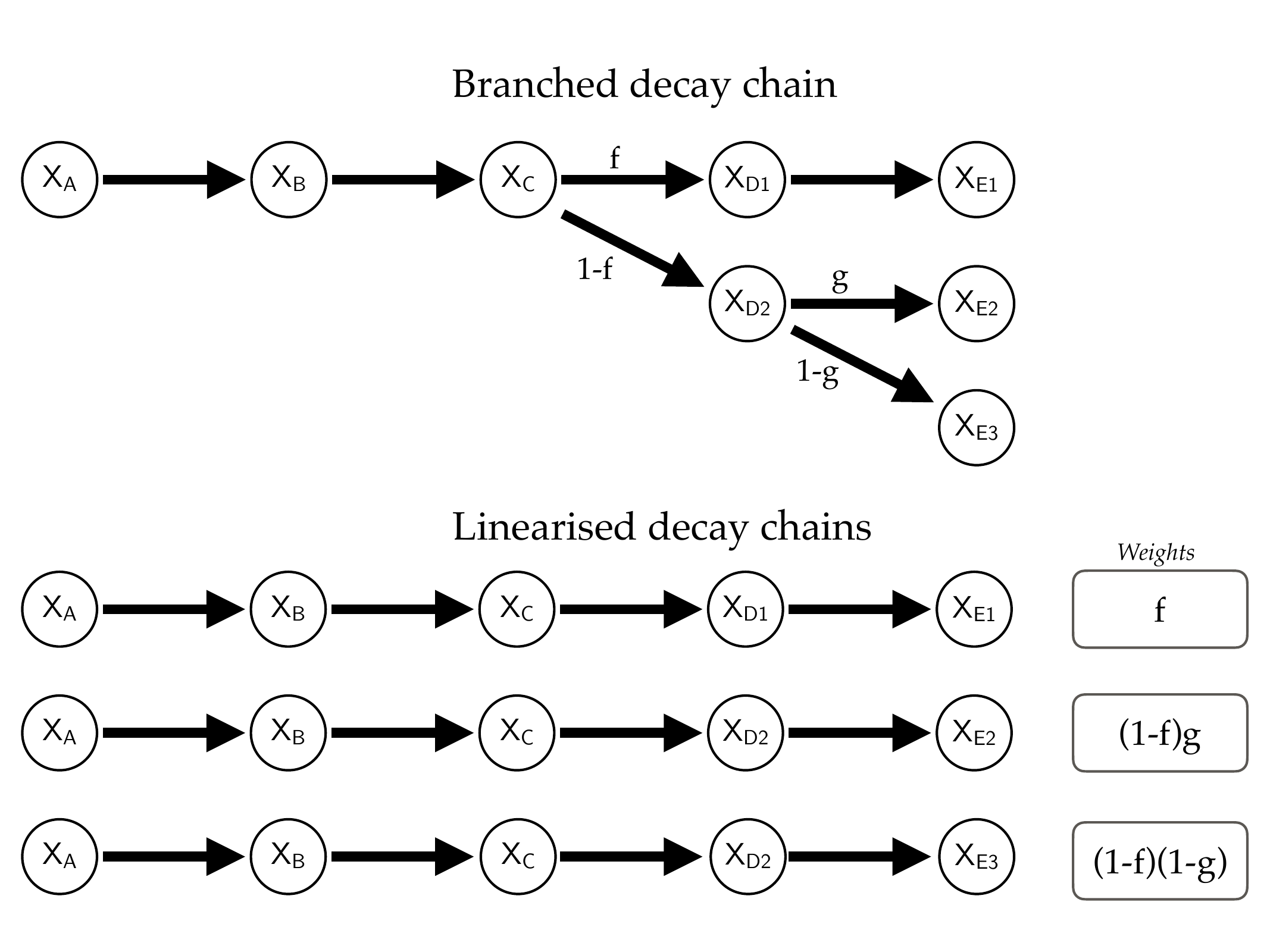}
    \caption{Linearization of decay chains. In the upper panel, a complex, branched decay chain is shown. $f$ and $g$ are the branching ratios. Lower panel: the corresponding linearized decay chains, with their associated total weight.}
    \label{fig:linearised-decay-chains}
\end{figure}

The algorithm proceeds as follows: for each independent isotope generated in the simulation, information on its decay modes is pulled from the Geant4 \texttt{RadioactiveDecay} database, and its decay chain is reconstructed.
Branched decay chains can be simplified by \emph{pruning} them, i.e., by removing the branches with very low probability (e.g., removing all the decay modes with less than 0.1\% probability to occur) and then by \emph{linearizing}, i.e., treating them as independent linear chains with a suitable weight, which is the product of the branching ratio for each path (Figure~\ref{fig:linearised-decay-chains}).

Consequently, the Bateman equations are solved for each (linearized) decay chain, assuming a normalized concentration of 1.0 for the initial isotope and a null population of daughters. As a result, normalized concentrations and thus activities are determined for each isotope and its daughters over a suitable array of $n$ times $\mathbf{t} = [t_0, t_1, ..., t_{n-1}]$. Note that one isotope can be generated through different decay chains: its activity at a given time is thus the sum of the activities from each possible channel.

Therefore, the algorithm determines, for each time $t_k \in \mathbf{t}$ and each volume, the type of isotopes present and their activity, normalized for the input irradiation fluence, and expressed in Bq/(protons/cm$^2$).

\subsubsection{Step 3}\label{s:step3}

The third step is the simulation of the effects of the decay of each isotope in each volume on the background rate. A simulation is performed, using the same mass model, by looping over every volume and every isotope present (at any given time) in it, generating at random the isotope in the volume and making it decay. To avoid double-counting, the simulation is stopped before letting any unstable daughter isotope decays in turn. The result of this step is a library of background spectra due to the decay of a given isotope in a given volume, and normalized for the input activity (expressed, e.g., in counts/s/keV/decay).

\bigskip

By putting together the results of these three steps, the expected background for any irradiation spectrum and time profile can be derived. In other words, the background spectrum that is recorded at a time $t_k$ after the irradiation is the cumulated effect of the contributions from each isotope that is possible to generate in each volume, weighted by the expected number of its decays (i.e., its activity) at the time $t_k$. This, in turn is a function of the irradiation fluence, spectrum and time profile/history.

For example, let's assume that the aim is to simulate an irradiation time profile $F_\mathrm{irr}(\tau)$ (say, a Gaussian shape, which is a good approximation of the flux encountered during a passage through the SAA) with a proton source having a differential fluence spectrum $S_\mathrm{src}(E)$ (e.g., a power-law shape), and we are interested in the background count rates and spectra for all times $t \in \textbf{t}$ (which can be later than the irradiation time profile, or can comprise the irradiation itself).

In this general case, the total proton fluence as a function of time is:
\begin{equation}
    \Phi(t) = \int_0^\infty \int_0^\infty S_\mathrm{src}(E)F_\mathrm{irr}(t-\tau) dEd\tau
\end{equation}

To transform the problem in a way easier to compute, the first step is to assume that the $F_\mathrm{irr}(\tau)$ irradiation time profile can be replaced by a sum of weighted, time-shifted \emph{instantaneous} irradiations:

\begin{equation}\label{e:discrete_irr}
    F_\mathrm{irr}(\tau) \longrightarrow \sum_{i} f_i\delta[\tau-\tau_i]
\end{equation}
where, in the time bin $\tau_i$, the integral proton fluence is considered constant and instantaneous (i.e., a Dirac delta), and $f_i = F_\mathrm{irr}(\tau_i)$.

Then, the input proton fluence spectrum $S_\mathrm{src}(E)$ (in protons/cm$^2$/MeV) can be discretized over an array of $m$ energies $\mathbf{E} = [E_0, E_1, ..., E_{m-1}]$:
\begin{equation}
    s_j = S_\mathrm{src}(E_j)
\end{equation}
and the total integral fluence provided by a monochromatic irradiation with protons of energy $E_j$ is $s_j\Delta E_j$ protons/cm$^2$,
where $\Delta E_j$ is the width of the $j$-th energy bin.

Let $a_{p,v}^j[t_k]$ be the normalized activity, in Bq/(protons/cm$^2$), of the $p$-th isotope in the $v$-th volume, upon an instantaneous irradiation with protons of energy $E_j$, calculated over the discretized time grid $\textbf{t}$.
The value of the normalized activity will depend on its decay scheme, and on the total number of the specific isotope (or of its mothers), generated by the instantaneous irradiation.

The total activity (in Bq) in the mass model $v$-th volume, at time $t_k$, after an instantaneous irradiation at time $t=0$ with spectrum $S_\mathrm{src}(E)$ is therefore:

\begin{equation}
    A_{p,v}[t_k] =  \sum_{j} s_j\Delta E_j a_{p,v}^j[t_k]
\end{equation}

The activity for a general discretized irradiation profile (Eq.~\ref{e:discrete_irr}) is therefore:
\begin{equation}
    A_{p,v}[t_k] = \sum_i\sum_{j} s_j\Delta E_j f_i a_{p,v}^j[\tau_i-t_k]
\end{equation}
that is, the sum of weighted, time-shifted activities calculated over a time $t_k$ with an instantaneous irradiation at a time $\tau_i$.

The total background spectrum on the detector, at a given time $t_k$, is the sum of the contributions of each isotope in each volume:
\begin{equation}
    \mathrm{Bkg}(\mathcal{E}, t_k) = \sum_v \sum_{p} A_{p,v}[t_k] \mathcal{F}_{p,v}(\mathcal{E})
\end{equation}
where $\mathcal{F}_{p,v}(\mathcal{E})$ is the background spectrum, normalized for unit decay of the isotope $p$ in the volume $v$, as a function of the recorded energy (or detector channel) $\mathcal{E}$.

\bigskip 

\begin{figure}
    \centering
    \includegraphics[width=0.9\textwidth]{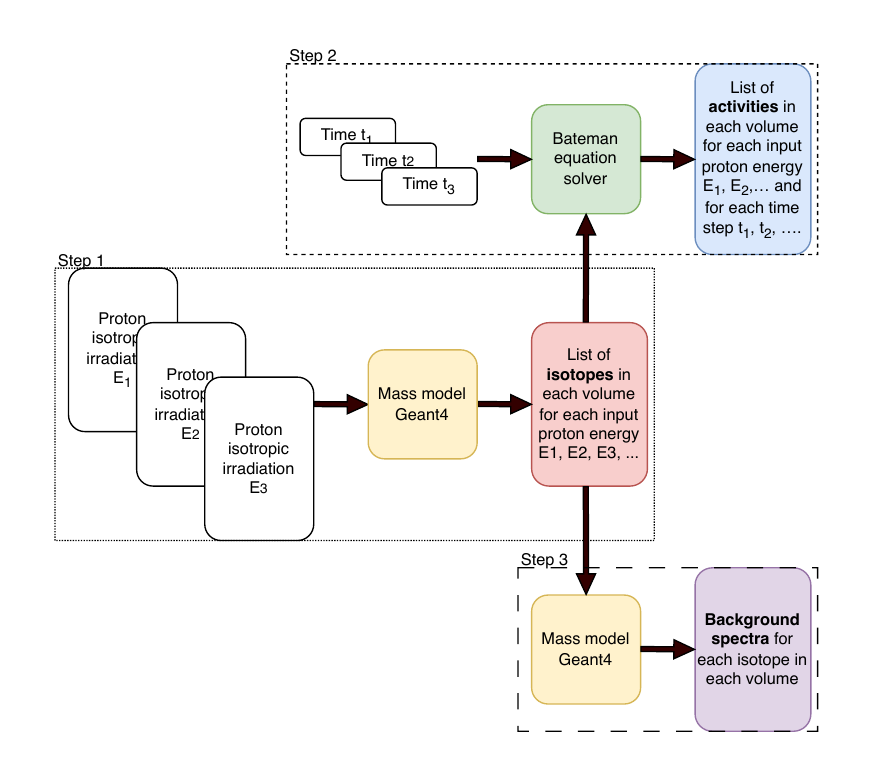}    
    \caption{Flow chart of the three-steps algorithm for activation simulations.}
    \label{fig:steps1-2-3}
\end{figure}

\begin{figure}
    \centering
    \includegraphics[width=0.9\textwidth]{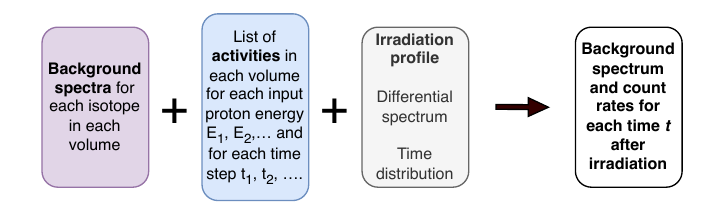}    
    \caption{The final synthesis of the outputs of the three steps algorithm for activation simulations.}
    \label{fig:step-final}
\end{figure}

The logical connections between each step and the final synthesis of the determination of the background spectrum and count rate for a given time after an arbitrary irradiation profile is schematically shown in Figures~\ref{fig:steps1-2-3} and~\ref{fig:step-final}. It is worth noticing that the first two blocks in the final synthesis shown in Figure~\ref{fig:step-final} are computed only once for a given mass model: this means that the activation-induced background spectrum can be very quickly calculated for different scenarios (e.g., when evaluating the impact of an orbital inclination change, or when comparing different trapped particle models).

\subsection{Validation}

A validation of the aforementioned schema can be carried out by comparing the results of a straightforward direct simulation (in which all the interactions and decays are followed at all time scales) to the predicted activities computed with the 3-step algorithm. 

\begin{figure}
    \centering
    \includegraphics[width=0.49\textwidth]{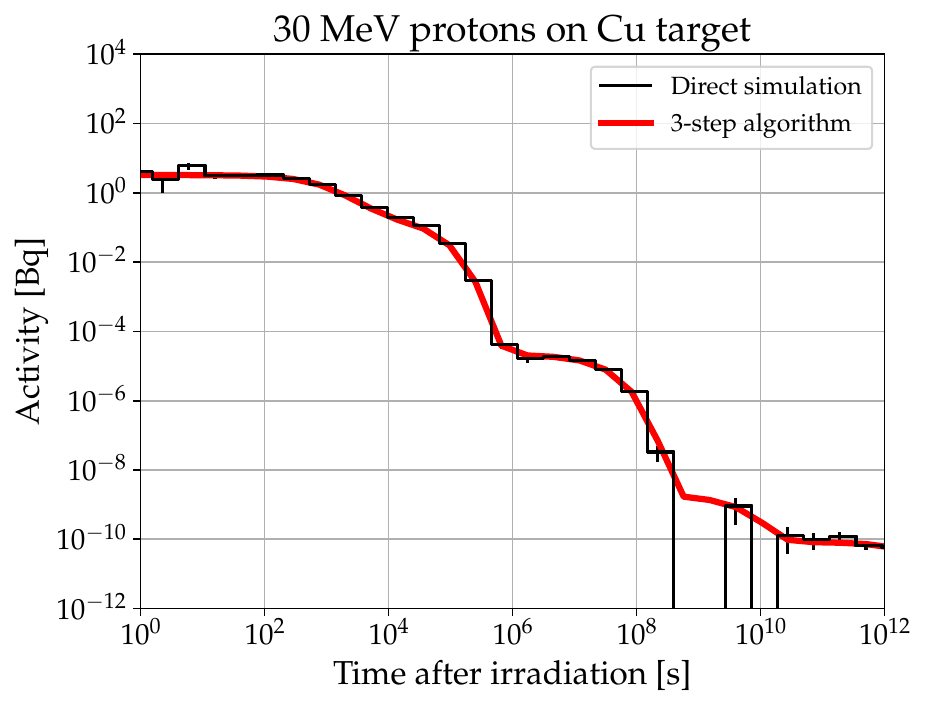}    \includegraphics[width=0.49\textwidth]{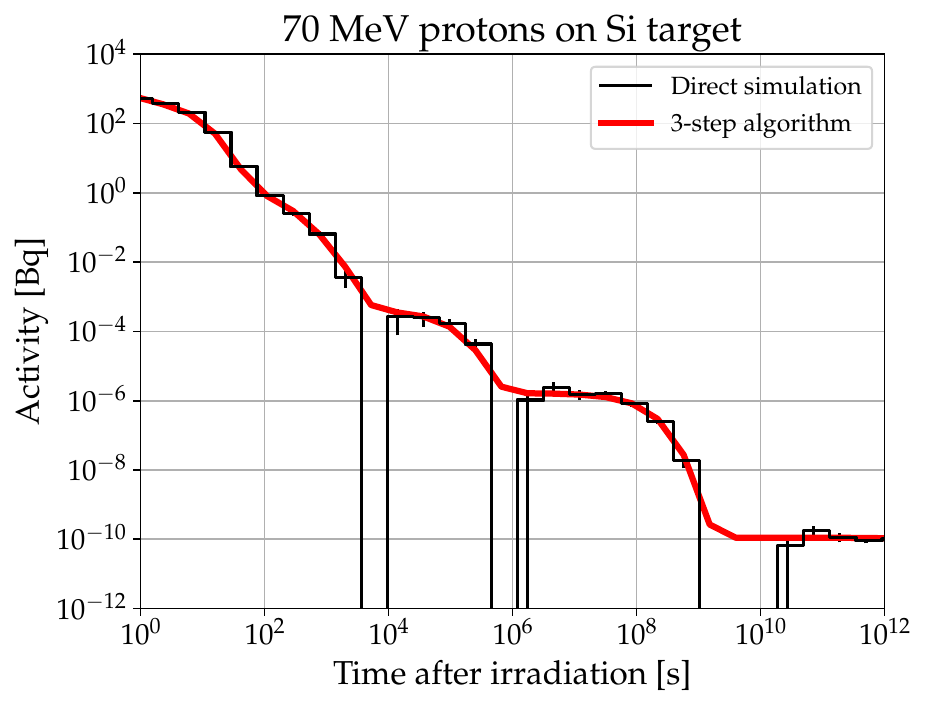}
    \caption{Activity after an instantaneous irradiation with a flux of 10$^6$ protons cm$^{-2}$ on a 1~cm thick target of natural copper (left panel, with 30~MeV protons) and silicon (right panel, 70~MeV protons). The black histogram is a direct simulation, while the red curve is the numerical prediction from the 3-step algorithm. Missing data at long times/low activity levels is due to the low statistics of the direct simulation.}
    \label{fig:activity_validation}
\end{figure}

In Figure~\ref{fig:activity_validation} the number of decays as a function of time is shown for two representative simple simulations. In this case, a 1~cm thick target of Cu (left panel) or Si (right panel) is irradiated with 10$^6$ monochromatic protons, with an energy of 30~MeV for the Cu target and 70~MeV for the Si target. The black line shows the activity determined with the direct simulation, while the red line is the prediction of the 3-step algorithm. In both cases the number of primary events is the same. The 3-step algorithm prediction follows closely the direct simulation, with different ``bumps'' corresponding to different isotopes (e.g., $^{26}$Al and $^{27}$Si, or $^{62}$Zn and $^{62}$Cu). Table~\ref{t:Cu30MeV_isotopes} reports, as an example, the various isotopes created in the 30~MeV protons on Cu target simulation.

\begin{table}[htbp]
\centering
\begin{tabular}{c|c|c|c|c}
Isotope & Efficiency & Half-life  & Decay mode & Decay scheme \\
\hline
$^{62}$Zn & $2.39 \cdot 10^{-3}$ & 9.193 h   & $\beta^+$/EC  & $^{62}$Zn $\rightarrow$ $^{62}$Cu $\rightarrow$ $^{62}$Ni (stable)       \\
$^{63}$Zn & $1.39 \cdot 10^{-3}$ & 38.38 min & $\beta^+$/EC  & $^{63}$Zn $\rightarrow$ $^{63}$Cu (stable)      \\
$^{62}$Cu & $1.25 \cdot 10^{-3}$ & 9.672 min & $\beta^+$/EC  & $^{62}$Cu $\rightarrow$ $^{62}$Ni (stable)     \\
$^{64}$Cu & $7.80 \cdot 10^{-4}$ & 12.70 h   & \makecell{$\beta^-$ (38\%) \\ EC (62\%)}  &    \makecell{$^{64}$Cu $\rightarrow$ $^{64}$Zn (stable) \\  $^{64}$Cu $\rightarrow$ $^{64}$Ni (stable)}  \\
$^{65}$Zn & $4.00 \cdot 10^{-4}$ & 243.93 d  &  $\beta^+$/EC  &  $^{65}$Zn  $\rightarrow$ $^{65}$Cu (stable)  \\
$^{61}$Cu & $2.46 \cdot 10^{-4}$ & 3.34 h    & $\beta^+$/EC  &   $^{61}$Cu  $\rightarrow$ $^{61}$Ni (stable)   \\
$^{59}$Ni & $1.79 \cdot 10^{-4}$ & 81000 y   &  $\beta^+$/EC &   $^{59}$Ni  $\rightarrow$ $^{59}$Co (stable)   \\
$^{63}$Ni & $5.00 \cdot 10^{-6}$ & 100.8 y   &  $\beta^-$ &     $^{63}$Ni  $\rightarrow$ $^{63}$Cu  (stable)  \\
$^{66}$Cu & $3.00 \cdot 10^{-6}$ & 5.09 min  &  $\beta^-$ &     $^{66}$Cu  $\rightarrow$ $^{66}$Zn  (stable) \\
$^{61}$Zn & $1.00 \cdot 10^{-6}$ & 88.8 s    &  $\beta^+$/EC &     $^{61}$Zn  $\rightarrow$ $^{61}$Cu (stable) \\
$^{61}$Co & $1.00 \cdot 10^{-6}$ & 1.65 h	 & $\beta^-$  &      $^{61}$Co  $\rightarrow$ $^{61}$Ni (stable)\\
\hline
\end{tabular}
\caption{Isotopes created in an irradiation of a 1~cm thick Cu target with 30~MeV protons. The efficiency is defined as the number of isotopes generated in the simulation, divided by the total number of primary protons simulated. For the half-life and decay scheme see also the NuDat database, \url{https://www.nndc.bnl.gov/nudat3/}.}
\label{t:Cu30MeV_isotopes}
\end{table}

\begin{figure}
    \centering
    \includegraphics[width=0.49\textwidth]{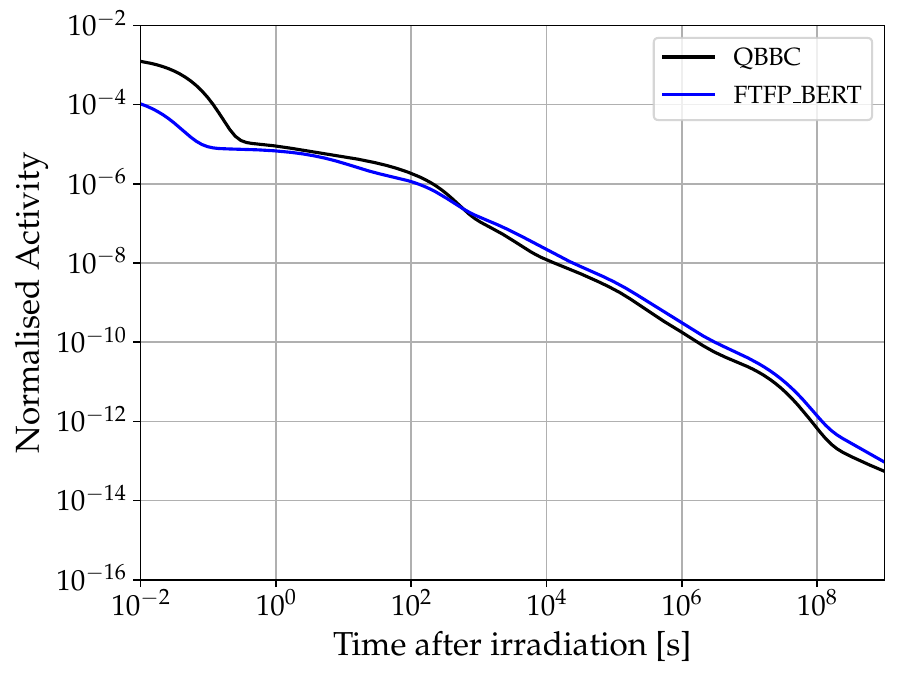}    
    \caption{Predicted normalized activity with the 3-step algorithm, for a simulation with 150~MeV protons incident along the long axis of a 1$\times$1$\times$4~cm GAGG:Ce block, for the \texttt{QBBC} and \texttt{FTFP\_BERT} physics lists.}
    \label{fig:comparison_physlist}
\end{figure}

In Geant4, the user can choose between different hadronic physics models, which simulate interactions involving hadrons, such as protons, over a wide energy range\footnote{See, e.g., the Geant4 Physics Reference Manual for more details \url{https://geant4-userdoc.web.cern.ch/UsersGuides/PhysicsReferenceManual/html/index.html}}. These models are categorized into parameterized, theoretical, and data-driven approaches. Parameterized models, like the Bertini cascade, are computationally efficient and suitable for low- to intermediate-energy interactions, typically below a few GeV. Theoretical models, such as the Fritiof (FTF) and Quark-Gluon String (QGS) models, describe high-energy interactions by considering partonic and string dynamics, and are appropriate for energies above several GeV. Data-driven models, like G4ParticleHP, rely on experimental cross-section data to simulate low-energy neutron interactions with high precision. The choice of a given model can be a compromise between computational time (usually, the more precise the model the larger the run time) and accuracy. In Table~\ref{t:runtime} simulation results with several different physics lists are compared. \texttt{FTFP\_BERT} is the default Geant4 hadronic physics list, and from a computational point of view gives the faster run time. The \texttt{QBBC} physics list is the one recommended for medical and space physics simulations, where accurate simulation for low-energy transport of protons and neutrons is needed. It runs only 6\% slower than \texttt{FTFP\_BERT}, but interestingly it generates a good 30\% more different types of isotopes (mostly excited and metastable states), although the total number of activated isotopes is lower. As shown in Figure~\ref{fig:comparison_physlist}, the former physics list produces a higher activity for short-lived isotopes (with half-lives below 0.1~s), while for mid- and long-lived isotopes the activities are similar. As a baseline, in the following \texttt{QBBC} will be used.

\begin{table}
    \centering
    \begin{tabular}{c|c|c|c}
        Physics list & Normalised run time & Total no. of isotopes & No. of different isotopes \\ \hline
       \texttt{FTFP\_BERT} & 1.00 & 3462 & 102 \\
       \texttt{QGSP\_BERT} & 1.01 & 3432 & 102 \\
       \texttt{QBBC} & 1.06 & 2685 & 137 \\
       \texttt{Shielding} & 1.46 & 3414 & 100 \\
       \texttt{QGSP\_BIC\_HP} & 1.57 & 3915 & 136\\\hline
    \end{tabular}
    \caption{Comparison of the results for different Geant4 physics lists. In this case, a beam of 150~MeV protons is incident along the long axis of a 1$\times$1$\times$4~cm GAGG:Ce block. 10$^6$ particles are generated in this simulation. Run time is normalized with respect to the \texttt{FTFP\_BERT} case (the absolute value on an Apple Silicon M3 Pro system at 3.57~GHz and 18~GB RAM, using 2 threads, is of 21.7 seconds). The last two columns report the total number of radioactive isotopes generated in the simulation, and the number of different species. \texttt{QGSP\_BERT} is essentially equivalent to \texttt{FTFP\_BERT} in the investigated energy range. The last two physics lists are recommended for deep shielding and/or accurate neutron transport simulations, and are significantly more computationally demanding.}
    \label{t:runtime}
\end{table}

\section{Case study I: HERMES}\label{s:hermes}

\subsection{Instrument overview}
The \emph{High Energy Rapid Modular Ensemble of Satellites} (HERMES) mission is a distributed space observatory consisting of a constellation of 3U CubeSats. The main mission goal is to validates a scalable, low-cost methodology for astrophysical observations, aiming to pioneer a low-cost, distributed approach to detecting and localizing high-energy astrophysical transients. The HERMES-Pathfinder project \cite{fiore21,fiore22}, composed of six identical 3U CubeSats, will observe the prompt emission of high-energy transients such as Gamma-Ray Bursts (GRBs), providing rapid localization (through the triangulation technique) and complementary observations to gravitational-wave detections. In particular, event-by-event time-tagging down to microsecond precision enables spectral-timing analysis of GRB prompt emission, probing the physics of relativistic jets and emission mechanisms on sub-millisecond timescales.

Each HERMES satellite hosts a compact, modular detector system \cite{evangelista22, evangelista24} occupying approximately 1U of the CubeSat volume. The scientific payload integrates a hybrid detector assembly, composed of Silicon Drift Detectors (SDDs) optically coupled to GAGG:Ce scintillator crystals, providing a wide energy band from a few keV to several MeV. 
The detector ``siswich'' architecture is based on the readout of 60 GAGG:Ce crystals (each with $6.94\times12.1\times15$~mm$^3$ dimensions) by two independent SDD cells. 
The latter act both as a direct X-ray detector (``X-mode'', $\sim$3--50~keV) and as photosensors for the optical scintillation light produced by the absorption of an higher-energy $\gamma$-ray in the crystal itself (``S-mode'', $\sim$20--2000~keV)
A custom low-noise readout Application-Specific Integrated Circuit (ASIC, \cite{gandola19}) and a dedicated Payload Data Handling Unit (PDHU, \cite{guzman21}) manage signal processing and event reconstruction. A chip-scale atomic clock (CSAC) ensures sub-microsecond timing accuracy across the constellation, which is essential for triangulation-based localization. The payload mass is approximately 1.5~kg with a nominal power consumption below 2~W, fully compatible with 3U CubeSat constraints. 
The payload design emphasizes manufacturability, modularity, and the use of commercial off-the-shelf (COTS) components.
 
Extensive ground calibration and environmental qualification campaigns \cite{dilillo24, campana24} have validated detector performance, thermal stability, and radiation tolerance.
The six satellites of the HERMES Pathfinder constellation have been launched on March 15, 2025. A seventh, identical, payload is hosted onboard the Australian satellite SpIRIT, launched on December 1st, 2023 \cite{trenti24}. Commissioning activities are currently undergoing (Baroni et al., 2026, submitted; Leone et al. 2026, in press).

\subsection{Activation simulations}

\begin{figure}
    \centering
    \includegraphics[width=0.8\textwidth]{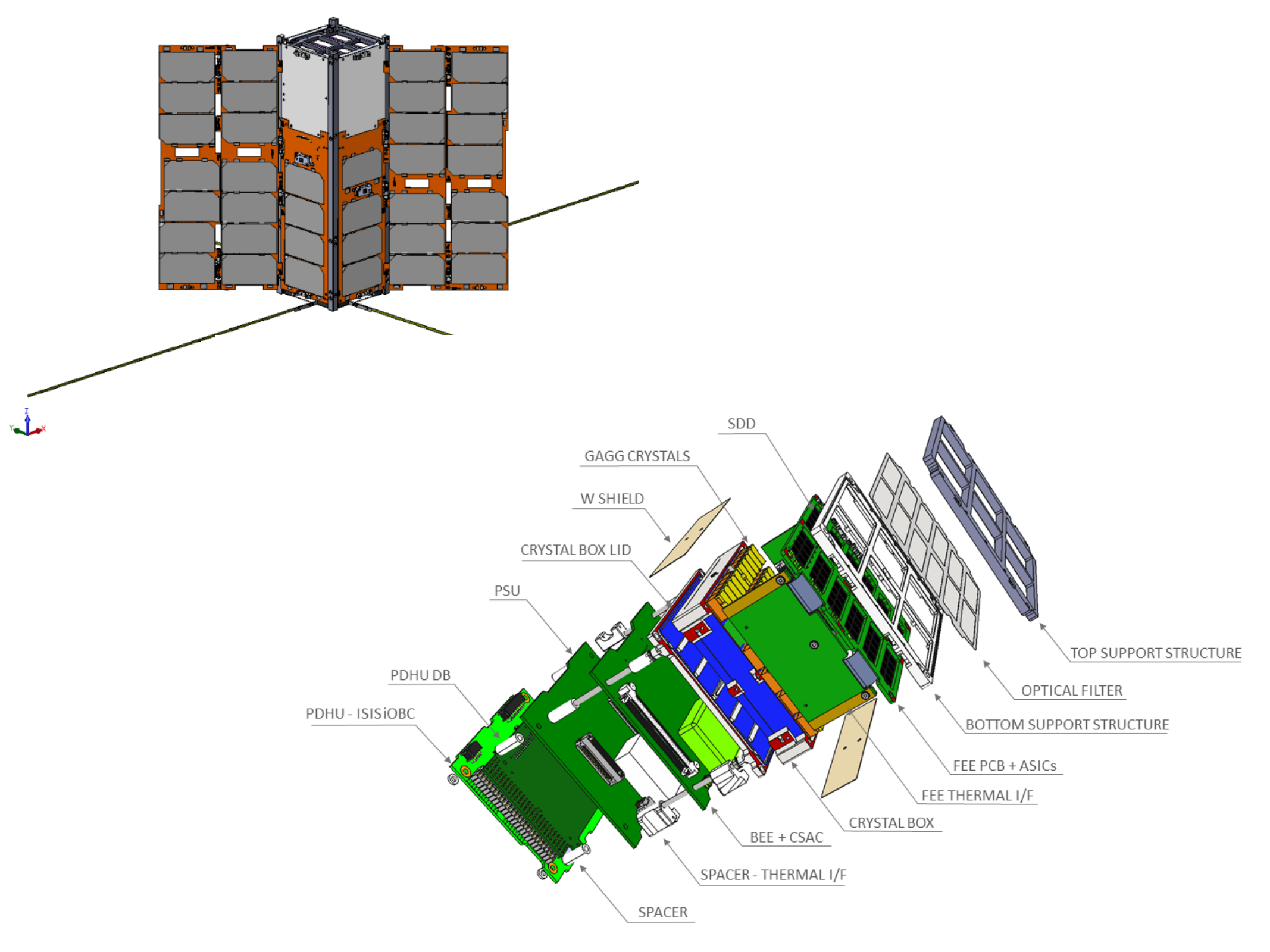}    
    \caption{The HERMES 3U CubeSat (upper left) and an exploded view of the payload (which composes the uppermost 1U of the satellite). Adapted from \cite{evangelista24}.}
    \label{fig:hermes}
\end{figure}

Figure~\ref{fig:hermes} shows an exploded view of the satellite design. 
The mass model is implemented in GDML format\footnote{\url{https://gdml.web.cern.ch/GDML/}} and background simulations \cite{campana21} of the non-activation sources \cite{campana22} have shown that the dominant background component is expected to be the cosmic diffuse X-ray background (CXB) up to energies of a few hundreds of keV, while particle-induced background becomes important for higher energies.

Although designed to operate in an equatorial, low Earth orbit, the constellation has been launched on a 515~km altitude Sun-synchronous orbit (SSO). 
For the HERMES mission, therefore, two possible scenarios are simulated: $(a)$ a 600 km, 5$^\circ$ inclination LEO, and $(b)$ a 550 km SSO (97$^\circ$ inclination). Note that the altitude in the latter scenario has been conservatively chosen significantly higher than the actual launch altitude.
For both scenarios, we consider two radiation models for the trapped proton population: the ``classic'' AP8 model evaluated at solar minimum (AP8MIN), and the ``new'' AP9 model, evaluated for a 90th percentile\footnote{In the AP9 trapped radiation model, percentiles represent statistical confidence levels that describe the variability and uncertainty in the predicted proton flux environment. The model generates flux estimates based on a large ensemble of simulated magnetic field and particle transport conditions. Lower percentiles (e.g., 5th) correspond to quieter space weather conditions with lower flux levels, indicating conservative or less severe radiation environments. Higher percentiles (e.g., 90th) indicate increasingly extreme radiation levels, useful for assessing absolute worst-case or design-limit scenarios.} and for ``mean'' results. In each case a 30-days long orbit starting from January 1st, 2025, sampled at one minute intervals, has been simulated using SPENVIS. The effect of atmospheric drag and orbital decay has been neglected, while the orbital perturbations have been propagated using standard algorithms. 

\begin{figure}
    \centering
    \includegraphics[width=0.7\textwidth]{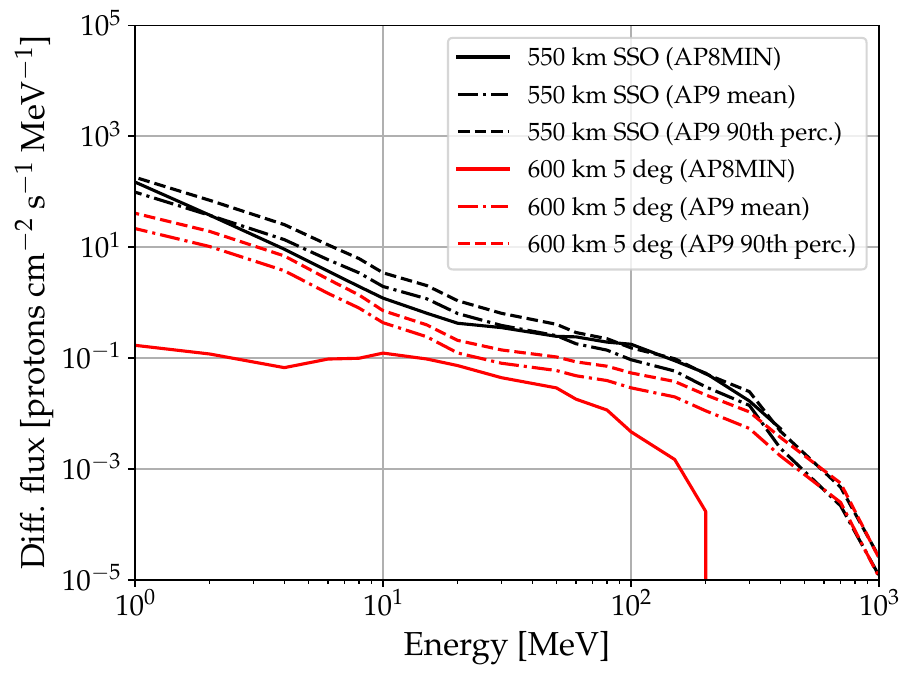}
    \caption{Orbit-averaged proton differential spectrum for the two orbital scenarios (LEO at 600 km altitude, 5$^\circ$ inclination or 550 km altitude Sun-synchronous orbit) for the AP8MIN and AP9 mean and 90th percentile models. }
    \label{fig:model_fluxes}
\end{figure}

Figure \ref{fig:model_fluxes} shows the orbit-averaged differential proton spectra predicted by these models. The difference between the prediction of the two trapped proton models is dramatic for low orbital inclinations, less so for polar orbits. It should be emphasized that the AP9 90th percentile model should be considered as an \emph{absolute worst case} from a simulation standpoint, and has been shown to be inaccurate at low orbital inclinations \cite{ripa21}.

\begin{figure}
    \centering
    \includegraphics[width=0.7\textwidth]{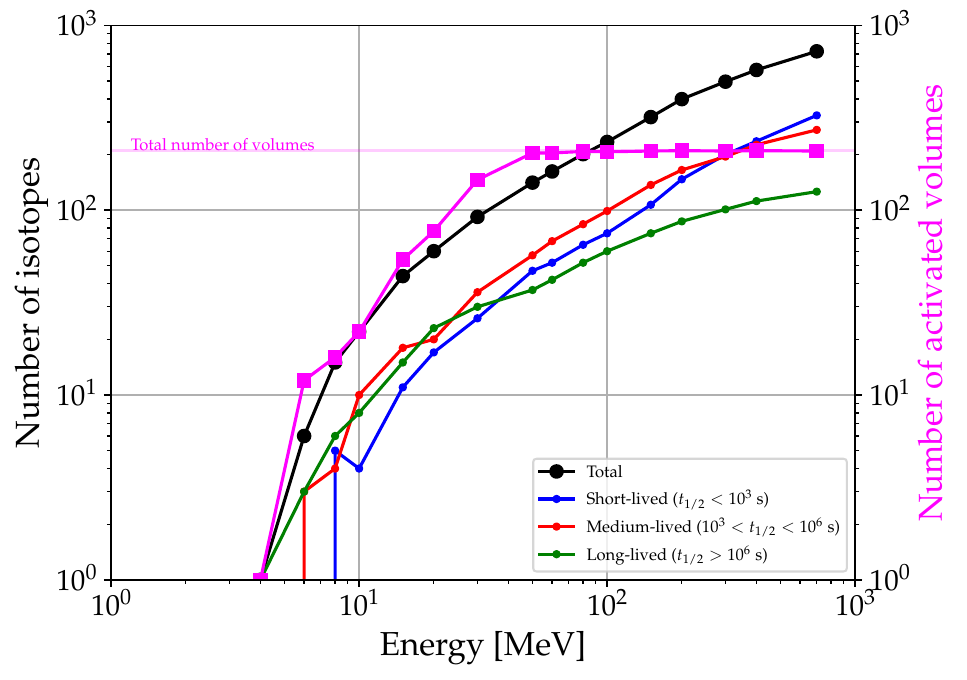} 
    \caption{Number of unstable isotopes created upon an irradiation of the HERMES mass model with different proton energies (left vertical axis) and number of volumes in the mass model which have at least one generated radioisotope (right vertical axis). The total number of different volumes in the GDML HERMES mass model employed is also shown.}
    \label{fig:hermes_n_iso_n_vol}
\end{figure}

Figure~\ref{fig:hermes_n_iso_n_vol} shows the output of the first step of the activation algorithm (Section~\ref{s:step1}). A minimum energy of about 60 MeV is necessary to reach and induce activation in every volume of the mass model (i.e., to reach also the innermost sectors of the detector or of the spacecraft), and the total number of isotopes which are generated is dramatically dependent on the energy, reaching up to the level of several hundreds of different isotopes.

\begin{figure}
    \centering
    \includegraphics[width=0.49\textwidth]{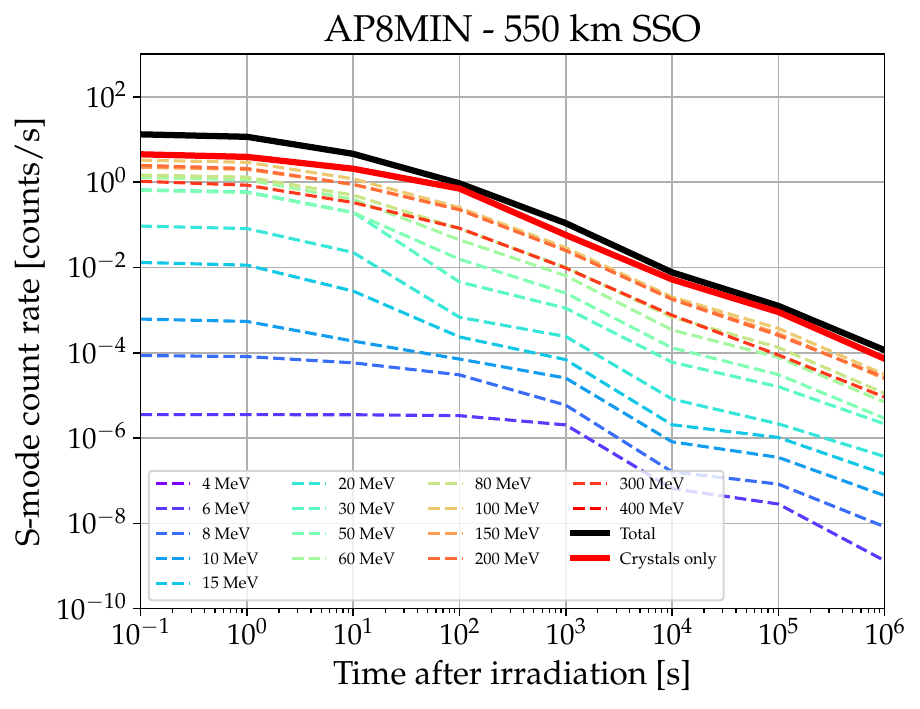} 
    \includegraphics[width=0.49\textwidth]{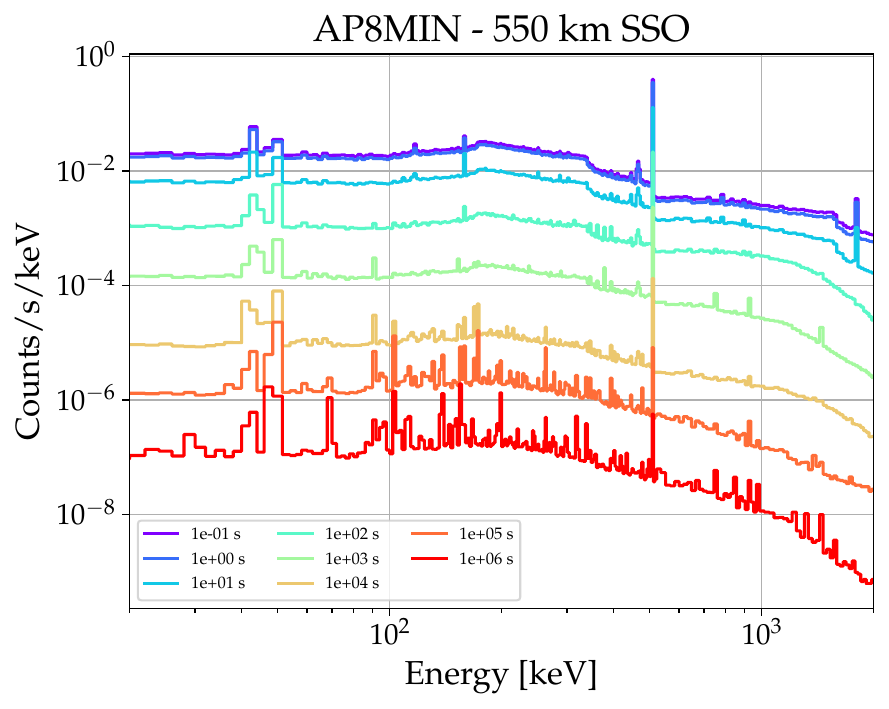} 
    \caption{\emph{Left panel}: Expected count rate in HERMES (S-mode, 20--2000~keV band) as a function of time, after an instantaneous irradiation with the orbit-averaged trapped proton spectrum predicted by the AP8MIN model for a 550~km altitude SSO. Besides the contribution of each input energy bin to the count rate, the plot shows the total count rate and the individual contribution of the GAGG:Ce scintillator crystals.  \emph{Right panel}: S-mode activation induced spectra spanning 0.1~s to 1~Ms after irradiation. Note the different lines present at different times. Spectra are not convoluted with the intrinsic detector spectral resolution (which is $\sim$5\% at 662 keV, see \cite{dilillo24}).}
    \label{fig:hermes_time_spectra}
\end{figure}

The left panel of Figure~\ref{fig:hermes_time_spectra} shows, as an example, the expected count rate (evaluated after simulation steps 2 and 3, Section~\ref{s:step2} and \ref{s:step3}), as a function of time, for an instantaneous irradiation with the AP8MIN orbit-averaged proton spectrum for a 550~km altitude SSO. Results are qualitatively similar for the other models; in particular, it is well apparent how the dominant contribution to the overall activation-induced background rate is due to radioisotopes generated inside the GAGG:Ce scintillator crystal themselves, at least for mid- to long-half lives ($t>100$~s). An analysis of the background spectra for several times after irradiation (Figure~\ref{fig:hermes_time_spectra}, right panel) clearly shows that different isotopes (with different characteristic $\gamma$-ray lines) dominate at different times. The 511~keV annihilation line is anyway the most prominent feature at all times.

\begin{figure}
    \centering
    \includegraphics[width=0.49\textwidth]{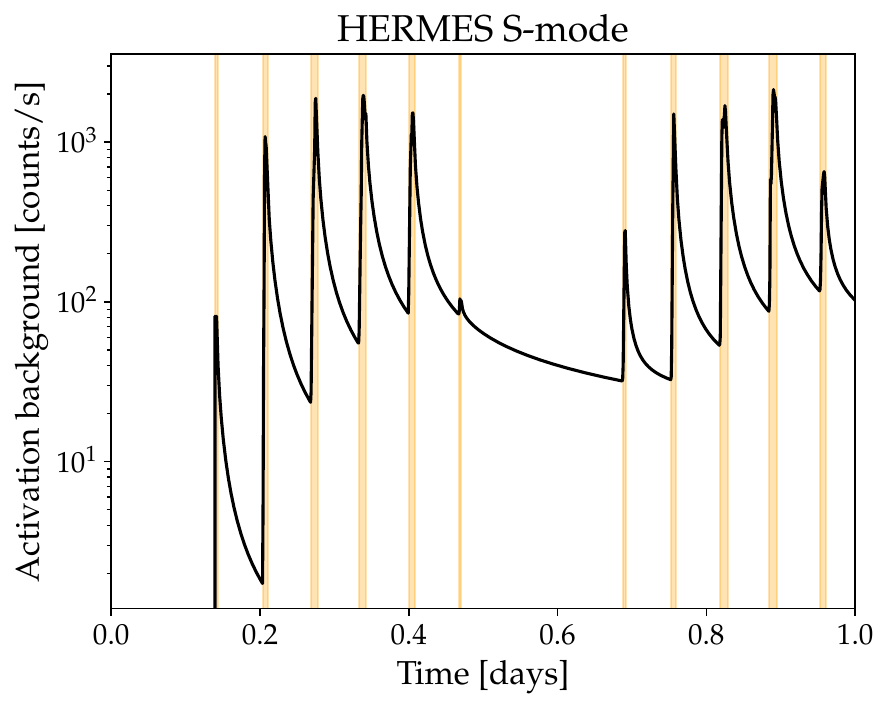} 
    \includegraphics[width=0.49\textwidth]{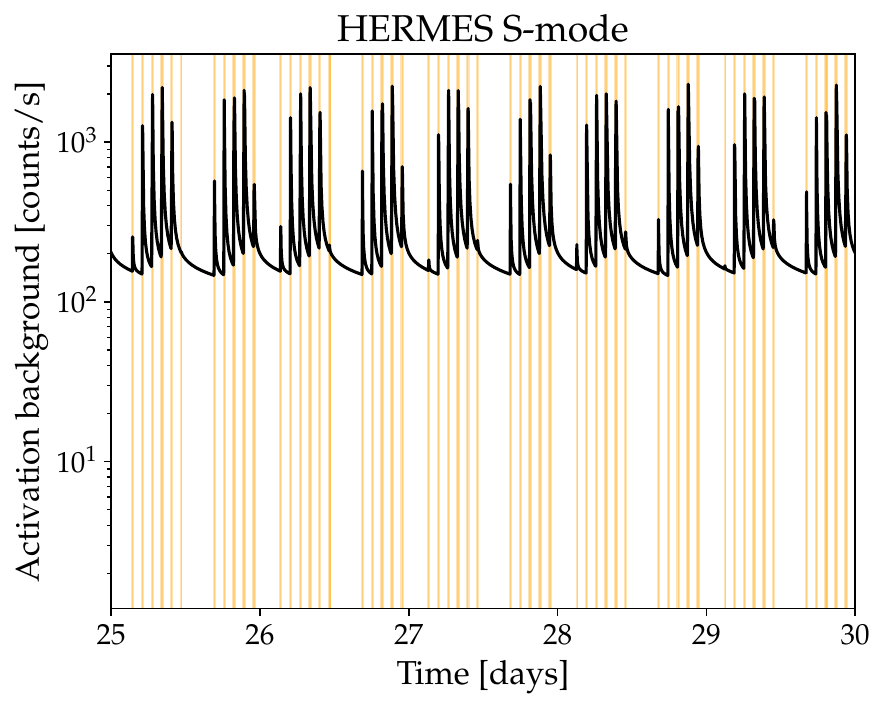} 
    \caption{\emph{Left panel}: HERMES S-mode (20--2000~keV) activation-induced background rate as a function of the time, for the first day in orbit. \emph{Right panel}: Activation-induced background rate after 25 to 30 days in orbit. The vertical shaded bands mark the SAA crossing periods. The curves shown are computed using a 550~km altitude SSO and the AP8MIN model.}
    \label{fig:hermes_orbit}
\end{figure}

\begin{figure}
    \centering
    \includegraphics[width=0.7\textwidth]{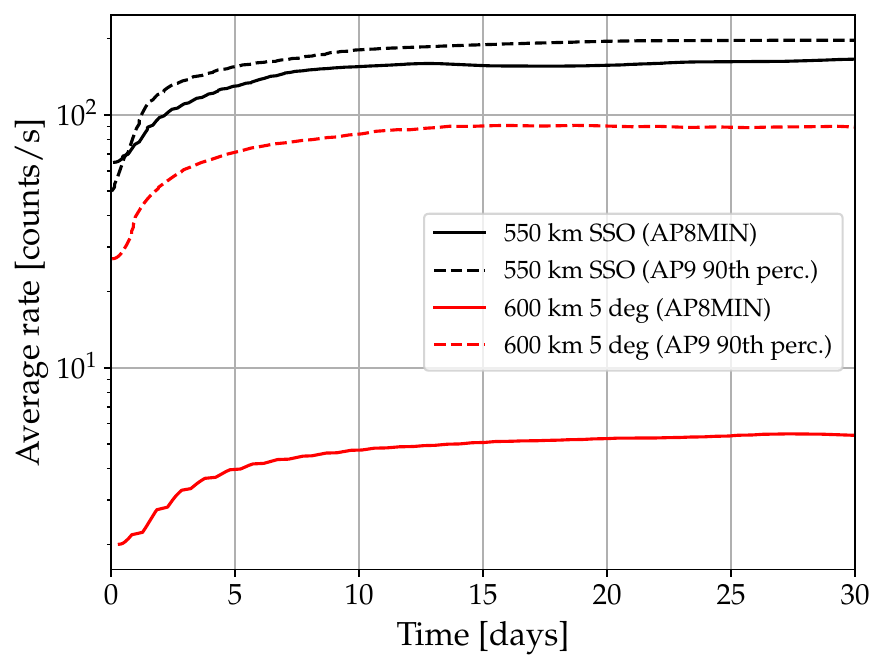}
    \caption{Long term steady-state averaged HERMES S-mode (20--2000~keV) activation-induced count rates for the two orbital scenarios (LEO at 600 km altitude, 5$^\circ$ inclination or 550 km altitude Sun-synchronous orbit) for the AP8MIN and AP9 90th percentile models. }
    \label{fig:avg_rates}
\end{figure}

Figure~\ref{fig:hermes_orbit} shows the activation-induced background count rate during the first day in orbit (left) and after 25--30 days (right). For a SSO, each day the satellite experiences two sets of irradiation events corresponding to the ascending and descending SAA transits, which differ in duration and intensity. The decay of short-lived isotopes causes a rapid drop (by one to two orders of magnitude) in the background rate, within minutes after each transit, while the accumulation of long-lived isotopes leads to a gradual increase that stabilizes after a few weeks.
Figure~\ref{fig:avg_rates} shows the long-term average of the ``steady-state'' background (i.e., the one induced by the accumulation of long-lived isotopes well outside SAA transits) as a function of the time for the four different scenarios. 
The resulting count rates can be compared with the expected non-activation background \cite{campana21}, which, depending on the orbital inclination, averages between 180 and 360 counts/s in the S-mode band (20--2000 keV). This shows that activation-induced background is a minor contributor to the overall background for HERMES in an equatorial orbit \cite{galgoczi2020,dellacasa24} but becomes significant at higher inclinations or in polar orbits.

The activation contribution for the X-mode band (3--50~keV), given the low interaction cross section on the thin (450~µm) silicon bulk of the SDDs, is on the contrary absolutely negligible with respect to the other background components (such as the diffuse X-ray emission collected in the large field of view).

\section{Case study II: eXTP}\label{s:extp}

\begin{figure}
    \centering
    \includegraphics[width=0.7\textwidth]{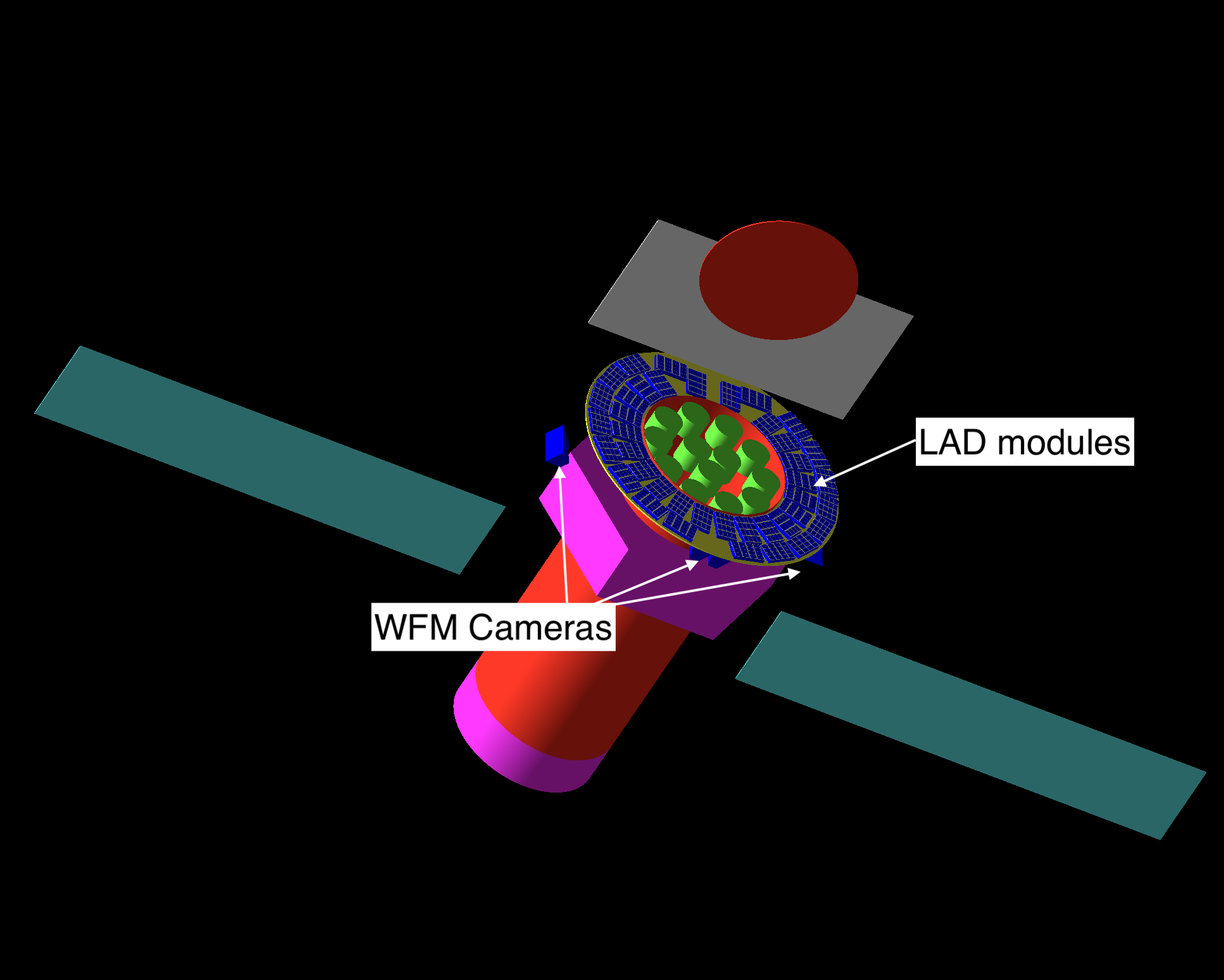} 
    \caption{The eXTP mass model implemented in Geant4, showing the location of the 40 LAD modules and of the 6 WFM cameras.}
    \label{fig:extp}
\end{figure}

\subsection{LAD}

The \textit{Large Area Detector} (LAD, \cite{feroci24}) is the high-throughput, spectral-timing instrument proposed for inclusion in the \emph{enhanced X-ray Timing and Polarimetry} (eXTP) mission \cite{zhang19}. Conceived as a European contribution to the Chinese Academy of Sciences flagship astrophysics project, the LAD is designed to probe matter under extreme gravitational, magnetic, and density conditions. Derived from the original LOFT mission concept \cite{feroci12}, the LAD combines a large effective area with fine spectral resolution to enable unprecedented spectral-timing studies of bright X-ray sources. The LAD underwent a Phase B study in 2018--22, but due to strategic and programmatic decisions is not currently foreseen for inclusion in the final flight configuration of the eXTP mission \cite{zhang25}.

The LAD is a collimated, modular detector array. Its nominal energy range is 2--30~keV (extendable up to 80~keV), providing an effective area exceeding 3~m$^2$ at 8~keV, the largest ever deployed in space for this energy band. The instrument achieves an energy resolution better than 260~eV (FWHM at 6~keV), a 1$^{\circ}$ field of view, and 10~µs timing resolution, with $<$1\% dead time at X-ray fluxes around 1~Crab (which corresponds to a total rate of $\sim$150,000 counts/s in 2--30~keV). 

The LAD structure is organized into 40 co-aligned Detector Modules (Figure~\ref{fig:extp}). Each module hosts 16 large-area Silicon Drift Detectors (SDDs), 16 lead-glass capillary plate collimators, and corresponding optical filters. The modules interface with four Panel Back-End Electronics (PBEE) units, which manage data collection, formatting, and transmission to the Instrument Control Unit (ICU). The ICU provides data handling, command, and power distribution, employing cold redundancy for reliability.

Each Detector Module consists of a \textit{Detector Tray} and a \textit{Collimator Tray}. The Detector Tray houses the SDDs, front-end electronics, and power supply units, with an integrated lead shield (300~µm thickness) and thermal radiator to control background and temperature. The Collimator Tray contains lead-glass capillary plates, 5~mm thick with 83~µm round pores, ensuring a 1$^{\circ}$ field of view and strong X-ray attenuation outside this range. Optical filters made from 1~µm polyimide films coated with 200~nm aluminum suppress UV, visible and infrared light while maintaining $>$90\% transmission at 2~keV. 

During Phase~B, the LAD consortium, led by INAF and supported by the Italian Space Agency and several European partners, developed a full-scale \textit{Demonstration Model} (DM) to validate technologies and integration procedures without U.S.-restricted components. The DM includes representative versions of all key subsystems: ICU, PBEE, Module Back-End Electronics (MBEE), Module Power Supply Units (MPSU), Detector Assemblies (DA), collimators, and filters. Bench integration at Thales Alenia Space (Italy) validated end-to-end functionality, demonstrating a Technology Readiness Level up to 5--6 \cite{feroci24}.

The eXTP mission is designed to fly at a quasi-equatorial LEO. As a baseline, for the background studies a 600~km altitude, 5$^\circ$ inclination orbit is assumed. In this orbit, the overall instrumental background is dominated by the cosmic X-ray diffuse emission collected in the field of view, and leaking through the lead-glass collimator plate \cite{campana13}. 

\begin{figure}
    \centering
    \includegraphics[width=0.7\textwidth]{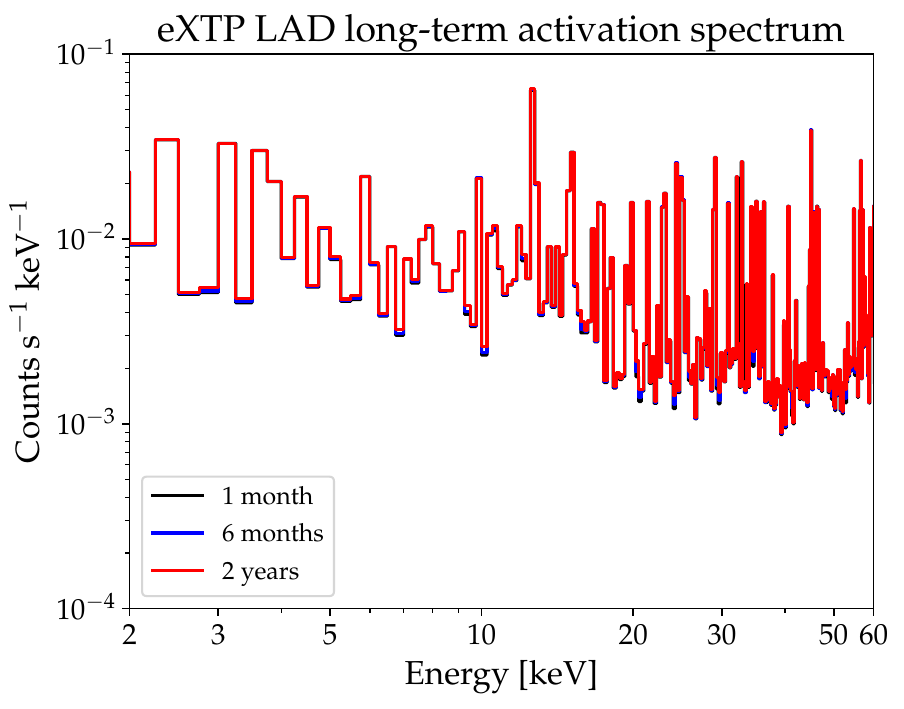} 
    \caption{eXTP LAD long-term activation-induced spectrum after 1 month, 6 months and 2 years in orbit, computed using the AP9 90th percentile trapped proton model. In order to compute the accumulation of long-lived isotopes and thus the long-term component of the activation background, the irradiation history in this case has been simplified as a sequence of instantaneous irradiations, one for each day elapsed in orbit, with a fluence corresponding to the daily orbit-averaged one.}
    \label{fig:eXTP_LAD_longterm_spectrum}
\end{figure}

The activation contribution to the overall background, as determined using the three-step algorithm applied to the Geant4 LAD mass model shown in Figure~\ref{fig:extp}, is found to be a significantly small fraction of the overall background. According to the trapped proton model used (AP8MIN up to AP9 90th percentile), the total activation-induced count rate ranges (between SAA transits) between 0.02\% to 2\% of the overall expected background \cite{campana13}. 
A significant fraction of the total activation originates from the lead contained in the glass collimator in front of the SDDs, and in the backshield on the bottom side of the module.
Accumulation of long-lived isotopes is reached pretty quickly once in orbit, achieving an almost steady-state residual activation-induced background level. 
The spectrum of this long-term activation component can be computed on time-scales of months-years. The irradiation profile can be simplified as a sequence of instantaneous irradiations, one for each day elapsed in orbit, with a fluence corresponding to the daily orbit-averaged one, and allowing for a suitable ``cool-down'' period to remove the short-lived isotopes. The results are shown in Figure~\ref{fig:eXTP_LAD_longterm_spectrum}: the spectra are rather flat, with some features arising from low-energy X-ray (fluorescence) and $\gamma$-ray lines, for example the Pb $L_\alpha$ and $L_\beta$ lines around 10.5--12.6~keV.

\subsection{WFM}

The Wide Field Monitor (WFM) is the other European-developed instrument originally foreseen for the eXTP mission. Designed to operate in the 2--50~keV band, the WFM provides wide-field X-ray imaging with arcminute-level source localization, thereby enabling real-time identification of transient and variable sources and delivering rapid triggers for follow-up observations with the narrow-field instruments (SFA, PFA, and LAD). 

The primary scientific objectives of the WFM include the detection of new and recurrent X-ray transients, monitoring the long-term evolution of bright X-ray sources, and providing rapid and accurate localizations of gamma-ray bursts (GRBs), gravitational-wave electromagnetic counterparts, and other fast high-energy phenomena. The instrument is designed to achieve a point-source localization accuracy better than 1~arcmin, an angular resolution of approximately 5~arcmin, and sensitivities down to a few mCrab for exposures of the order of tens of kiloseconds. High temporal resolution (10~µs for event data) and a very large instantaneous field of view (greater than 3~sr) enable continuous monitoring of roughly one-third of the sky.

The WFM consists of six identical coded-mask cameras arranged in three orthogonal pairs (Figure~\ref{fig:extp}), allowing full two-dimensional imaging through the combination of perpendicular one-dimensional projections. Each camera provides a field of view of $90^\circ \times 90^\circ$ at zero response and $30^\circ \times 30^\circ$ fully coded; together, the six cameras cover $180^\circ \times 90^\circ$ and are oriented to maximize overlap with the sky accessible to the narrow-field instruments.

Each camera employs large-area Silicon Drift Detectors (SDDs) with an anode pitch of 169~µm, providing asymmetric spatial resolution, better than 60~µm along the anode direction and better than 8~mm along the drift direction. These properties are matched to a tungsten coded mask with a 250~µm pitch and an effective open fraction of 21\,\%. The mask is tensioned within a mechanical frame to maintain flatness and is temperature-stabilized by a dedicated thermal foil. A carbon-fiber reinforced polymer collimator, radiation-shielded with tungsten, and coated with thin layers of molybdenum and copper to exploit their fluorescence lines as on-board calibrators, ensures mechanical stability during launch and operation.

The front-end electronics (FEE) perform the analog readout of the SDDs, while the back-end electronics (BEE) handle digitization, charge-cloud fitting, noise suppression, event reconstruction, and time tagging. Two redundant Instrument Control Units (ICUs) oversee configuration, on-board time management, image accumulation, and burst triggering.

Following a successful Instrument Systems Requirements Review in 2023, the WFM entered an advanced phase of subsystem demonstration. Full-size SDD detectors, coded masks, detector mechanical structures, and mock-up assemblies have been manufactured and tested, confirming the feasibility of the proposed architecture. Although the WFM is not part of the current Chinese-approved eXTP baseline, its design has reached a high level of maturity and remains under active development within the European consortium.

\begin{figure}
    \centering
    \includegraphics[width=0.49\textwidth]{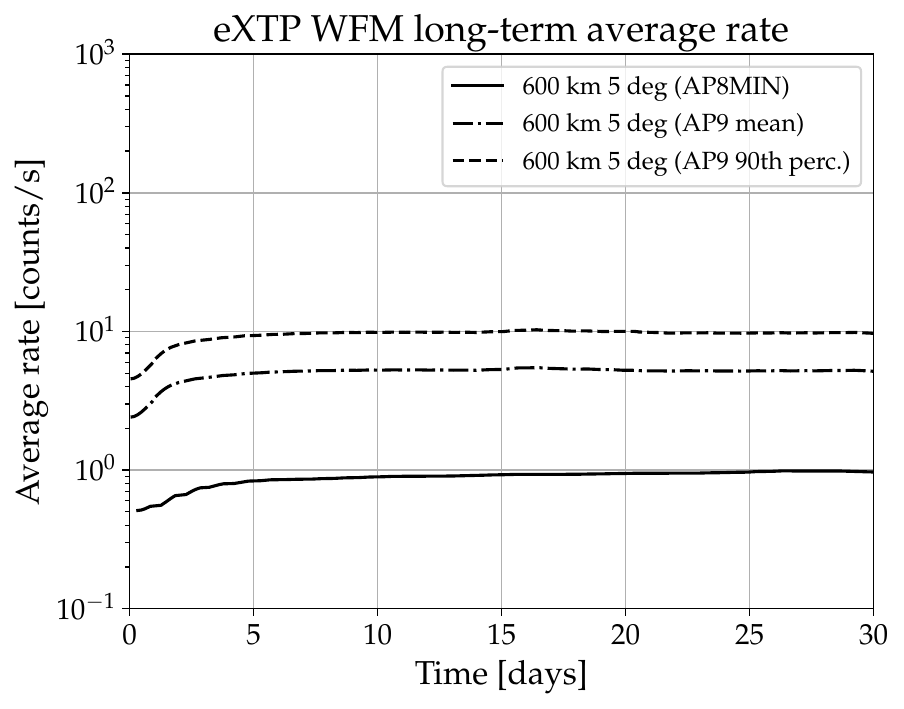}     \includegraphics[width=0.49\textwidth]{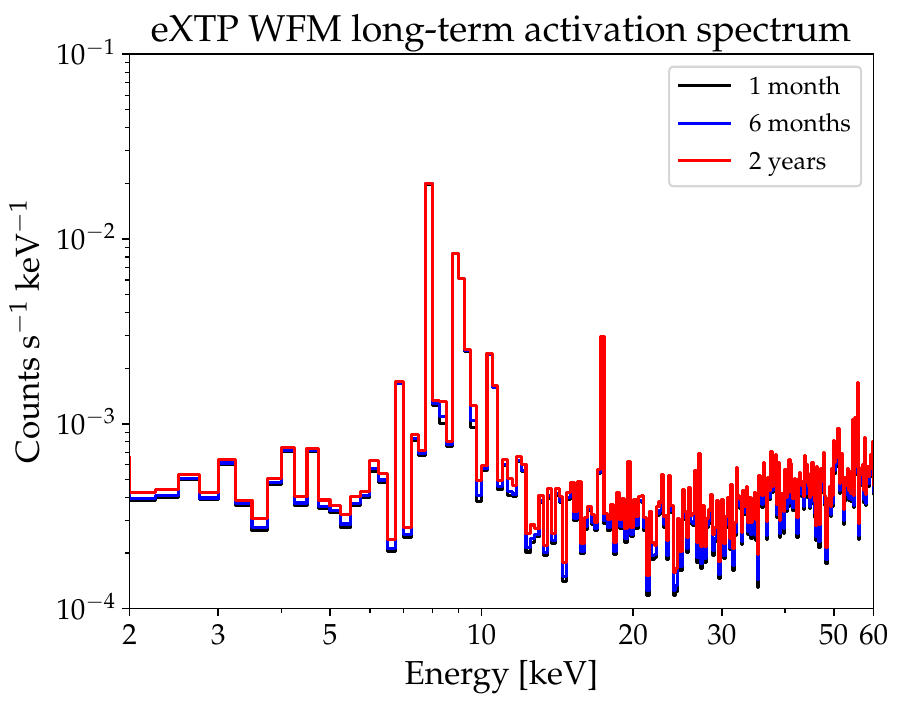} 
    \caption{Left panel: Average long-term (outside SAA transits) activation induced background count rate for one WFM camera, as a function of the time elapsed in orbit, and for three different trapped proton models. Right panel: Long-term activation spectrum for 1 month, 6 months and 2 years in orbit.}
    \label{fig:extp_WFM_sim_res}
\end{figure}

Some representative simulation results are shown in Figure~\ref{fig:extp_WFM_sim_res}. Most of the induced activation in WFM is due to the tungsten in the collimators and in the coded mask, but the overall effect, even for the worst-case models, is significantly smaller than the other sources of background (most notably, the diffuse X-ray background radiation collected in the large WFM field of view). From a spectral standpoint, the counts are rather flat in the 2--30~keV nominal energy range, and the most noteworthy effect of the activation is to enhance the X-ray fluorescence lines of Cu ($K_\alpha$ and $K_\beta$ at 8.05 and 8.90~keV) Mo (17.48 and 19.61~keV) and W ($L_\alpha$ and $L_\beta$ lines at 8.40 and 9.67~keV).

\section{Conclusions}
In this work we have presented a new implementation of the efficient three-step algorithm for the simulation of proton-induced activation and the resulting delayed instrumental background in high-energy astrophysics missions operating in low-Earth orbit. By decoupling the production of radioactive isotopes, the time evolution of their decay chains, and the synthesis of the detector response to their decay products, the method overcomes the intrinsic inefficiencies of direct Monte Carlo simulations, which require extremely large statistics to properly sample both rare activation events and long-lived decay processes.

The proposed framework combines: \emph{(i)} Geant4-based simulations to build a comprehensive database of radionuclide production as a function of proton energy and mass model individual volume; \emph{(ii)} a numerical solution of the Bateman equations, including the treatment of branched decay chains through linearization and weighting; and \emph{(iii)} dedicated simulations of the detector response to each isotope decay in each relevant volume. Once these elements are computed, the activation-induced background for arbitrary irradiation histories, proton spectra, and time profiles can be reconstructed. This makes the approach particularly well suited for design trade-off studies, background budgeting, and sensitivity analyses in the early and intermediate phases of mission development.

The validation against direct Geant4 simulations demonstrates that the three-step algorithm reproduces the time evolution of the activity with excellent accuracy over several orders of magnitude in time and activity. The agreement holds for different materials, proton energies, and choices of Geant4 hadronic physics lists, confirming the robustness of the approach. At the same time, the computational gain is substantial: once the databases are built, the synthesis of background spectra and count rates for new orbital scenarios or irradiation models becomes essentially instantaneous when compared to full end-to-end simulations.

The application to realistic and complex instruments highlights the versatility of the method. For the HERMES CubeSat payloads, it is found that activation-induced background is a minor contributor in equatorial or low-inclination orbits, but can become significant in Sun-synchronous or polar configurations, especially when conservative trapped-proton models such as AP9 at high percentiles are adopted. The simulations clearly show the interplay between short-lived isotopes, which dominate the overall background immediately after SAA transits, and long-lived isotopes, which build up to a steady-state level on timescales of weeks. For the specific case of HERMES, activation is largely dominated by radionuclides produced directly in the GAGG:Ce scintillators, while the contribution in the low-energy direct SDD X-ray detection mode ($\sim$2--30~keV) remains negligible.

For the eXTP mission, both for the LAD and the WFM instruments, the results indicate that activation-induced background represents only a small fraction of the total background in the baseline quasi-equatorial orbit. In the LAD, even under worst-case assumptions for the trapped proton environment, activation contributes at most a few percent of the total background, confirming that other components, such as the cosmic diffuse X-ray emission leaking through the collimators, remain dominant. In the WFM, activation is mainly associated with heavy materials such as W, Mo, and Cu in the coded mask and collimators, and manifests primarily as an enhancement of fluorescence and nuclear lines superimposed on an otherwise flat continuum. In both cases, the accumulation of long-lived isotopes reaches a steady state relatively quickly, and the long-term activation spectrum remains stable on mission timescales.

Beyond the specific case studies presented here, the methodology is general and can be readily applied to other instruments, orbital configurations, and radiation environments. Its modular structure allows straightforward updates when new trapped-particle models, improved cross-section databases, or refined Geant4 physics lists become available. In addition, the separation between physics processes and mission-specific irradiation histories enables fast comparisons between different design options, such as alternative materials, shielding strategies, or orbital choices.

In conclusion, the three-step algorithm provides a powerful and flexible tool for the quantitative assessment of activation-induced background in space-borne high-energy detectors. It achieves a favorable compromise between physical accuracy and computational efficiency, making it suitable not only for detailed background studies of existing missions but also for rapid feasibility analyses and optimization studies of future X-ray and $\gamma$-ray experiments in low-Earth orbit.

\section*{Acknowledgements}
Partial support from the Italian Space Agency (under agreements n. 2020-3-HH.0, 2020-3-HH.1-2021, ASI-INAF n.2017-14-H.O) and INFN (project XRO, X-ray Observatories) is acknowledged. The author is grateful to Gábor Galgóczi for several in-depth discussions on the topic of activation simulations, and to Yuri Evangelista and Edoardo Borciani for many useful suggestions.

\begin{appendices}

\section{The Bateman equations of radioactive decay}\label{a:bateman}

Consider a simple decay scheme in which an isotope $X_1$ decays into a stable isotope $X_2$:
\begin{equation}
X_1 \rightarrow X_2 \mbox{\,\, (+ other products)}
\end{equation}
If we assume that at time $t=0$ we don't have any daughter element, $N_2(t=0) = 0$, then:
\begin{equation}\label{e:simpledecay}
\frac{dN_1}{dt} = -\lambda N_1
\end{equation}
where $\lambda$ is the \emph{decay constant}. The solution of the simple radioactive decay equation is:

\begin{equation}
    N_1(t) = N_1(0)e^{-\lambda t} 
\end{equation}

The population of isotopes of type $X_1$ decays exponentially with an \emph{half-life} $t_{1/2} = \log(2)/\lambda$. The \emph{activity} of the isotope $X_1$ (number of disintegrations per seconds, which in the SI is expressed in becquerels, Bq) is $A_1 = \lambda N_1$.

If we have a \emph{decay chain}, in which the isotope $X_1$ decays in an unstable isotope $X_2$, which in turn decays into a stable isotope $X_3$:
\begin{equation}
X_1 \rightarrow X_2 \mbox{\,\, (+ other products)} \rightarrow X_3 \mbox{\,\, (+ other products)}
\end{equation}
and assuming that at $t=0$ the population of daughters is void, i.e. $N_2(0)=N_3(0)=0$, the previous decay equation \ref{e:simpledecay} generalizes into a system of equations:

\begin{equation}
\begin{aligned}
\left\{
\begin{aligned}
    \frac{dN_1}{dt} &= -\lambda_1 N_1 \\
    \frac{dN_2}{dt} &= -\lambda_2 N_2 + \lambda_1 N_1 
\end{aligned}
\right.
\end{aligned}
\end{equation}

The solution is:

\begin{equation}
\begin{aligned}
\left\{
\begin{aligned}
    N_1(t) &= N_1(0)e^{-\lambda t} \\
    N_2(t) &= \frac{N_1(0)\lambda_1}{\lambda_2-\lambda_1}(e^{-\lambda_1 t}-e^{-\lambda_2 t})
\end{aligned}
\right.
\end{aligned}
\end{equation}

Generalizing the previous approach to a decay chain with $k$ elements, we arrive at a system of first-order differential equations, the so-called Bateman equations \cite{bateman1910}:
\begin{equation}\label{e:bateman}
\begin{aligned}
\left\{
\begin{aligned}
    \frac{dN_1}{dt} &= -\lambda_1 N_1 \\
    \frac{dN_j}{dt} &= -\lambda_j N_j + \lambda_{j-1} N_{j-1} \mbox{\,\,\, for $1 < j \le k$}
\end{aligned}
\right.
\end{aligned}
\end{equation}

The general solution can be found by means of a Laplace transform \cite{pressyanov2002}, and is:
\begin{equation}
N_k = \frac{N_1(0)}{\lambda_k} \sum_{i=1}^{k} \lambda_i c_i e^{-\lambda_i t}
\end{equation}
where:
\begin{equation}
c_i = \prod_{j=1,i\neq j}^{k} \frac{\lambda_j}{\lambda_j-\lambda_i}
\end{equation}

However, from a computational point of view, given the presence of the $\lambda_j-\lambda_i$ term in the expression for the coefficient $c_i$, this solution is numerically unstable when two isotopes have very close half-lives ($\lambda_j \approx \lambda_i$) or when they have drastically different half-lives ($\lambda_j \gg \lambda_i$).

A different approach, more numerically robust, is to recast the Bateman equations \ref{e:bateman} in a matrix form \cite{onega1969, moralpacheco2003, amaku2010, hauf13}:
\begin{equation}
\frac{d\mathbf{N}}{dt} = \Lambda \mathbf{N}
\end{equation}
where $\mathbf{N}$ is the vector
\begin{equation}
\mathbf{N} =
\begin{bmatrix}
    N_1(t) \\
    N_2(t) \\
    \vdots \\
    N_k(t)
\end{bmatrix}
\end{equation}
and $\Lambda$ is a bidiagonal triangular matrix constructed from the decay constants:
\begin{equation}
    \Lambda =
\begin{bmatrix}
    -\lambda_1 & 0          & 0          & \cdots & 0 \\
    \lambda_1  & -\lambda_2 & 0          & \cdots & 0 \\
    0          & \lambda_2  & -\lambda_3 & \cdots & 0 \\
    0          & 0          & \lambda_3  & \cdots & 0 \\
    \vdots     & \vdots     & \vdots     & \ddots & \vdots \\
    0          & 0          & 0          & \cdots & -\lambda_k
\end{bmatrix}
\end{equation}

The solution of the matrix equation is:
\begin{equation}
\mathbf{N}(t) = C e^{Dt} C^{-1}\mathbf{N}(0)
\end{equation}
where $C$ is the (column) matrix of the eigenvectors of $\Lambda$ and $D$ is a diagonal matrix whose elements are the eigenvalues of $\Lambda$. These are easily computable with numerically stable libraries (e.g., using the \texttt{linalg} Python package), or by the recurrence expressions found by Amaku et al. \cite{amaku2010}. This matricial approach can be easily extended to consider also isotope activation production rates (e.g., from proton or neutron continuous irradiation) and/or branched decay chains, by means of an opportune redefinition of the $\Lambda$ matrix.

\end{appendices}

\bibliography{bibliography}


\begin{thebibliography}{35}
\ifx \bisbn   \undefined \def \bisbn  #1{ISBN #1}\fi
\ifx \binits  \undefined \def \binits#1{#1}\fi
\ifx \bauthor  \undefined \def \bauthor#1{#1}\fi
\ifx \batitle  \undefined \def \batitle#1{#1}\fi
\ifx \bjtitle  \undefined \def \bjtitle#1{#1}\fi
\ifx \bvolume  \undefined \def \bvolume#1{\textbf{#1}}\fi
\ifx \byear  \undefined \def \byear#1{#1}\fi
\ifx \bissue  \undefined \def \bissue#1{#1}\fi
\ifx \bfpage  \undefined \def \bfpage#1{#1}\fi
\ifx \blpage  \undefined \def \blpage #1{#1}\fi
\ifx \burl  \undefined \def \burl#1{\textsf{#1}}\fi
\ifx \doiurl  \undefined \def \doiurl#1{\url{https://doi.org/#1}}\fi
\ifx \betal  \undefined \def \betal{\textit{et al.}}\fi
\ifx \binstitute  \undefined \def \binstitute#1{#1}\fi
\ifx \binstitutionaled  \undefined \def \binstitutionaled#1{#1}\fi
\ifx \bctitle  \undefined \def \bctitle#1{#1}\fi
\ifx \beditor  \undefined \def \beditor#1{#1}\fi
\ifx \bpublisher  \undefined \def \bpublisher#1{#1}\fi
\ifx \bbtitle  \undefined \def \bbtitle#1{#1}\fi
\ifx \bedition  \undefined \def \bedition#1{#1}\fi
\ifx \bseriesno  \undefined \def \bseriesno#1{#1}\fi
\ifx \blocation  \undefined \def \blocation#1{#1}\fi
\ifx \bsertitle  \undefined \def \bsertitle#1{#1}\fi
\ifx \bsnm \undefined \def \bsnm#1{#1}\fi
\ifx \bsuffix \undefined \def \bsuffix#1{#1}\fi
\ifx \bparticle \undefined \def \bparticle#1{#1}\fi
\ifx \barticle \undefined \def \barticle#1{#1}\fi
\bibcommenthead
\ifx \bconfdate \undefined \def \bconfdate #1{#1}\fi
\ifx \botherref \undefined \def \botherref #1{#1}\fi
\ifx \url \undefined \def \url#1{\textsf{#1}}\fi
\ifx \bchapter \undefined \def \bchapter#1{#1}\fi
\ifx \bbook \undefined \def \bbook#1{#1}\fi
\ifx \bcomment \undefined \def \bcomment#1{#1}\fi
\ifx \oauthor \undefined \def \oauthor#1{#1}\fi
\ifx \citeauthoryear \undefined \def \citeauthoryear#1{#1}\fi
\ifx \endbibitem  \undefined \def \endbibitem {}\fi
\ifx \bconflocation  \undefined \def \bconflocation#1{#1}\fi
\ifx \arxivurl  \undefined \def \arxivurl#1{\textsf{#1}}\fi
\csname PreBibitemsHook\endcsname

\bibitem[\protect\citeauthoryear{{Campana}}{2022}]{campana22}
\begin{bbook}
\bauthor{\bsnm{{Campana}}, \binits{R.}}:
In: \beditor{\bsnm{{Bambi}}, \binits{C.}},
\beditor{\bsnm{{Santangelo}}, \binits{A.}} (eds.)
\bbtitle{{In-Orbit Background for X-Ray Detectors}},
p. \bfpage{39}.
\bpublisher{{Springer Nature}},
\blocation{{Singapore}}
(\byear{2022}).
\doiurl{10.1007/978-981-16-4544-0_28-1}
\end{bbook}
\endbibitem

\bibitem[\protect\citeauthoryear{{Agostinelli} et~al.}{2003}]{agostinelli2003}
\begin{barticle}
\bauthor{\bsnm{{Agostinelli}}, \binits{S.}},
\bauthor{\bsnm{{Allison}}, \binits{J.}},
\bauthor{\bsnm{{Amako}}, \binits{K.}},
\bauthor{\bsnm{{Apostolakis}}, \binits{J.}},
\bauthor{\bsnm{{Araujo}}, \binits{H.}},
\bauthor{\bsnm{{Arce}}, \binits{P.}},
\bauthor{\bsnm{{Asai}}, \binits{M.}},
\bauthor{\bsnm{{Axen}}, \binits{D.}},
\bauthor{\bsnm{{Banerjee}}, \binits{S.}},
\bauthor{\bsnm{{Barrand}}, \binits{G.}},
\bauthor{\bsnm{{Behner}}, \binits{F.}},
\bauthor{\bsnm{{Bellagamba}}, \binits{L.}},
\bauthor{\bsnm{{Boudreau}}, \binits{J.}},
\bauthor{\bsnm{{Broglia}}, \binits{L.}},
\bauthor{\bsnm{{Brunengo}}, \binits{A.}},
\bauthor{\bsnm{{Burkhardt}}, \binits{H.}},
\bauthor{\bsnm{{Chauvie}}, \binits{S.}},
\bauthor{\bsnm{{Chuma}}, \binits{J.}},
\bauthor{\bsnm{{Chytracek}}, \binits{R.}},
\bauthor{\bsnm{{Cooperman}}, \binits{G.}},
\bauthor{\bsnm{{Cosmo}}, \binits{G.}},
\bauthor{\bsnm{{Degtyarenko}}, \binits{P.}},
\bauthor{\bsnm{{Dell'Acqua}}, \binits{A.}},
\bauthor{\bsnm{{Depaola}}, \binits{G.}},
\bauthor{\bsnm{{Dietrich}}, \binits{D.}},
\bauthor{\bsnm{{Enami}}, \binits{R.}},
\bauthor{\bsnm{{Feliciello}}, \binits{A.}},
\bauthor{\bsnm{{Ferguson}}, \binits{C.}},
\bauthor{\bsnm{{Fesefeldt}}, \binits{H.}},
\bauthor{\bsnm{{Folger}}, \binits{G.}},
\bauthor{\bsnm{{Foppiano}}, \binits{F.}},
\bauthor{\bsnm{{Forti}}, \binits{A.}},
\bauthor{\bsnm{{Garelli}}, \binits{S.}},
\bauthor{\bsnm{{Giani}}, \binits{S.}},
\bauthor{\bsnm{{Giannitrapani}}, \binits{R.}},
\bauthor{\bsnm{{Gibin}}, \binits{D.}},
\bauthor{\bsnm{{G{\'o}mez Cadenas}}, \binits{J.J.}},
\bauthor{\bsnm{{Gonz{\'a}lez}}, \binits{I.}},
\bauthor{\bsnm{{Gracia Abril}}, \binits{G.}},
\bauthor{\bsnm{{Greeniaus}}, \binits{G.}},
\bauthor{\bsnm{{Greiner}}, \binits{W.}},
\bauthor{\bsnm{{Grichine}}, \binits{V.}},
\bauthor{\bsnm{{Grossheim}}, \binits{A.}},
\bauthor{\bsnm{{Guatelli}}, \binits{S.}},
\bauthor{\bsnm{{Gumplinger}}, \binits{P.}},
\bauthor{\bsnm{{Hamatsu}}, \binits{R.}},
\bauthor{\bsnm{{Hashimoto}}, \binits{K.}},
\bauthor{\bsnm{{Hasui}}, \binits{H.}},
\bauthor{\bsnm{{Heikkinen}}, \binits{A.}},
\bauthor{\bsnm{{Howard}}, \binits{A.}},
\bauthor{\bsnm{{Ivanchenko}}, \binits{V.}},
\bauthor{\bsnm{{Johnson}}, \binits{A.}},
\bauthor{\bsnm{{Jones}}, \binits{F.W.}},
\bauthor{\bsnm{{Kallenbach}}, \binits{J.}},
\bauthor{\bsnm{{Kanaya}}, \binits{N.}},
\bauthor{\bsnm{{Kawabata}}, \binits{M.}},
\bauthor{\bsnm{{Kawabata}}, \binits{Y.}},
\bauthor{\bsnm{{Kawaguti}}, \binits{M.}},
\bauthor{\bsnm{{Kelner}}, \binits{S.}},
\bauthor{\bsnm{{Kent}}, \binits{P.}},
\bauthor{\bsnm{{Kimura}}, \binits{A.}},
\bauthor{\bsnm{{Kodama}}, \binits{T.}},
\bauthor{\bsnm{{Kokoulin}}, \binits{R.}},
\bauthor{\bsnm{{Kossov}}, \binits{M.}},
\bauthor{\bsnm{{Kurashige}}, \binits{H.}},
\bauthor{\bsnm{{Lamanna}}, \binits{E.}},
\bauthor{\bsnm{{Lamp{\'e}n}}, \binits{T.}},
\bauthor{\bsnm{{Lara}}, \binits{V.}},
\bauthor{\bsnm{{Lefebure}}, \binits{V.}},
\bauthor{\bsnm{{Lei}}, \binits{F.}},
\bauthor{\bsnm{{Liendl}}, \binits{M.}},
\bauthor{\bsnm{{Lockman}}, \binits{W.}},
\bauthor{\bsnm{{Longo}}, \binits{F.}},
\bauthor{\bsnm{{Magni}}, \binits{S.}},
\bauthor{\bsnm{{Maire}}, \binits{M.}},
\bauthor{\bsnm{{Medernach}}, \binits{E.}},
\bauthor{\bsnm{{Minamimoto}}, \binits{K.}},
\bauthor{\bsnm{{Mora de Freitas}}, \binits{P.}},
\bauthor{\bsnm{{Morita}}, \binits{Y.}},
\bauthor{\bsnm{{Murakami}}, \binits{K.}},
\bauthor{\bsnm{{Nagamatu}}, \binits{M.}},
\bauthor{\bsnm{{Nartallo}}, \binits{R.}},
\bauthor{\bsnm{{Nieminen}}, \binits{P.}},
\bauthor{\bsnm{{Nishimura}}, \binits{T.}},
\bauthor{\bsnm{{Ohtsubo}}, \binits{K.}},
\bauthor{\bsnm{{Okamura}}, \binits{M.}},
\bauthor{\bsnm{{O'Neale}}, \binits{S.}},
\bauthor{\bsnm{{Oohata}}, \binits{Y.}},
\bauthor{\bsnm{{Paech}}, \binits{K.}},
\bauthor{\bsnm{{Perl}}, \binits{J.}},
\bauthor{\bsnm{{Pfeiffer}}, \binits{A.}},
\bauthor{\bsnm{{Pia}}, \binits{M.G.}},
\bauthor{\bsnm{{Ranjard}}, \binits{F.}},
\bauthor{\bsnm{{Rybin}}, \binits{A.}},
\bauthor{\bsnm{{Sadilov}}, \binits{S.}},
\bauthor{\bsnm{{Di Salvo}}, \binits{E.}},
\bauthor{\bsnm{{Santin}}, \binits{G.}},
\bauthor{\bsnm{{Sasaki}}, \binits{T.}},
\bauthor{\bsnm{{Savvas}}, \binits{N.}},
\bauthor{\bsnm{{Sawada}}, \binits{Y.}},
\bauthor{\bsnm{{Scherer}}, \binits{S.}},
\bauthor{\bsnm{{Sei}}, \binits{S.}},
\bauthor{\bsnm{{Sirotenko}}, \binits{V.}},
\bauthor{\bsnm{{Smith}}, \binits{D.}},
\bauthor{\bsnm{{Starkov}}, \binits{N.}},
\bauthor{\bsnm{{Stoecker}}, \binits{H.}},
\bauthor{\bsnm{{Sulkimo}}, \binits{J.}},
\bauthor{\bsnm{{Takahata}}, \binits{M.}},
\bauthor{\bsnm{{Tanaka}}, \binits{S.}},
\bauthor{\bsnm{{Tcherniaev}}, \binits{E.}},
\bauthor{\bsnm{{Safai Tehrani}}, \binits{E.}},
\bauthor{\bsnm{{Tropeano}}, \binits{M.}},
\bauthor{\bsnm{{Truscott}}, \binits{P.}},
\bauthor{\bsnm{{Uno}}, \binits{H.}},
\bauthor{\bsnm{{Urban}}, \binits{L.}},
\bauthor{\bsnm{{Urban}}, \binits{P.}},
\bauthor{\bsnm{{Verderi}}, \binits{M.}},
\bauthor{\bsnm{{Walkden}}, \binits{A.}},
\bauthor{\bsnm{{Wander}}, \binits{W.}},
\bauthor{\bsnm{{Weber}}, \binits{H.}},
\bauthor{\bsnm{{Wellisch}}, \binits{J.P.}},
\bauthor{\bsnm{{Wenaus}}, \binits{T.}},
\bauthor{\bsnm{{Williams}}, \binits{D.C.}},
\bauthor{\bsnm{{Wright}}, \binits{D.}},
\bauthor{\bsnm{{Yamada}}, \binits{T.}},
\bauthor{\bsnm{{Yoshida}}, \binits{H.}},
\bauthor{\bsnm{{Zschiesche}}, \binits{D.}},
\bauthor{\bsnm{{GEANT4 Collaboration}}}:
\batitle{{GEANT4{\textemdash}a simulation toolkit}}.
\bjtitle{Nuclear Instruments and Methods in Physics Research A}
\bvolume{506}(\bissue{3}),
\bfpage{250}--\blpage{303}
(\byear{2003})
\doiurl{10.1016/S0168-9002(03)01368-8}
\end{barticle}
\endbibitem

\bibitem[\protect\citeauthoryear{{Weidenspointner}
  et~al.}{2005}]{weidenspointner2005}
\begin{barticle}
\bauthor{\bsnm{{Weidenspointner}}, \binits{G.}},
\bauthor{\bsnm{{Harris}}, \binits{M.J.}},
\bauthor{\bsnm{{Sturner}}, \binits{S.}},
\bauthor{\bsnm{{Teegarden}}, \binits{B.J.}},
\bauthor{\bsnm{{Ferguson}}, \binits{C.}}:
\batitle{{MGGPOD: a Monte Carlo Suite for Modeling Instrumental Line and
  Continuum Backgrounds in Gamma-Ray Astronomy}}.
\bjtitle{The Astrophysical Journal Supplement Series}
\bvolume{156}(\bissue{1}),
\bfpage{69}--\blpage{91}
(\byear{2005})
\doiurl{10.1086/425577}
{\href{https://arxiv.org/abs/astro-ph/0408399}{{arXiv:astro-ph/0408399}}}
{[astro-ph]}
\end{barticle}
\endbibitem

\bibitem[\protect\citeauthoryear{{Zoglauer} et~al.}{2006}]{zoglauer2006}
\begin{barticle}
\bauthor{\bsnm{{Zoglauer}}, \binits{A.}},
\bauthor{\bsnm{{Andritschke}}, \binits{R.}},
\bauthor{\bsnm{{Schopper}}, \binits{F.}}:
\batitle{{MEGAlib The Medium Energy Gamma-ray Astronomy Library}}.
\bjtitle{New Astronomy Review}
\bvolume{50}(\bissue{7-8}),
\bfpage{629}--\blpage{632}
(\byear{2006})
\doiurl{10.1016/j.newar.2006.06.049}
\end{barticle}
\endbibitem

\bibitem[\protect\citeauthoryear{{Odaka} et~al.}{2018}]{odaka2018}
\begin{barticle}
\bauthor{\bsnm{{Odaka}}, \binits{H.}},
\bauthor{\bsnm{{Asai}}, \binits{M.}},
\bauthor{\bsnm{{Hagino}}, \binits{K.}},
\bauthor{\bsnm{{Koi}}, \binits{T.}},
\bauthor{\bsnm{{Madejski}}, \binits{G.}},
\bauthor{\bsnm{{Mizuno}}, \binits{T.}},
\bauthor{\bsnm{{Ohno}}, \binits{M.}},
\bauthor{\bsnm{{Saito}}, \binits{S.}},
\bauthor{\bsnm{{Sato}}, \binits{T.}},
\bauthor{\bsnm{{Wright}}, \binits{D.H.}},
\bauthor{\bsnm{{Enoto}}, \binits{T.}},
\bauthor{\bsnm{{Fukazawa}}, \binits{Y.}},
\bauthor{\bsnm{{Hayashi}}, \binits{K.}},
\bauthor{\bsnm{{Kataoka}}, \binits{J.}},
\bauthor{\bsnm{{Katsuta}}, \binits{J.}},
\bauthor{\bsnm{{Kawaharada}}, \binits{M.}},
\bauthor{\bsnm{{Kobayashi}}, \binits{S.B.}},
\bauthor{\bsnm{{Kokubun}}, \binits{M.}},
\bauthor{\bsnm{{Laurent}}, \binits{P.}},
\bauthor{\bsnm{{Lebrun}}, \binits{F.}},
\bauthor{\bsnm{{Limousin}}, \binits{O.}},
\bauthor{\bsnm{{Maier}}, \binits{D.}},
\bauthor{\bsnm{{Makishima}}, \binits{K.}},
\bauthor{\bsnm{{Mimura}}, \binits{T.}},
\bauthor{\bsnm{{Miyake}}, \binits{K.}},
\bauthor{\bsnm{{Mori}}, \binits{K.}},
\bauthor{\bsnm{{Murakami}}, \binits{H.}},
\bauthor{\bsnm{{Nakamori}}, \binits{T.}},
\bauthor{\bsnm{{Nakano}}, \binits{T.}},
\bauthor{\bsnm{{Nakazawa}}, \binits{K.}},
\bauthor{\bsnm{{Noda}}, \binits{H.}},
\bauthor{\bsnm{{Ohta}}, \binits{M.}},
\bauthor{\bsnm{{Ozaki}}, \binits{M.}},
\bauthor{\bsnm{{Sato}}, \binits{G.}},
\bauthor{\bsnm{{Sato}}, \binits{R.}},
\bauthor{\bsnm{{Tajima}}, \binits{H.}},
\bauthor{\bsnm{{Takahashi}}, \binits{H.}},
\bauthor{\bsnm{{Takahashi}}, \binits{T.}},
\bauthor{\bsnm{{Takeda}}, \binits{S.}},
\bauthor{\bsnm{{Tanaka}}, \binits{T.}},
\bauthor{\bsnm{{Tanaka}}, \binits{Y.}},
\bauthor{\bsnm{{Terada}}, \binits{Y.}},
\bauthor{\bsnm{{Uchiyama}}, \binits{H.}},
\bauthor{\bsnm{{Uchiyama}}, \binits{Y.}},
\bauthor{\bsnm{{Watanabe}}, \binits{S.}},
\bauthor{\bsnm{{Yamaoka}}, \binits{K.}},
\bauthor{\bsnm{{Yasuda}}, \binits{T.}},
\bauthor{\bsnm{{Yatsu}}, \binits{Y.}},
\bauthor{\bsnm{{Yuasa}}, \binits{T.}},
\bauthor{\bsnm{{Zoglauer}}, \binits{A.}}:
\batitle{{Modeling of proton-induced radioactivation background in hard X-ray
  telescopes: Geant4-based simulation and its demonstration by Hitomi's
  measurement in a low Earth orbit}}.
\bjtitle{Nuclear Instruments and Methods in Physics Research A}
\bvolume{891},
\bfpage{92}--\blpage{105}
(\byear{2018})
\doiurl{10.1016/j.nima.2018.02.071}
{\href{https://arxiv.org/abs/1804.00827}{{arXiv:1804.00827}}}
{[astro-ph.IM]}
\end{barticle}
\endbibitem

\bibitem[\protect\citeauthoryear{{Galgoczi} et~al.}{2020}]{galgoczi2020}
\begin{bchapter}
\bauthor{\bsnm{{Galgoczi}}, \binits{G.}},
\bauthor{\bsnm{{Ripa}}, \binits{J.}},
\bauthor{\bsnm{{Dilillo}}, \binits{P.}},
\bauthor{\bsnm{{Ohno}}, \binits{M.}},
\bauthor{\bsnm{{Campana}}, \binits{R.}},
\bauthor{\bsnm{{Werner}}, \binits{N.}}:
\bctitle{{A software toolkit to simulate activation background for high energy
  detectors onboard satellites}}.
In: \beditor{\bsnm{{den Herder}}, \binits{J.-W.A.}},
\beditor{\bsnm{{Nikzad}}, \binits{S.}},
\beditor{\bsnm{{Nakazawa}}, \binits{K.}} (eds.)
\bbtitle{Space Telescopes and Instrumentation 2020: Ultraviolet to Gamma Ray}.
\bsertitle{Society of Photo-Optical Instrumentation Engineers (SPIE) Conference
  Series},
vol. \bseriesno{11444},
p. \bfpage{1144492}
(\byear{2020}).
\doiurl{10.1117/12.2560829}
\end{bchapter}
\endbibitem

\bibitem[\protect\citeauthoryear{{Fiore} et~al.}{2021}]{fiore21}
\begin{bchapter}
\bauthor{\bsnm{{Fiore}}, \binits{F.}},
\bauthor{\bsnm{{Burderi}}, \binits{L.}},
\bauthor{\bsnm{{Lavagna}}, \binits{M.}},
\bauthor{\bsnm{{Bertacin}}, \binits{R.}},
\bauthor{\bsnm{{Evangelista}}, \binits{Y.}},
\bauthor{\bsnm{{Campana}}, \binits{R.}},
\bauthor{\bsnm{{Fuschino}}, \binits{F.}},
\bauthor{\bsnm{{Lunghi}}, \binits{P.}},
\bauthor{\bsnm{{Monge}}, \binits{A.}},
\bauthor{\bsnm{{Negri}}, \binits{B.}},
\bauthor{\bsnm{{Pirrotta}}, \binits{S.}},
\bauthor{\bsnm{{Puccetti}}, \binits{S.}},
\bauthor{\bsnm{{Sanna}}, \binits{A.}},
\bauthor{\bsnm{{Amarilli}}, \binits{F.}},
\bauthor{\bsnm{{Ambrosino}}, \binits{F.}},
\bauthor{\bsnm{{Amelino-Camelia}}, \binits{G.}},
\bauthor{\bsnm{{Anitra}}, \binits{A.}},
\bauthor{\bsnm{{Auricchio}}, \binits{N.}},
\bauthor{\bsnm{{Barbera}}, \binits{M.}},
\bauthor{\bsnm{{Bechini}}, \binits{M.}},
\bauthor{\bsnm{{Bellutti}}, \binits{P.}},
\bauthor{\bsnm{{Bertuccio}}, \binits{G.}},
\bauthor{\bsnm{{Cao}}, \binits{J.}},
\bauthor{\bsnm{{Ceraudo}}, \binits{F.}},
\bauthor{\bsnm{{Chen}}, \binits{T.}},
\bauthor{\bsnm{{Cinelli}}, \binits{M.}},
\bauthor{\bsnm{{Citossi}}, \binits{M.}},
\bauthor{\bsnm{{Clerici}}, \binits{A.}},
\bauthor{\bsnm{{Colagrossi}}, \binits{A.}},
\bauthor{\bsnm{{Curzel}}, \binits{S.}},
\bauthor{\bsnm{{Della Casa}}, \binits{G.}},
\bauthor{\bsnm{{Demenev}}, \binits{E.}},
\bauthor{\bsnm{{Del Santo}}, \binits{M.}},
\bauthor{\bsnm{{Dilillo}}, \binits{G.}},
\bauthor{\bsnm{{Di Salvo}}, \binits{T.}},
\bauthor{\bsnm{{Efremov}}, \binits{P.}},
\bauthor{\bsnm{{Feroci}}, \binits{M.}},
\bauthor{\bsnm{{Feruglio}}, \binits{C.}},
\bauthor{\bsnm{{Ferrandi}}, \binits{F.}},
\bauthor{\bsnm{{Fiorini}}, \binits{M.}},
\bauthor{\bsnm{{Fiorito}}, \binits{M.}},
\bauthor{\bsnm{{Frontera}}, \binits{F.}},
\bauthor{\bsnm{{Gacnik}}, \binits{D.}},
\bauthor{\bsnm{{Galgoczi}}, \binits{G.}},
\bauthor{\bsnm{{Gao}}, \binits{N.}},
\bauthor{\bsnm{{Gambino}}, \binits{A.F.}},
\bauthor{\bsnm{{Gandola}}, \binits{M.}},
\bauthor{\bsnm{{Ghirlanda}}, \binits{G.}},
\bauthor{\bsnm{{Gomboc}}, \binits{A.}},
\bauthor{\bsnm{{Grassi}}, \binits{M.}},
\bauthor{\bsnm{{Guzman}}, \binits{A.}},
\bauthor{\bsnm{{Karlica}}, \binits{M.}},
\bauthor{\bsnm{{Kostic}}, \binits{U.}},
\bauthor{\bsnm{{Labanti}}, \binits{C.}},
\bauthor{\bsnm{{La Rosa}}, \binits{G.}},
\bauthor{\bsnm{{Lo Cicero}}, \binits{U.}},
\bauthor{\bsnm{{Lopez-Fernandez}}, \binits{B.}},
\bauthor{\bsnm{{Malcovati}}, \binits{P.}},
\bauthor{\bsnm{{Maselli}}, \binits{A.}},
\bauthor{\bsnm{{Manca}}, \binits{A.}},
\bauthor{\bsnm{{Mele}}, \binits{F.}},
\bauthor{\bsnm{{Milankovich}}, \binits{D.}},
\bauthor{\bsnm{{Morgante}}, \binits{G.}},
\bauthor{\bsnm{{Nava}}, \binits{L.}},
\bauthor{\bsnm{{Nogara}}, \binits{P.}},
\bauthor{\bsnm{{Ohno}}, \binits{M.}},
\bauthor{\bsnm{{Ottolina}}, \binits{D.}},
\bauthor{\bsnm{{Pasquale}}, \binits{A.}},
\bauthor{\bsnm{{Pal}}, \binits{A.}},
\bauthor{\bsnm{{Perri}}, \binits{M.}},
\bauthor{\bsnm{{Piazzolla}}, \binits{R.}},
\bauthor{\bsnm{{Piccinin}}, \binits{M.}},
\bauthor{\bsnm{{Pliego-Caballero}}, \binits{S.}},
\bauthor{\bsnm{{Prinetto}}, \binits{J.}},
\bauthor{\bsnm{{Pucacco}}, \binits{G.}},
\bauthor{\bsnm{{Rashevsky}}, \binits{A.}},
\bauthor{\bsnm{{Rashevskaya}}, \binits{I.}},
\bauthor{\bsnm{{Riggio}}, \binits{A.}},
\bauthor{\bsnm{{Ripa}}, \binits{J.}},
\bauthor{\bsnm{{Russo}}, \binits{F.}},
\bauthor{\bsnm{{Papitto}}, \binits{A.}},
\bauthor{\bsnm{{Piranomonte}}, \binits{S.}},
\bauthor{\bsnm{{Santangelo}}, \binits{A.}},
\bauthor{\bsnm{{Scala}}, \binits{F.}},
\bauthor{\bsnm{{Sciarrone}}, \binits{G.}},
\bauthor{\bsnm{{Selcan}}, \binits{D.}},
\bauthor{\bsnm{{Silvestrini}}, \binits{S.}},
\bauthor{\bsnm{{Sottile}}, \binits{G.}},
\bauthor{\bsnm{{Rotovnik}}, \binits{T.}},
\bauthor{\bsnm{{Tenzer}}, \binits{C.}},
\bauthor{\bsnm{{Troisi}}, \binits{I.}},
\bauthor{\bsnm{{Vacchi}}, \binits{A.}},
\bauthor{\bsnm{{Virgilli}}, \binits{E.}},
\bauthor{\bsnm{{Werner}}, \binits{N.}},
\bauthor{\bsnm{{Wang}}, \binits{L.}},
\bauthor{\bsnm{{Xu}}, \binits{Y.}},
\bauthor{\bsnm{{Zampa}}, \binits{G.}},
\bauthor{\bsnm{{Zampa}}, \binits{N.}},
\bauthor{\bsnm{{Zanotti}}, \binits{G.}}:
\bctitle{{The HERMES-technologic and scientific pathfinder}}.
In: \beditor{\bsnm{{den Herder}}, \binits{J.-W.A.}},
\beditor{\bsnm{{Nikzad}}, \binits{S.}},
\beditor{\bsnm{{Nakazawa}}, \binits{K.}} (eds.)
\bbtitle{Society of Photo-Optical Instrumentation Engineers (SPIE) Conference
  Series}.
\bsertitle{Society of Photo-Optical Instrumentation Engineers (SPIE) Conference
  Series},
vol. \bseriesno{11444},
p. \bfpage{114441}
(\byear{2021}).
\doiurl{10.1117/12.2560680}
\end{bchapter}
\endbibitem

\bibitem[\protect\citeauthoryear{{Feroci} et~al.}{2024}]{feroci24}
\begin{bchapter}
\bauthor{\bsnm{{Feroci}}, \binits{M.}},
\bauthor{\bsnm{{Ambrosi}}, \binits{G.}},
\bauthor{\bsnm{{Antonelli}}, \binits{M.}},
\bauthor{\bsnm{{Argan}}, \binits{A.}},
\bauthor{\bsnm{{Babinec}}, \binits{V.}},
\bauthor{\bsnm{{Barbera}}, \binits{M.}},
\bauthor{\bsnm{{Bastia}}, \binits{P.}},
\bauthor{\bsnm{{Bayer}}, \binits{J.}},
\bauthor{\bsnm{{Bellutti}}, \binits{P.}},
\bauthor{\bsnm{{Bertucci}}, \binits{B.}},
\bauthor{\bsnm{{Bertuccio}}, \binits{G.}},
\bauthor{\bsnm{{Bonfitto}}, \binits{F.}},
\bauthor{\bsnm{{Bonvicini}}, \binits{V.}},
\bauthor{\bsnm{{Bozzo}}, \binits{E.}},
\bauthor{\bsnm{{Baudin}}, \binits{D.}},
\bauthor{\bsnm{{Bouyjou}}, \binits{F.}},
\bauthor{\bsnm{{Brienza}}, \binits{D.}},
\bauthor{\bsnm{{Cadoux}}, \binits{F.}},
\bauthor{\bsnm{{Campana}}, \binits{R.}},
\bauthor{\bsnm{{Candia}}, \binits{R.}},
\bauthor{\bsnm{{Cao}}, \binits{J.}},
\bauthor{\bsnm{{Cavazzuti}}, \binits{E.}},
\bauthor{\bsnm{{Ceraudo}}, \binits{F.}},
\bauthor{\bsnm{{Chen}}, \binits{T.}},
\bauthor{\bsnm{{Chen}}, \binits{W.}},
\bauthor{\bsnm{{Coimbra}}, \binits{A.}},
\bauthor{\bsnm{{Colucci}}, \binits{A.}},
\bauthor{\bsnm{{Cong}}, \binits{X.}},
\bauthor{\bsnm{{Cirrincione}}, \binits{D.}},
\bauthor{\bsnm{{De Angelis}}, \binits{N.}},
\bauthor{\bsnm{{De Rosa}}, \binits{A.}},
\bauthor{\bsnm{{Della Casa}}, \binits{G.}},
\bauthor{\bsnm{{Del Monte}}, \binits{E.}},
\bauthor{\bsnm{{Di Cicca}}, \binits{G.}},
\bauthor{\bsnm{{Di Cosimo}}, \binits{S.}},
\bauthor{\bsnm{{Dilillo}}, \binits{G.}},
\bauthor{\bsnm{{Dohnal}}, \binits{R.}},
\bauthor{\bsnm{{Donnarumma}}, \binits{I.}},
\bauthor{\bsnm{{Evangelista}}, \binits{Y.}},
\bauthor{\bsnm{{Fan}}, \binits{P.}},
\bauthor{\bsnm{{Fan}}, \binits{Q.}},
\bauthor{\bsnm{{Favre}}, \binits{Y.}},
\bauthor{\bsnm{{Fiandrini}}, \binits{E.}},
\bauthor{\bsnm{{Ficorella}}, \binits{F.}},
\bauthor{\bsnm{{Gao}}, \binits{N.}},
\bauthor{\bsnm{{Gevin}}, \binits{O.}},
\bauthor{\bsnm{{Grassi}}, \binits{M.}},
\bauthor{\bsnm{{Guedel}}, \binits{M.}},
\bauthor{\bsnm{{Guzman Cabrera}}, \binits{A.}},
\bauthor{\bsnm{{Han}}, \binits{D.}},
\bauthor{\bsnm{{He}}, \binits{H.}},
\bauthor{\bsnm{{Hedderman}}, \binits{P.}},
\bauthor{\bsnm{{den Herder}}, \binits{J.-W.}},
\bauthor{\bsnm{{Hynek}}, \binits{R.}},
\bauthor{\bsnm{{Reynaard Kole}}, \binits{M.}},
\bauthor{\bsnm{{Karas}}, \binits{V.}},
\bauthor{\bsnm{{Komarek}}, \binits{M.}},
\bauthor{\bsnm{{Labanti}}, \binits{C.}},
\bauthor{\bsnm{{La Marra}}, \binits{D.}},
\bauthor{\bsnm{{Lesci}}, \binits{G.}},
\bauthor{\bsnm{{Li}}, \binits{G.}},
\bauthor{\bsnm{{Li}}, \binits{L.}},
\bauthor{\bsnm{{Li}}, \binits{T.}},
\bauthor{\bsnm{{Limousin}}, \binits{O.}},
\bauthor{\bsnm{{Liu}}, \binits{H.}},
\bauthor{\bsnm{{Liu}}, \binits{R.}},
\bauthor{\bsnm{{Liu}}, \binits{Y.}},
\bauthor{\bsnm{{Liu}}, \binits{X.}},
\bauthor{\bsnm{{Lo Cicero}}, \binits{U.}},
\bauthor{\bsnm{{Loehring}}, \binits{J.}},
\bauthor{\bsnm{{Lombardi}}, \binits{G.}},
\bauthor{\bsnm{{Lu}}, \binits{F.-J.}},
\bauthor{\bsnm{{Luo}}, \binits{T.}},
\bauthor{\bsnm{{Malcovati}}, \binits{P.}},
\bauthor{\bsnm{{Marinucci}}, \binits{A.}},
\bauthor{\bsnm{{Mele}}, \binits{F.}},
\bauthor{\bsnm{{Mendes}}, \binits{V.}},
\bauthor{\bsnm{{Merkl}}, \binits{M.}},
\bauthor{\bsnm{{Meuris}}, \binits{A.}},
\bauthor{\bsnm{{Michalska}}, \binits{M.}},
\bauthor{\bsnm{{Morbidini}}, \binits{A.}},
\bauthor{\bsnm{{Morgante}}, \binits{G.}},
\bauthor{\bsnm{{Muleri}}, \binits{F.}},
\bauthor{\bsnm{{Munini}}, \binits{R.}},
\bauthor{\bsnm{{Mussolin}}, \binits{L.}},
\bauthor{\bsnm{{Negri}}, \binits{B.}},
\bauthor{\bsnm{{Nov{\'a}k}}, \binits{P.}},
\bauthor{\bsnm{{Nowosielski}}, \binits{W.}},
\bauthor{\bsnm{{Nuti}}, \binits{A.}},
\bauthor{\bsnm{{Orleanski}}, \binits{P.}},
\bauthor{\bsnm{{Ottensamer}}, \binits{R.}},
\bauthor{\bsnm{{Pacciani}}, \binits{L.}},
\bauthor{\bsnm{{Paltani}}, \binits{S.}},
\bauthor{\bsnm{{Pan}}, \binits{T.}},
\bauthor{\bsnm{{Pepponi}}, \binits{G.}},
\bauthor{\bsnm{{Perinati}}, \binits{E.}},
\bauthor{\bsnm{{Piazzolla}}, \binits{R.}},
\bauthor{\bsnm{{Picciotto}}, \binits{A.}},
\bauthor{\bsnm{{Pliego}}, \binits{S.}},
\bauthor{\bsnm{{Putz}}, \binits{A.}},
\bauthor{\bsnm{{Rachevski}}, \binits{A.}},
\bauthor{\bsnm{{Rashevskaia}}, \binits{I.}},
\bauthor{\bsnm{{Samusenko}}, \binits{A.}},
\bauthor{\bsnm{{Santangelo}}, \binits{A.}},
\bauthor{\bsnm{{Schanne}}, \binits{S.}},
\bauthor{\bsnm{{Serafinelli}}, \binits{R.}},
\bauthor{\bsnm{{Skup}}, \binits{K.}},
\bauthor{\bsnm{{Sveda}}, \binits{L.}},
\bauthor{\bsnm{{Svoboda}}, \binits{J.}},
\bauthor{\bsnm{{Tenzer}}, \binits{C.}},
\bauthor{\bsnm{{Todaro}}, \binits{M.}},
\bauthor{\bsnm{{Torok}}, \binits{G.}},
\bauthor{\bsnm{{Trois}}, \binits{A.}},
\bauthor{\bsnm{{Vacchi}}, \binits{A.}},
\bauthor{\bsnm{{Varisco}}, \binits{S.}},
\bauthor{\bsnm{{Villa}}, \binits{F.}},
\bauthor{\bsnm{{Virgilli}}, \binits{E.}},
\bauthor{\bsnm{{Xiang}}, \binits{H.}},
\bauthor{\bsnm{{Xiong}}, \binits{H.}},
\bauthor{\bsnm{{Wang}}, \binits{J.}},
\bauthor{\bsnm{{Wang}}, \binits{X.}},
\bauthor{\bsnm{{Winter}}, \binits{B.}},
\bauthor{\bsnm{{Wu}}, \binits{Z.}},
\bauthor{\bsnm{{Wu}}, \binits{X.}},
\bauthor{\bsnm{{Xu}}, \binits{Y.}},
\bauthor{\bsnm{{Zampa}}, \binits{G.}},
\bauthor{\bsnm{{Zampa}}, \binits{N.}},
\bauthor{\bsnm{{Zdziarski}}, \binits{A.A.}},
\bauthor{\bsnm{{Zhang}}, \binits{L.}},
\bauthor{\bsnm{{Zhang}}, \binits{S.}},
\bauthor{\bsnm{{Zhang}}, \binits{S.-N.}},
\bauthor{\bsnm{{Zhang}}, \binits{Y.}},
\bauthor{\bsnm{{Zhang}}, \binits{W.}},
\bauthor{\bsnm{{Zhao}}, \binits{Q.}},
\bauthor{\bsnm{{Zhu}}, \binits{C.}},
\bauthor{\bsnm{{Zhu}}, \binits{X.}},
\bauthor{\bsnm{{Zorzi}}, \binits{N.}}:
\bctitle{{The Large Area Detector for the eXTP mission}}.
In: \beditor{\bsnm{{den Herder}}, \binits{J.-W.A.}},
\beditor{\bsnm{{Nikzad}}, \binits{S.}},
\beditor{\bsnm{{Nakazawa}}, \binits{K.}} (eds.)
\bbtitle{Space Telescopes and Instrumentation 2024: Ultraviolet to Gamma Ray}.
\bsertitle{Society of Photo-Optical Instrumentation Engineers (SPIE) Conference
  Series},
vol. \bseriesno{13093},
p. \bfpage{130931}
(\byear{2024}).
\doiurl{10.1117/12.3019868}
\end{bchapter}
\endbibitem

\bibitem[\protect\citeauthoryear{{Hernanz} et~al.}{2024}]{hernanz24}
\begin{bchapter}
\bauthor{\bsnm{{Hernanz}}, \binits{M.}},
\bauthor{\bsnm{{Feroci}}, \binits{M.}},
\bauthor{\bsnm{{Evangelista}}, \binits{Y.}},
\bauthor{\bsnm{{Meuris}}, \binits{A.}},
\bauthor{\bsnm{{Schanne}}, \binits{S.}},
\bauthor{\bsnm{{Zampa}}, \binits{G.}},
\bauthor{\bsnm{{Tenzer}}, \binits{C.}},
\bauthor{\bsnm{{Bayer}}, \binits{J.}},
\bauthor{\bsnm{{Nowosielski}}, \binits{W.}},
\bauthor{\bsnm{{Michalska}}, \binits{M.}},
\bauthor{\bsnm{{Kalemci}}, \binits{E.}},
\bauthor{\bsnm{{Sungur}}, \binits{M.}},
\bauthor{\bsnm{{Brandt}}, \binits{S.}},
\bauthor{\bsnm{{Kuvvetli}}, \binits{I.}},
\bauthor{\bsnm{{Alvarez Franco}}, \binits{D.}},
\bauthor{\bsnm{{Carmona}}, \binits{A.}},
\bauthor{\bsnm{{G{\'a}lvez}}, \binits{J.-L.}},
\bauthor{\bsnm{{Patruno}}, \binits{A.}},
\bauthor{\bsnm{{In't Zand}}, \binits{J.}},
\bauthor{\bsnm{{Zwart}}, \binits{F.}},
\bauthor{\bsnm{{Santangelo}}, \binits{A.}},
\bauthor{\bsnm{{Bozzo}}, \binits{E.}},
\bauthor{\bsnm{{Zhang}}, \binits{S.-N.}},
\bauthor{\bsnm{{Lu}}, \binits{F.}},
\bauthor{\bsnm{{Xu}}, \binits{Y.}},
\bauthor{\bsnm{{Campana}}, \binits{R.}},
\bauthor{\bsnm{{Del Monte}}, \binits{E.}},
\bauthor{\bsnm{{Ceraudo}}, \binits{F.}},
\bauthor{\bsnm{{Nuti}}, \binits{A.}},
\bauthor{\bsnm{{Della Casa}}, \binits{G.}},
\bauthor{\bsnm{{Argan}}, \binits{A.}},
\bauthor{\bsnm{{Minervini}}, \binits{G.}},
\bauthor{\bsnm{{Antonelli}}, \binits{M.}},
\bauthor{\bsnm{{Bonvicini}}, \binits{V.}},
\bauthor{\bsnm{{Boezio}}, \binits{M.}},
\bauthor{\bsnm{{Cirrincione}}, \binits{D.}},
\bauthor{\bsnm{{Munini}}, \binits{R.}},
\bauthor{\bsnm{{Rachevski}}, \binits{A.}},
\bauthor{\bsnm{{Vacchi}}, \binits{A.}},
\bauthor{\bsnm{{Zampa}}, \binits{N.}},
\bauthor{\bsnm{{Rashevskaya}}, \binits{I.}},
\bauthor{\bsnm{{Ficorella}}, \binits{F.}},
\bauthor{\bsnm{{Picciotto}}, \binits{A.}},
\bauthor{\bsnm{{Zorzi}}, \binits{N.}},
\bauthor{\bsnm{{Baudin}}, \binits{D.}},
\bauthor{\bsnm{{Bouyjou}}, \binits{F.}},
\bauthor{\bsnm{{Gevin}}, \binits{O.}},
\bauthor{\bsnm{{Limousin}}, \binits{O.}},
\bauthor{\bsnm{{Hedderman}}, \binits{P.}},
\bauthor{\bsnm{{Pliego}}, \binits{S.}},
\bauthor{\bsnm{{Xiong}}, \binits{H.}},
\bauthor{\bsnm{{de la Rie}}, \binits{R.}},
\bauthor{\bsnm{{Laubert}}, \binits{P.}},
\bauthor{\bsnm{{Aitink-Kroes}}, \binits{G.}},
\bauthor{\bsnm{{Kuiper}}, \binits{L.}},
\bauthor{\bsnm{{Orleanski}}, \binits{P.}},
\bauthor{\bsnm{{Skup}}, \binits{K.}},
\bauthor{\bsnm{{Tcherniak}}, \binits{D.}},
\bauthor{\bsnm{{Turhan}}, \binits{O.}},
\bauthor{\bsnm{{Bozkurt}}, \binits{A.}},
\bauthor{\bsnm{{Onat}}, \binits{A.}}:
\bctitle{{The Wide Field Monitor (WFM) of the China-Europe eXTP (enhanced X-ray
  Timing and Polarimetry) mission}}.
In: \beditor{\bsnm{{den Herder}}, \binits{J.-W.A.}},
\beditor{\bsnm{{Nikzad}}, \binits{S.}},
\beditor{\bsnm{{Nakazawa}}, \binits{K.}} (eds.)
\bbtitle{Space Telescopes and Instrumentation 2024: Ultraviolet to Gamma Ray}.
\bsertitle{Society of Photo-Optical Instrumentation Engineers (SPIE) Conference
  Series},
vol. \bseriesno{13093},
p. \bfpage{130931}
(\byear{2024}).
\doiurl{10.1117/12.3020020}
\end{bchapter}
\endbibitem

\bibitem[\protect\citeauthoryear{{Van Allen} et~al.}{1958}]{vanallen58}
\begin{barticle}
\bauthor{\bsnm{{Van Allen}}, \binits{J.A.}},
\bauthor{\bsnm{{Ludwig}}, \binits{G.H.}},
\bauthor{\bsnm{{Ray}}, \binits{E.C.}},
\bauthor{\bsnm{{McIlwain}}, \binits{C.E.}}:
\batitle{Observation of high intensity radiation by satellites 1958 alpha and
  gamma}.
\bjtitle{Journal of Jet Propulsion}
\bvolume{28}(\bissue{9}),
\bfpage{588}--\blpage{592}
(\byear{1958})
\doiurl{10.2514/8.7396}
\end{barticle}
\endbibitem

\bibitem[\protect\citeauthoryear{{Sawyer} and {Vette}}{1976}]{sawyer76}
\begin{botherref}
\oauthor{\bsnm{{Sawyer}}, \binits{D.M.}},
\oauthor{\bsnm{{Vette}}, \binits{J.I.}}:
{AP-8 trapped proton model environment for solar maximum and minimum}.
Technical Report NSSDC/WDC-A-R\&S 76-06,
Natl. Space Sci. Data Cent
(1976)
\end{botherref}
\endbibitem

\bibitem[\protect\citeauthoryear{{Vette}}{1991}]{vette91}
\begin{botherref}
\oauthor{\bsnm{{Vette}}, \binits{J.I.}}:
{ The AE-8 trapped electron model environment}.
Technical Report NSSDC/WDC-A-R\&S 91-24,
NASA Goddard Space Flight Center
(1991)
\end{botherref}
\endbibitem

\bibitem[\protect\citeauthoryear{{Heynderickx} et~al.}{1999}]{heynderickx99}
\begin{barticle}
\bauthor{\bsnm{{Heynderickx}}, \binits{D.}},
\bauthor{\bsnm{{Kruglanski}}, \binits{M.}},
\bauthor{\bsnm{{Pierrard}}, \binits{V.}},
\bauthor{\bsnm{{Lemaire}}, \binits{J.}},
\bauthor{\bsnm{{Looper}}, \binits{M.D.}},
\bauthor{\bsnm{{Blake}}, \binits{J.B.}}:
\batitle{{A low altitude trapped proton model for solar minimum conditions
  based on SAMPEX/PET data}}.
\bjtitle{IEEE Transactions on Nuclear Science}
\bvolume{46}(\bissue{6}),
\bfpage{1475}--\blpage{1480}
(\byear{1999})
\doiurl{10.1109/23.819110}
\end{barticle}
\endbibitem

\bibitem[\protect\citeauthoryear{{Ginet} et~al.}{2013}]{ginet13}
\begin{barticle}
\bauthor{\bsnm{{Ginet}}, \binits{G.P.}},
\bauthor{\bsnm{{O'Brien}}, \binits{T.P.}},
\bauthor{\bsnm{{Huston}}, \binits{S.L.}},
\bauthor{\bsnm{{Johnston}}, \binits{W.R.}},
\bauthor{\bsnm{{Guild}}, \binits{T.B.}},
\bauthor{\bsnm{{Friedel}}, \binits{R.}},
\bauthor{\bsnm{{Lindstrom}}, \binits{C.D.}},
\bauthor{\bsnm{{Roth}}, \binits{C.J.}},
\bauthor{\bsnm{{Whelan}}, \binits{P.}},
\bauthor{\bsnm{{Quinn}}, \binits{R.A.}},
\bauthor{\bsnm{{Madden}}, \binits{D.}},
\bauthor{\bsnm{{Morley}}, \binits{S.}},
\bauthor{\bsnm{{Su}}, \binits{Y.-J.}}:
\batitle{{AE9, AP9 and SPM: New Models for Specifying the Trapped Energetic
  Particle and Space Plasma Environment}}.
\bjtitle{Space Science Review}
\bvolume{179}(\bissue{1-4}),
\bfpage{579}--\blpage{615}
(\byear{2013})
\doiurl{10.1007/s11214-013-9964-y}
\end{barticle}
\endbibitem

\bibitem[\protect\citeauthoryear{{Fiore} et~al.}{2022}]{fiore22}
\begin{bchapter}
\bauthor{\bsnm{{Fiore}}, \binits{F.}},
\bauthor{\bsnm{{Guzman}}, \binits{A.}},
\bauthor{\bsnm{{Campana}}, \binits{R.}},
\bauthor{\bsnm{{Evangelista}}, \binits{Y.}}:
\bctitle{{HERMES-Pathfinder}}.
In: \beditor{\bsnm{{Bambi}}, \binits{C.}},
\beditor{\bsnm{{Sangangelo}}, \binits{A.}} (eds.)
\bbtitle{Handbook of X-ray and Gamma-ray Astrophysics},
p. \bfpage{38}
(\byear{2022}).
\doiurl{10.1007/978-981-16-4544-0_35-1}
\end{bchapter}
\endbibitem

\bibitem[\protect\citeauthoryear{{Evangelista} et~al.}{2022}]{evangelista22}
\begin{bchapter}
\bauthor{\bsnm{{Evangelista}}, \binits{Y.}},
\bauthor{\bsnm{{Fiore}}, \binits{F.}},
\bauthor{\bsnm{{Campana}}, \binits{R.}},
\bauthor{\bsnm{{Ceraudo}}, \binits{F.}},
\bauthor{\bsnm{{Della Casa}}, \binits{G.}},
\bauthor{\bsnm{{Demenev}}, \binits{E.}},
\bauthor{\bsnm{{Dilillo}}, \binits{G.}},
\bauthor{\bsnm{{Fiorini}}, \binits{M.}},
\bauthor{\bsnm{{Grassi}}, \binits{M.}},
\bauthor{\bsnm{{Guzman}}, \binits{A.}},
\bauthor{\bsnm{{Hedderman}}, \binits{P.}},
\bauthor{\bsnm{{Marchesini}}, \binits{E.J.}},
\bauthor{\bsnm{{Morgante}}, \binits{G.}},
\bauthor{\bsnm{{Mele}}, \binits{F.}},
\bauthor{\bsnm{{Nogara}}, \binits{P.}},
\bauthor{\bsnm{{Nuti}}, \binits{A.}},
\bauthor{\bsnm{{Piazzolla}}, \binits{R.}},
\bauthor{\bsnm{{Pliego Caballero}}, \binits{S.}},
\bauthor{\bsnm{{Rashevskaya}}, \binits{I.}},
\bauthor{\bsnm{{Russo}}, \binits{F.}},
\bauthor{\bsnm{{Sottile}}, \binits{G.}},
\bauthor{\bsnm{{Labanti}}, \binits{C.}},
\bauthor{\bsnm{{Baroni}}, \binits{G.}},
\bauthor{\bsnm{{Bellutti}}, \binits{P.}},
\bauthor{\bsnm{{Bertuccio}}, \binits{G.}},
\bauthor{\bsnm{{Cao}}, \binits{J.}},
\bauthor{\bsnm{{Chen}}, \binits{T.}},
\bauthor{\bsnm{{Dedolli}}, \binits{I.}},
\bauthor{\bsnm{{Feroci}}, \binits{M.}},
\bauthor{\bsnm{{Fuschino}}, \binits{F.}},
\bauthor{\bsnm{{Gandola}}, \binits{M.}},
\bauthor{\bsnm{{Gao}}, \binits{N.}},
\bauthor{\bsnm{{Ficorella}}, \binits{F.}},
\bauthor{\bsnm{{Malcovati}}, \binits{P.}},
\bauthor{\bsnm{{Picciotto}}, \binits{A.}},
\bauthor{\bsnm{{Rachevski}}, \binits{A.}},
\bauthor{\bsnm{{Santangelo}}, \binits{A.}},
\bauthor{\bsnm{{Tenzer}}, \binits{C.}},
\bauthor{\bsnm{{Vacchi}}, \binits{A.}},
\bauthor{\bsnm{{Wang}}, \binits{L.}},
\bauthor{\bsnm{{Xu}}, \binits{Y.}},
\bauthor{\bsnm{{Zampa}}, \binits{G.}},
\bauthor{\bsnm{{Zampa}}, \binits{N.}},
\bauthor{\bsnm{{Zorzi}}, \binits{N.}}:
\bctitle{{Design, integration, and test of the scientific payloads on-board the
  HERMES constellation and the SpIRIT mission}}.
In: \beditor{\bsnm{{den Herder}}, \binits{J.-W.A.}},
\beditor{\bsnm{{Nikzad}}, \binits{S.}},
\beditor{\bsnm{{Nakazawa}}, \binits{K.}} (eds.)
\bbtitle{Space Telescopes and Instrumentation 2022: Ultraviolet to Gamma Ray}.
\bsertitle{Society of Photo-Optical Instrumentation Engineers (SPIE) Conference
  Series},
vol. \bseriesno{12181},
p. \bfpage{121811}
(\byear{2022}).
\doiurl{10.1117/12.2628978}
\end{bchapter}
\endbibitem

\bibitem[\protect\citeauthoryear{{Evangelista} et~al.}{2024}]{evangelista24}
\begin{bchapter}
\bauthor{\bsnm{{Evangelista}}, \binits{Y.}},
\bauthor{\bsnm{{Fiore}}, \binits{F.}},
\bauthor{\bsnm{{Campana}}, \binits{R.}},
\bauthor{\bsnm{{Baroni}}, \binits{G.}},
\bauthor{\bsnm{{Ceraudo}}, \binits{F.}},
\bauthor{\bsnm{{Della Casa}}, \binits{G.}},
\bauthor{\bsnm{{Demenev}}, \binits{E.}},
\bauthor{\bsnm{{Dilillo}}, \binits{G.}},
\bauthor{\bsnm{{Fiorini}}, \binits{M.}},
\bauthor{\bsnm{{Ghirlanda}}, \binits{G.}},
\bauthor{\bsnm{{Grassi}}, \binits{M.}},
\bauthor{\bsnm{{Guzm{\'a}n}}, \binits{A.}},
\bauthor{\bsnm{{Hedderman}}, \binits{P.}},
\bauthor{\bsnm{{Marchesini}}, \binits{E.J.}},
\bauthor{\bsnm{{Morgante}}, \binits{G.}},
\bauthor{\bsnm{{Mele}}, \binits{F.}},
\bauthor{\bsnm{{Nava}}, \binits{L.}},
\bauthor{\bsnm{{Nogara}}, \binits{P.}},
\bauthor{\bsnm{{Nuti}}, \binits{A.}},
\bauthor{\bsnm{{Pliego Caballero}}, \binits{S.}},
\bauthor{\bsnm{{Rashevskaya}}, \binits{I.}},
\bauthor{\bsnm{{Russo}}, \binits{F.}},
\bauthor{\bsnm{{Sottile}}, \binits{G.}},
\bauthor{\bsnm{{Lavagna}}, \binits{M.}},
\bauthor{\bsnm{{Colagrossi}}, \binits{A.}},
\bauthor{\bsnm{{Silvestrini}}, \binits{S.}},
\bauthor{\bsnm{{Quirino}}, \binits{M.}},
\bauthor{\bsnm{{Bechini}}, \binits{M.}},
\bauthor{\bsnm{{Brandonisio}}, \binits{A.}},
\bauthor{\bsnm{{De Cecio}}, \binits{F.}},
\bauthor{\bsnm{{Dottori}}, \binits{A.}},
\bauthor{\bsnm{{Troisi}}, \binits{I.}},
\bauthor{\bsnm{{Bertacin}}, \binits{R.}},
\bauthor{\bsnm{{Bellutti}}, \binits{P.}},
\bauthor{\bsnm{{Bertuccio}}, \binits{G.}},
\bauthor{\bsnm{{Burderi}}, \binits{L.}},
\bauthor{\bsnm{{Chen}}, \binits{T.}},
\bauthor{\bsnm{{Citossi}}, \binits{M.}},
\bauthor{\bsnm{{Di Salvo}}, \binits{T.}},
\bauthor{\bsnm{{Feroci}}, \binits{M.}},
\bauthor{\bsnm{{Ficorella}}, \binits{F.}},
\bauthor{\bsnm{{Gao}}, \binits{N.}},
\bauthor{\bsnm{{Grappasonni}}, \binits{C.}},
\bauthor{\bsnm{{Labanti}}, \binits{C.}},
\bauthor{\bsnm{{La Rosa}}, \binits{G.}},
\bauthor{\bsnm{{Leone}}, \binits{W.}},
\bauthor{\bsnm{{Malcovati}}, \binits{P.}},
\bauthor{\bsnm{{Negri}}, \binits{B.}},
\bauthor{\bsnm{{Pepponi}}, \binits{G.}},
\bauthor{\bsnm{{Perri}}, \binits{M.}},
\bauthor{\bsnm{{Piazzolla}}, \binits{R.}},
\bauthor{\bsnm{{Picciotto}}, \binits{A.}},
\bauthor{\bsnm{{Pirrotta}}, \binits{S.}},
\bauthor{\bsnm{{Puccetti}}, \binits{S.}},
\bauthor{\bsnm{{Rashevsky}}, \binits{A.}},
\bauthor{\bsnm{{Riggio}}, \binits{A.}},
\bauthor{\bsnm{{Rinaldi}}, \binits{M.}},
\bauthor{\bsnm{{Sanna}}, \binits{A.}},
\bauthor{\bsnm{{Santangelo}}, \binits{A.}},
\bauthor{\bsnm{{Tenzer}}, \binits{C.}},
\bauthor{\bsnm{{Tiberia}}, \binits{A.}},
\bauthor{\bsnm{{Trenti}}, \binits{M.}},
\bauthor{\bsnm{{Trevisan}}, \binits{S.}},
\bauthor{\bsnm{{Vacchi}}, \binits{A.}},
\bauthor{\bsnm{{Xiong}}, \binits{S.}},
\bauthor{\bsnm{{Zampa}}, \binits{G.}},
\bauthor{\bsnm{{Zampa}}, \binits{N.}},
\bauthor{\bsnm{{Zhang}}, \binits{S.}},
\bauthor{\bsnm{{Zorzi}}, \binits{N.}},
\bauthor{\bsnm{{Ripa}}, \binits{J.}},
\bauthor{\bsnm{{Werner}}, \binits{N.}}:
\bctitle{{The HERMES (High Energy Rapid Modular Ensemble of Satellites)
  Pathfinder mission}}.
In: \beditor{\bsnm{{den Herder}}, \binits{J.-W.A.}},
\beditor{\bsnm{{Nikzad}}, \binits{S.}},
\beditor{\bsnm{{Nakazawa}}, \binits{K.}} (eds.)
\bbtitle{Space Telescopes and Instrumentation 2024: Ultraviolet to Gamma Ray}.
\bsertitle{Society of Photo-Optical Instrumentation Engineers (SPIE) Conference
  Series},
vol. \bseriesno{13093},
p. \bfpage{130931}
(\byear{2024}).
\doiurl{10.1117/12.3018259}
\end{bchapter}
\endbibitem

\bibitem[\protect\citeauthoryear{Gandola et~al.}{2019}]{gandola19}
\begin{bchapter}
\bauthor{\bsnm{Gandola}, \binits{M.}},
\bauthor{\bsnm{Grassi}, \binits{M.}},
\bauthor{\bsnm{Mele}, \binits{F.}},
\bauthor{\bsnm{Malcovati}, \binits{P.}},
\bauthor{\bsnm{Bertuccio}, \binits{G.}}:
\bctitle{{LYRA: A Multi-Chip ASIC Designed for HERMES X and Gamma Ray
  Detector}}.
In: \bbtitle{2019 IEEE Nuclear Science Symposium and Medical Imaging Conference
  (NSS/MIC)},
pp. \bfpage{1}--\blpage{3}
(\byear{2019}).
\doiurl{10.1109/NSS/MIC42101.2019.9059616}
\end{bchapter}
\endbibitem

\bibitem[\protect\citeauthoryear{{Guzman} et~al.}{2021}]{guzman21}
\begin{bchapter}
\bauthor{\bsnm{{Guzman}}, \binits{A.}},
\bauthor{\bsnm{{Pliego}}, \binits{S.}},
\bauthor{\bsnm{{Bayer}}, \binits{J.}},
\bauthor{\bsnm{{Evangelista}}, \binits{Y.}},
\bauthor{\bsnm{{La Rosa}}, \binits{G.}},
\bauthor{\bsnm{{Sottile}}, \binits{G.}},
\bauthor{\bsnm{{Curzel}}, \binits{S.}},
\bauthor{\bsnm{{Campana}}, \binits{R.}},
\bauthor{\bsnm{{Fiore}}, \binits{F.}},
\bauthor{\bsnm{{Fuschino}}, \binits{F.}},
\bauthor{\bsnm{{Colagrossi}}, \binits{A.}},
\bauthor{\bsnm{{Fiorito}}, \binits{M.}},
\bauthor{\bsnm{{Nogara}}, \binits{P.}},
\bauthor{\bsnm{{Piazzolla}}, \binits{R.}},
\bauthor{\bsnm{{Russo}}, \binits{F.}},
\bauthor{\bsnm{{Santangelo}}, \binits{A.}},
\bauthor{\bsnm{{Tenzer}}, \binits{C.}}:
\bctitle{{The Payload Data Handling Unit (PDHU) on-board the HERMES-TP and
  HERMES-SP CubeSat Missions}}.
In: \beditor{\bsnm{{den Herder}}, \binits{J.-W.A.}},
\beditor{\bsnm{{Nikzad}}, \binits{S.}},
\beditor{\bsnm{{Nakazawa}}, \binits{K.}} (eds.)
\bbtitle{Society of Photo-Optical Instrumentation Engineers (SPIE) Conference
  Series}.
\bsertitle{Society of Photo-Optical Instrumentation Engineers (SPIE) Conference
  Series},
vol. \bseriesno{11444},
p. \bfpage{1144450}
(\byear{2021}).
\doiurl{10.1117/12.2562325}
\end{bchapter}
\endbibitem

\bibitem[\protect\citeauthoryear{{Dilillo} et~al.}{2024}]{dilillo24}
\begin{barticle}
\bauthor{\bsnm{{Dilillo}}, \binits{G.}},
\bauthor{\bsnm{{Marchesini}}, \binits{E.J.}},
\bauthor{\bsnm{{Baroni}}, \binits{G.}},
\bauthor{\bsnm{{Della Casa}}, \binits{G.}},
\bauthor{\bsnm{{Campana}}, \binits{R.}},
\bauthor{\bsnm{{Evangelista}}, \binits{Y.}},
\bauthor{\bsnm{{Guzm{\'a}n}}, \binits{A.}},
\bauthor{\bsnm{{Hedderman}}, \binits{P.}},
\bauthor{\bsnm{{Bellutti}}, \binits{P.}},
\bauthor{\bsnm{{Bertuccio}}, \binits{G.}},
\bauthor{\bsnm{{Ceraudo}}, \binits{F.}},
\bauthor{\bsnm{{Citossi}}, \binits{M.}},
\bauthor{\bsnm{{Cirrincione}}, \binits{D.}},
\bauthor{\bsnm{{Dedolli}}, \binits{I.}},
\bauthor{\bsnm{{Demenev}}, \binits{E.}},
\bauthor{\bsnm{{Feroci}}, \binits{M.}},
\bauthor{\bsnm{{Ficorella}}, \binits{F.}},
\bauthor{\bsnm{{Fiorini}}, \binits{M.}},
\bauthor{\bsnm{{Gandola}}, \binits{M.}},
\bauthor{\bsnm{{Grassi}}, \binits{M.}},
\bauthor{\bsnm{{Rosa}}, \binits{G.L.}},
\bauthor{\bsnm{{Lombardi}}, \binits{G.}},
\bauthor{\bsnm{{Malcovati}}, \binits{P.}},
\bauthor{\bsnm{{Mele}}, \binits{F.}},
\bauthor{\bsnm{{Nogara}}, \binits{P.}},
\bauthor{\bsnm{{Nuti}}, \binits{A.}},
\bauthor{\bsnm{{Perri}}, \binits{M.}},
\bauthor{\bsnm{{Pliego-Caballero}}, \binits{S.}},
\bauthor{\bsnm{{Pirrotta}}, \binits{S.}},
\bauthor{\bsnm{{Puccetti}}, \binits{S.}},
\bauthor{\bsnm{{Rashevskaya}}, \binits{I.}},
\bauthor{\bsnm{{Russo}}, \binits{F.}},
\bauthor{\bsnm{{Sottile}}, \binits{G.}},
\bauthor{\bsnm{{Tenzer}}, \binits{C.}},
\bauthor{\bsnm{{Trenti}}, \binits{M.}},
\bauthor{\bsnm{{Trevisan}}, \binits{S.}},
\bauthor{\bsnm{{Vacchi}}, \binits{A.}},
\bauthor{\bsnm{{Zampa}}, \binits{G.}},
\bauthor{\bsnm{{Zampa}}, \binits{N.}},
\bauthor{\bsnm{{Fiore}}, \binits{F.}}:
\batitle{{The ground calibration of the HERMES-Pathfinder payload flight
  models}}.
\bjtitle{Experimental Astronomy}
\bvolume{58}(\bissue{3}),
\bfpage{14}
(\byear{2024})
\doiurl{10.1007/s10686-024-09958-4}
{\href{https://arxiv.org/abs/2410.06056}{{arXiv:2410.06056}}}
{[astro-ph.IM]}
\end{barticle}
\endbibitem

\bibitem[\protect\citeauthoryear{{Campana} et~al.}{2024}]{campana24}
\begin{bchapter}
\bauthor{\bsnm{{Campana}}, \binits{R.}},
\bauthor{\bsnm{{Evangelista}}, \binits{Y.}},
\bauthor{\bsnm{{Fiore}}, \binits{F.}},
\bauthor{\bsnm{{Guzm{\'a}n}}, \binits{A.}},
\bauthor{\bsnm{{Baroni}}, \binits{G.}},
\bauthor{\bsnm{{Della Casa}}, \binits{G.}},
\bauthor{\bsnm{{Dilillo}}, \binits{G.}},
\bauthor{\bsnm{{Hedderman}}, \binits{P.}},
\bauthor{\bsnm{{Marchesini}}, \binits{E.J.}},
\bauthor{\bsnm{{Bertuccio}}, \binits{G.}},
\bauthor{\bsnm{{Ceraudo}}, \binits{F.}},
\bauthor{\bsnm{{Demenev}}, \binits{E.}},
\bauthor{\bsnm{{Fiorini}}, \binits{M.}},
\bauthor{\bsnm{{Grassi}}, \binits{M.}},
\bauthor{\bsnm{{Malcovati}}, \binits{P.}},
\bauthor{\bsnm{{Mele}}, \binits{F.}},
\bauthor{\bsnm{{Nogara}}, \binits{P.}},
\bauthor{\bsnm{{Nuti}}, \binits{A.}},
\bauthor{\bsnm{{Perri}}, \binits{M.}},
\bauthor{\bsnm{{Pirrotta}}, \binits{S.}},
\bauthor{\bsnm{{Pliego-Caballero}}, \binits{S.}},
\bauthor{\bsnm{{Puccetti}}, \binits{S.}},
\bauthor{\bsnm{{Sottile}}, \binits{G.}},
\bauthor{\bsnm{{Russo}}, \binits{F.}},
\bauthor{\bsnm{{Trevisan}}, \binits{S.}}:
\bctitle{{Design and development of the HERMES Pathfinder payloads}}.
In: \beditor{\bsnm{{den Herder}}, \binits{J.-W.A.}},
\beditor{\bsnm{{Nikzad}}, \binits{S.}},
\beditor{\bsnm{{Nakazawa}}, \binits{K.}} (eds.)
\bbtitle{Space Telescopes and Instrumentation 2024: Ultraviolet to Gamma Ray}.
\bsertitle{Society of Photo-Optical Instrumentation Engineers (SPIE) Conference
  Series},
vol. \bseriesno{13093},
p. \bfpage{130936}
(\byear{2024}).
\doiurl{10.1117/12.3018007}
\end{bchapter}
\endbibitem

\bibitem[\protect\citeauthoryear{{Trenti} et~al.}{2024}]{trenti24}
\begin{botherref}
\oauthor{\bsnm{{Trenti}}, \binits{M.}},
\oauthor{\bsnm{{Ortiz del Castillo}}, \binits{M.}},
\oauthor{\bsnm{{Mearns}}, \binits{R.}},
\oauthor{\bsnm{{McRobbie}}, \binits{J.}},
\oauthor{\bsnm{{Therakam}}, \binits{C.}},
\oauthor{\bsnm{{Chapman}}, \binits{A.}},
\oauthor{\bsnm{{Woods}}, \binits{A.}},
\oauthor{\bsnm{{Morgan}}, \binits{J.}},
\oauthor{\bsnm{{Barraclough}}, \binits{S.}},
\oauthor{\bsnm{{Rodriguez Mallo}}, \binits{I.}},
\oauthor{\bsnm{{Baroni}}, \binits{G.}},
\oauthor{\bsnm{{Fiore}}, \binits{F.}},
\oauthor{\bsnm{{Evangelista}}, \binits{Y.}},
\oauthor{\bsnm{{Campana}}, \binits{R.}},
\oauthor{\bsnm{{Guzman}}, \binits{A.}},
\oauthor{\bsnm{{Hedderman}}, \binits{P.}}:
{SpIRIT Mission: In-Orbit Results and Technology Demonstrations}.
arXiv e-prints,
2407--14034
(2024)
\doiurl{10.48550/arXiv.2407.14034}
{\href{https://arxiv.org/abs/2407.14034}{{arXiv:2407.14034}}}
{[astro-ph.IM]}
\end{botherref}
\endbibitem

\bibitem[\protect\citeauthoryear{{Campana} et~al.}{2021}]{campana21}
\begin{bchapter}
\bauthor{\bsnm{{Campana}}, \binits{R.}},
\bauthor{\bsnm{{Fuschino}}, \binits{F.}},
\bauthor{\bsnm{{Evangelista}}, \binits{Y.}},
\bauthor{\bsnm{{Dilillo}}, \binits{G.}},
\bauthor{\bsnm{{Fiore}}, \binits{F.}}:
\bctitle{{The HERMES-TP/SP background and response simulations}}.
In: \beditor{\bsnm{{den Herder}}, \binits{J.-W.A.}},
\beditor{\bsnm{{Nikzad}}, \binits{S.}},
\beditor{\bsnm{{Nakazawa}}, \binits{K.}} (eds.)
\bbtitle{Society of Photo-Optical Instrumentation Engineers (SPIE) Conference
  Series}.
\bsertitle{Society of Photo-Optical Instrumentation Engineers (SPIE) Conference
  Series},
vol. \bseriesno{11444},
p. \bfpage{114444}
(\byear{2021}).
\doiurl{10.1117/12.2560365}
\end{bchapter}
\endbibitem

\bibitem[\protect\citeauthoryear{{{\v{R}}{\'\i}pa} et~al.}{2021}]{ripa21}
\begin{bchapter}
\bauthor{\bsnm{{{\v{R}}{\'\i}pa}}, \binits{J.}},
\bauthor{\bsnm{{Dilillo}}, \binits{G.}},
\bauthor{\bsnm{{Campana}}, \binits{R.}},
\bauthor{\bsnm{{Galg{\'o}czi}}, \binits{G.}}:
\bctitle{{A comparison of trapped particle models in low Earth orbit}}.
In: \beditor{\bsnm{{den Herder}}, \binits{J.-W.A.}},
\beditor{\bsnm{{Nikzad}}, \binits{S.}},
\beditor{\bsnm{{Nakazawa}}, \binits{K.}} (eds.)
\bbtitle{Society of Photo-Optical Instrumentation Engineers (SPIE) Conference
  Series}.
\bsertitle{Society of Photo-Optical Instrumentation Engineers (SPIE) Conference
  Series},
vol. \bseriesno{11444},
p. \bfpage{114443}
(\byear{2021}).
\doiurl{10.1117/12.2561011}
\end{bchapter}
\endbibitem

\bibitem[\protect\citeauthoryear{{Della Casa} et~al.}{2024}]{dellacasa24}
\begin{barticle}
\bauthor{\bsnm{{Della Casa}}, \binits{G.}},
\bauthor{\bsnm{Zampa}, \binits{N.}},
\bauthor{\bsnm{Cirrincione}, \binits{D.}},
\bauthor{\bsnm{Monzani}, \binits{S.}},
\bauthor{\bsnm{Baruzzo}, \binits{M.}},
\bauthor{\bsnm{Campana}, \binits{R.}},
\bauthor{\bsnm{Cauz}, \binits{D.}},
\bauthor{\bsnm{Citossi}, \binits{M.}},
\bauthor{\bsnm{Crupi}, \binits{R.}},
\bauthor{\bsnm{Dilillo}, \binits{G.}},
\bauthor{\bsnm{Pauletta}, \binits{G.}},
\bauthor{\bsnm{Fiore}, \binits{F.}},
\bauthor{\bsnm{Vacchi}, \binits{A.}}:
\batitle{{New detailed characterization of the residual luminescence emitted by
  the GAGG:Ce scintillator crystals for the HERMES Pathfinder mission}}.
\bjtitle{Nuclear Instruments and Methods in Physics Research Section A:
  Accelerators, Spectrometers, Detectors and Associated Equipment}
\bvolume{1058},
\bfpage{168825}
(\byear{2024})
\doiurl{10.1016/j.nima.2023.168825}
\end{barticle}
\endbibitem

\bibitem[\protect\citeauthoryear{{Zhang} et~al.}{2019}]{zhang19}
\begin{barticle}
\bauthor{\bsnm{{Zhang}}, \binits{S.}},
\bauthor{\bsnm{{Santangelo}}, \binits{A.}},
\bauthor{\bsnm{{Feroci}}, \binits{M.}},
\bauthor{\bsnm{{Xu}}, \binits{Y.}},
\bauthor{\bsnm{{Lu}}, \binits{F.}},
\bauthor{\bsnm{{Chen}}, \binits{Y.}},
\bauthor{\bsnm{{Feng}}, \binits{H.}},
\bauthor{\bsnm{{Zhang}}, \binits{S.}},
\bauthor{\bsnm{{Brandt}}, \binits{S.}},
\bauthor{\bsnm{{Hernanz}}, \binits{M.}},
\bauthor{\bsnm{{Baldini}}, \binits{L.}},
\bauthor{\bsnm{{Bozzo}}, \binits{E.}},
\bauthor{\bsnm{{Campana}}, \binits{R.}},
\bauthor{\bsnm{{De Rosa}}, \binits{A.}},
\bauthor{\bsnm{{Dong}}, \binits{Y.}},
\bauthor{\bsnm{{Evangelista}}, \binits{Y.}},
\bauthor{\bsnm{{Karas}}, \binits{V.}},
\bauthor{\bsnm{{Meidinger}}, \binits{N.}},
\bauthor{\bsnm{{Meuris}}, \binits{A.}},
\bauthor{\bsnm{{Nandra}}, \binits{K.}},
\bauthor{\bsnm{{Pan}}, \binits{T.}},
\bauthor{\bsnm{{Pareschi}}, \binits{G.}},
\bauthor{\bsnm{{Orleanski}}, \binits{P.}},
\bauthor{\bsnm{{Huang}}, \binits{Q.}},
\bauthor{\bsnm{{Schanne}}, \binits{S.}},
\bauthor{\bsnm{{Sironi}}, \binits{G.}},
\bauthor{\bsnm{{Spiga}}, \binits{D.}},
\bauthor{\bsnm{{Svoboda}}, \binits{J.}},
\bauthor{\bsnm{{Tagliaferri}}, \binits{G.}},
\bauthor{\bsnm{{Tenzer}}, \binits{C.}},
\bauthor{\bsnm{{Vacchi}}, \binits{A.}},
\bauthor{\bsnm{{Zane}}, \binits{S.}},
\bauthor{\bsnm{{Walton}}, \binits{D.}},
\bauthor{\bsnm{{Wang}}, \binits{Z.}},
\bauthor{\bsnm{{Winter}}, \binits{B.}},
\bauthor{\bsnm{{Wu}}, \binits{X.}},
\bauthor{\bsnm{{in't Zand}}, \binits{J.J.M.}},
\bauthor{\bsnm{{Ahangarianabhari}}, \binits{M.}},
\bauthor{\bsnm{{Ambrosi}}, \binits{G.}},
\bauthor{\bsnm{{Ambrosino}}, \binits{F.}},
\bauthor{\bsnm{{Barbera}}, \binits{M.}},
\bauthor{\bsnm{{Basso}}, \binits{S.}},
\bauthor{\bsnm{{Bayer}}, \binits{J.}},
\bauthor{\bsnm{{Bellazzini}}, \binits{R.}},
\bauthor{\bsnm{{Bellutti}}, \binits{P.}},
\bauthor{\bsnm{{Bertucci}}, \binits{B.}},
\bauthor{\bsnm{{Bertuccio}}, \binits{G.}},
\bauthor{\bsnm{{Borghi}}, \binits{G.}},
\bauthor{\bsnm{{Cao}}, \binits{X.}},
\bauthor{\bsnm{{Cadoux}}, \binits{F.}},
\bauthor{\bsnm{{Campana}}, \binits{R.}},
\bauthor{\bsnm{{Ceraudo}}, \binits{F.}},
\bauthor{\bsnm{{Chen}}, \binits{T.}},
\bauthor{\bsnm{{Chen}}, \binits{Y.}},
\bauthor{\bsnm{{Chevenez}}, \binits{J.}},
\bauthor{\bsnm{{Civitani}}, \binits{M.}},
\bauthor{\bsnm{{Cui}}, \binits{W.}},
\bauthor{\bsnm{{Cui}}, \binits{W.}},
\bauthor{\bsnm{{Dauser}}, \binits{T.}},
\bauthor{\bsnm{{Del Monte}}, \binits{E.}},
\bauthor{\bsnm{{Di Cosimo}}, \binits{S.}},
\bauthor{\bsnm{{Diebold}}, \binits{S.}},
\bauthor{\bsnm{{Doroshenko}}, \binits{V.}},
\bauthor{\bsnm{{Dovciak}}, \binits{M.}},
\bauthor{\bsnm{{Du}}, \binits{Y.}},
\bauthor{\bsnm{{Ducci}}, \binits{L.}},
\bauthor{\bsnm{{Fan}}, \binits{Q.}},
\bauthor{\bsnm{{Favre}}, \binits{Y.}},
\bauthor{\bsnm{{Fuschino}}, \binits{F.}},
\bauthor{\bsnm{{G{\'a}lvez}}, \binits{J.L.}},
\bauthor{\bsnm{{Gao}}, \binits{M.}},
\bauthor{\bsnm{{Ge}}, \binits{M.}},
\bauthor{\bsnm{{Gevin}}, \binits{O.}},
\bauthor{\bsnm{{Grassi}}, \binits{M.}},
\bauthor{\bsnm{{Gu}}, \binits{Q.}},
\bauthor{\bsnm{{Gu}}, \binits{Y.}},
\bauthor{\bsnm{{Han}}, \binits{D.}},
\bauthor{\bsnm{{Hong}}, \binits{B.}},
\bauthor{\bsnm{{Hu}}, \binits{W.}},
\bauthor{\bsnm{{Ji}}, \binits{L.}},
\bauthor{\bsnm{{Jia}}, \binits{S.}},
\bauthor{\bsnm{{Jiang}}, \binits{W.}},
\bauthor{\bsnm{{Kennedy}}, \binits{T.}},
\bauthor{\bsnm{{Kreykenbohm}}, \binits{I.}},
\bauthor{\bsnm{{Kuvvetli}}, \binits{I.}},
\bauthor{\bsnm{{Labanti}}, \binits{C.}},
\bauthor{\bsnm{{Latronico}}, \binits{L.}},
\bauthor{\bsnm{{Li}}, \binits{G.}},
\bauthor{\bsnm{{Li}}, \binits{M.}},
\bauthor{\bsnm{{Li}}, \binits{X.}},
\bauthor{\bsnm{{Li}}, \binits{W.}},
\bauthor{\bsnm{{Li}}, \binits{Z.}},
\bauthor{\bsnm{{Limousin}}, \binits{O.}},
\bauthor{\bsnm{{Liu}}, \binits{H.}},
\bauthor{\bsnm{{Liu}}, \binits{X.}},
\bauthor{\bsnm{{Lu}}, \binits{B.}},
\bauthor{\bsnm{{Luo}}, \binits{T.}},
\bauthor{\bsnm{{Macera}}, \binits{D.}},
\bauthor{\bsnm{{Malcovati}}, \binits{P.}},
\bauthor{\bsnm{{Martindale}}, \binits{A.}},
\bauthor{\bsnm{{Michalska}}, \binits{M.}},
\bauthor{\bsnm{{Meng}}, \binits{B.}},
\bauthor{\bsnm{{Minuti}}, \binits{M.}},
\bauthor{\bsnm{{Morbidini}}, \binits{A.}},
\bauthor{\bsnm{{Muleri}}, \binits{F.}},
\bauthor{\bsnm{{Paltani}}, \binits{S.}},
\bauthor{\bsnm{{Perinati}}, \binits{E.}},
\bauthor{\bsnm{{Picciotto}}, \binits{A.}},
\bauthor{\bsnm{{Piemonte}}, \binits{C.}},
\bauthor{\bsnm{{Qu}}, \binits{J.}},
\bauthor{\bsnm{{Rachevski}}, \binits{A.}},
\bauthor{\bsnm{{Rashevskaya}}, \binits{I.}},
\bauthor{\bsnm{{Rodriguez}}, \binits{J.}},
\bauthor{\bsnm{{Schanz}}, \binits{T.}},
\bauthor{\bsnm{{Shen}}, \binits{Z.}},
\bauthor{\bsnm{{Sheng}}, \binits{L.}},
\bauthor{\bsnm{{Song}}, \binits{J.}},
\bauthor{\bsnm{{Song}}, \binits{L.}},
\bauthor{\bsnm{{Sgro}}, \binits{C.}},
\bauthor{\bsnm{{Sun}}, \binits{L.}},
\bauthor{\bsnm{{Tan}}, \binits{Y.}},
\bauthor{\bsnm{{Uttley}}, \binits{P.}},
\bauthor{\bsnm{{Wang}}, \binits{B.}},
\bauthor{\bsnm{{Wang}}, \binits{D.}},
\bauthor{\bsnm{{Wang}}, \binits{G.}},
\bauthor{\bsnm{{Wang}}, \binits{J.}},
\bauthor{\bsnm{{Wang}}, \binits{L.}},
\bauthor{\bsnm{{Wang}}, \binits{Y.}},
\bauthor{\bsnm{{Watts}}, \binits{A.L.}},
\bauthor{\bsnm{{Wen}}, \binits{X.}},
\bauthor{\bsnm{{Wilms}}, \binits{J.}},
\bauthor{\bsnm{{Xiong}}, \binits{S.}},
\bauthor{\bsnm{{Yang}}, \binits{J.}},
\bauthor{\bsnm{{Yang}}, \binits{S.}},
\bauthor{\bsnm{{Yang}}, \binits{Y.}},
\bauthor{\bsnm{{Yu}}, \binits{N.}},
\bauthor{\bsnm{{Zhang}}, \binits{W.}},
\bauthor{\bsnm{{Zampa}}, \binits{G.}},
\bauthor{\bsnm{{Zampa}}, \binits{N.}},
\bauthor{\bsnm{{Zdziarski}}, \binits{A.A.}},
\bauthor{\bsnm{{Zhang}}, \binits{A.}},
\bauthor{\bsnm{{Zhang}}, \binits{C.}},
\bauthor{\bsnm{{Zhang}}, \binits{F.}},
\bauthor{\bsnm{{Zhang}}, \binits{L.}},
\bauthor{\bsnm{{Zhang}}, \binits{T.}},
\bauthor{\bsnm{{Zhang}}, \binits{Y.}},
\bauthor{\bsnm{{Zhang}}, \binits{X.}},
\bauthor{\bsnm{{Zhang}}, \binits{Z.}},
\bauthor{\bsnm{{Zhao}}, \binits{B.}},
\bauthor{\bsnm{{Zheng}}, \binits{S.}},
\bauthor{\bsnm{{Zhou}}, \binits{Y.}},
\bauthor{\bsnm{{Zorzi}}, \binits{N.}},
\bauthor{\bsnm{{Zwart}}, \binits{J.F.}}:
\batitle{{The enhanced X-ray Timing and Polarimetry mission{\textemdash}eXTP}}.
\bjtitle{Science China Physics, Mechanics, and Astronomy}
\bvolume{62}(\bissue{2}),
\bfpage{29502}
(\byear{2019})
\doiurl{10.1007/s11433-018-9309-2}
{\href{https://arxiv.org/abs/1812.04020}{{arXiv:1812.04020}}}
{[astro-ph.IM]}
\end{barticle}
\endbibitem

\bibitem[\protect\citeauthoryear{{Feroci} et~al.}{2012}]{feroci12}
\begin{barticle}
\bauthor{\bsnm{{Feroci}}, \binits{M.}},
\bauthor{\bsnm{{Stella}}, \binits{L.}},
\bauthor{\bsnm{{van der Klis}}, \binits{M.}},
\bauthor{\bsnm{{Courvoisier}}, \binits{T.J.-L.}},
\bauthor{\bsnm{{Hernanz}}, \binits{M.}},
\bauthor{\bsnm{{Hudec}}, \binits{R.}},
\bauthor{\bsnm{{Santangelo}}, \binits{A.}},
\bauthor{\bsnm{{Walton}}, \binits{D.}},
\bauthor{\bsnm{{Zdziarski}}, \binits{A.}},
\bauthor{\bsnm{{Barret}}, \binits{D.}},
\bauthor{\bsnm{{Belloni}}, \binits{T.}},
\bauthor{\bsnm{{Braga}}, \binits{J.}},
\bauthor{\bsnm{{Brandt}}, \binits{S.}},
\bauthor{\bsnm{{Budtz-J{\o}rgensen}}, \binits{C.}},
\bauthor{\bsnm{{Campana}}, \binits{S.}},
\bauthor{\bsnm{{den Herder}}, \binits{J.-W.}},
\bauthor{\bsnm{{Huovelin}}, \binits{J.}},
\bauthor{\bsnm{{Israel}}, \binits{G.L.}},
\bauthor{\bsnm{{Pohl}}, \binits{M.}},
\bauthor{\bsnm{{Ray}}, \binits{P.}},
\bauthor{\bsnm{{Vacchi}}, \binits{A.}},
\bauthor{\bsnm{{Zane}}, \binits{S.}},
\bauthor{\bsnm{{Argan}}, \binits{A.}},
\bauthor{\bsnm{{Attin{\`a}}}, \binits{P.}},
\bauthor{\bsnm{{Bertuccio}}, \binits{G.}},
\bauthor{\bsnm{{Bozzo}}, \binits{E.}},
\bauthor{\bsnm{{Campana}}, \binits{R.}},
\bauthor{\bsnm{{Chakrabarty}}, \binits{D.}},
\bauthor{\bsnm{{Costa}}, \binits{E.}},
\bauthor{\bsnm{{De Rosa}}, \binits{A.}},
\bauthor{\bsnm{{Del Monte}}, \binits{E.}},
\bauthor{\bsnm{{Di Cosimo}}, \binits{S.}},
\bauthor{\bsnm{{Donnarumma}}, \binits{I.}},
\bauthor{\bsnm{{Evangelista}}, \binits{Y.}},
\bauthor{\bsnm{{Haas}}, \binits{D.}},
\bauthor{\bsnm{{Jonker}}, \binits{P.}},
\bauthor{\bsnm{{Korpela}}, \binits{S.}},
\bauthor{\bsnm{{Labanti}}, \binits{C.}},
\bauthor{\bsnm{{Malcovati}}, \binits{P.}},
\bauthor{\bsnm{{Mignani}}, \binits{R.}},
\bauthor{\bsnm{{Muleri}}, \binits{F.}},
\bauthor{\bsnm{{Rapisarda}}, \binits{M.}},
\bauthor{\bsnm{{Rashevsky}}, \binits{A.}},
\bauthor{\bsnm{{Rea}}, \binits{N.}},
\bauthor{\bsnm{{Rubini}}, \binits{A.}},
\bauthor{\bsnm{{Tenzer}}, \binits{C.}},
\bauthor{\bsnm{{Wilson-Hodge}}, \binits{C.}},
\bauthor{\bsnm{{Winter}}, \binits{B.}},
\bauthor{\bsnm{{Wood}}, \binits{K.}},
\bauthor{\bsnm{{Zampa}}, \binits{G.}},
\bauthor{\bsnm{{Zampa}}, \binits{N.}},
\bauthor{\bsnm{{Abramowicz}}, \binits{M.A.}},
\bauthor{\bsnm{{Alpar}}, \binits{M.A.}},
\bauthor{\bsnm{{Altamirano}}, \binits{D.}},
\bauthor{\bsnm{{Alvarez}}, \binits{J.M.}},
\bauthor{\bsnm{{Amati}}, \binits{L.}},
\bauthor{\bsnm{{Amoros}}, \binits{C.}},
\bauthor{\bsnm{{Antonelli}}, \binits{L.A.}},
\bauthor{\bsnm{{Artigue}}, \binits{R.}},
\bauthor{\bsnm{{Azzarello}}, \binits{P.}},
\bauthor{\bsnm{{Bachetti}}, \binits{M.}},
\bauthor{\bsnm{{Baldazzi}}, \binits{G.}},
\bauthor{\bsnm{{Barbera}}, \binits{M.}},
\bauthor{\bsnm{{Barbieri}}, \binits{C.}},
\bauthor{\bsnm{{Basa}}, \binits{S.}},
\bauthor{\bsnm{{Baykal}}, \binits{A.}},
\bauthor{\bsnm{{Belmont}}, \binits{R.}},
\bauthor{\bsnm{{Boirin}}, \binits{L.}},
\bauthor{\bsnm{{Bonvicini}}, \binits{V.}},
\bauthor{\bsnm{{Burderi}}, \binits{L.}},
\bauthor{\bsnm{{Bursa}}, \binits{M.}},
\bauthor{\bsnm{{Cabanac}}, \binits{C.}},
\bauthor{\bsnm{{Cackett}}, \binits{E.}},
\bauthor{\bsnm{{Caliandro}}, \binits{G.A.}},
\bauthor{\bsnm{{Casella}}, \binits{P.}},
\bauthor{\bsnm{{Chaty}}, \binits{S.}},
\bauthor{\bsnm{{Chenevez}}, \binits{J.}},
\bauthor{\bsnm{{Coe}}, \binits{M.J.}},
\bauthor{\bsnm{{Collura}}, \binits{A.}},
\bauthor{\bsnm{{Corongiu}}, \binits{A.}},
\bauthor{\bsnm{{Covino}}, \binits{S.}},
\bauthor{\bsnm{{Cusumano}}, \binits{G.}},
\bauthor{\bsnm{{D'Amico}}, \binits{F.}},
\bauthor{\bsnm{{Dall'Osso}}, \binits{S.}},
\bauthor{\bsnm{{De Martino}}, \binits{D.}},
\bauthor{\bsnm{{De Paris}}, \binits{G.}},
\bauthor{\bsnm{{Di Persio}}, \binits{G.}},
\bauthor{\bsnm{{Di Salvo}}, \binits{T.}},
\bauthor{\bsnm{{Done}}, \binits{C.}},
\bauthor{\bsnm{{Dov{\v{c}}iak}}, \binits{M.}},
\bauthor{\bsnm{{Drago}}, \binits{A.}},
\bauthor{\bsnm{{Ertan}}, \binits{U.}},
\bauthor{\bsnm{{Fabiani}}, \binits{S.}},
\bauthor{\bsnm{{Falanga}}, \binits{M.}},
\bauthor{\bsnm{{Fender}}, \binits{R.}},
\bauthor{\bsnm{{Ferrando}}, \binits{P.}},
\bauthor{\bsnm{{Della Monica Ferreira}}, \binits{D.}},
\bauthor{\bsnm{{Fraser}}, \binits{G.}},
\bauthor{\bsnm{{Frontera}}, \binits{F.}},
\bauthor{\bsnm{{Fuschino}}, \binits{F.}},
\bauthor{\bsnm{{Galvez}}, \binits{J.L.}},
\bauthor{\bsnm{{Gandhi}}, \binits{P.}},
\bauthor{\bsnm{{Giommi}}, \binits{P.}},
\bauthor{\bsnm{{Godet}}, \binits{O.}},
\bauthor{\bsnm{{G{\"o}{\v{g}}{\"u}{\c{s}}}}, \binits{E.}},
\bauthor{\bsnm{{Goldwurm}}, \binits{A.}},
\bauthor{\bsnm{{G{\"o}tz}}, \binits{D.}},
\bauthor{\bsnm{{Grassi}}, \binits{M.}},
\bauthor{\bsnm{{Guttridge}}, \binits{P.}},
\bauthor{\bsnm{{Hakala}}, \binits{P.}},
\bauthor{\bsnm{{Henri}}, \binits{G.}},
\bauthor{\bsnm{{Hermsen}}, \binits{W.}},
\bauthor{\bsnm{{Horak}}, \binits{J.}},
\bauthor{\bsnm{{Hornstrup}}, \binits{A.}},
\bauthor{\bsnm{{in't Zand}}, \binits{J.J.M.}},
\bauthor{\bsnm{{Isern}}, \binits{J.}},
\bauthor{\bsnm{{Kalemci}}, \binits{E.}},
\bauthor{\bsnm{{Kanbach}}, \binits{G.}},
\bauthor{\bsnm{{Karas}}, \binits{V.}},
\bauthor{\bsnm{{Kataria}}, \binits{D.}},
\bauthor{\bsnm{{Kennedy}}, \binits{T.}},
\bauthor{\bsnm{{Klochkov}}, \binits{D.}},
\bauthor{\bsnm{{Klu{\'z}niak}}, \binits{W.}},
\bauthor{\bsnm{{Kokkotas}}, \binits{K.}},
\bauthor{\bsnm{{Kreykenbohm}}, \binits{I.}},
\bauthor{\bsnm{{Krolik}}, \binits{J.}},
\bauthor{\bsnm{{Kuiper}}, \binits{L.}},
\bauthor{\bsnm{{Kuvvetli}}, \binits{I.}},
\bauthor{\bsnm{{Kylafis}}, \binits{N.}},
\bauthor{\bsnm{{Lattimer}}, \binits{J.M.}},
\bauthor{\bsnm{{Lazzarotto}}, \binits{F.}},
\bauthor{\bsnm{{Leahy}}, \binits{D.}},
\bauthor{\bsnm{{Lebrun}}, \binits{F.}},
\bauthor{\bsnm{{Lin}}, \binits{D.}},
\bauthor{\bsnm{{Lund}}, \binits{N.}},
\bauthor{\bsnm{{Maccarone}}, \binits{T.}},
\bauthor{\bsnm{{Malzac}}, \binits{J.}},
\bauthor{\bsnm{{Marisaldi}}, \binits{M.}},
\bauthor{\bsnm{{Martindale}}, \binits{A.}},
\bauthor{\bsnm{{Mastropietro}}, \binits{M.}},
\bauthor{\bsnm{{McClintock}}, \binits{J.}},
\bauthor{\bsnm{{McHardy}}, \binits{I.}},
\bauthor{\bsnm{{Mendez}}, \binits{M.}},
\bauthor{\bsnm{{Mereghetti}}, \binits{S.}},
\bauthor{\bsnm{{Miller}}, \binits{M.C.}},
\bauthor{\bsnm{{Mineo}}, \binits{T.}},
\bauthor{\bsnm{{Morelli}}, \binits{E.}},
\bauthor{\bsnm{{Morsink}}, \binits{S.}},
\bauthor{\bsnm{{Motch}}, \binits{C.}},
\bauthor{\bsnm{{Motta}}, \binits{S.}},
\bauthor{\bsnm{{Mu{\~n}oz-Darias}}, \binits{T.}},
\bauthor{\bsnm{{Naletto}}, \binits{G.}},
\bauthor{\bsnm{{Neustroev}}, \binits{V.}},
\bauthor{\bsnm{{Nevalainen}}, \binits{J.}},
\bauthor{\bsnm{{Olive}}, \binits{J.F.}},
\bauthor{\bsnm{{Orio}}, \binits{M.}},
\bauthor{\bsnm{{Orlandini}}, \binits{M.}},
\bauthor{\bsnm{{Orleanski}}, \binits{P.}},
\bauthor{\bsnm{{Ozel}}, \binits{F.}},
\bauthor{\bsnm{{Pacciani}}, \binits{L.}},
\bauthor{\bsnm{{Paltani}}, \binits{S.}},
\bauthor{\bsnm{{Papadakis}}, \binits{I.}},
\bauthor{\bsnm{{Papitto}}, \binits{A.}},
\bauthor{\bsnm{{Patruno}}, \binits{A.}},
\bauthor{\bsnm{{Pellizzoni}}, \binits{A.}},
\bauthor{\bsnm{{Petr{\'a}{\v{c}}ek}}, \binits{V.}},
\bauthor{\bsnm{{Petri}}, \binits{J.}},
\bauthor{\bsnm{{Petrucci}}, \binits{P.O.}},
\bauthor{\bsnm{{Phlips}}, \binits{B.}},
\bauthor{\bsnm{{Picolli}}, \binits{L.}},
\bauthor{\bsnm{{Possenti}}, \binits{A.}},
\bauthor{\bsnm{{Psaltis}}, \binits{D.}},
\bauthor{\bsnm{{Rambaud}}, \binits{D.}},
\bauthor{\bsnm{{Reig}}, \binits{P.}},
\bauthor{\bsnm{{Remillard}}, \binits{R.}},
\bauthor{\bsnm{{Rodriguez}}, \binits{J.}},
\bauthor{\bsnm{{Romano}}, \binits{P.}},
\bauthor{\bsnm{{Romanova}}, \binits{M.}},
\bauthor{\bsnm{{Schanz}}, \binits{T.}},
\bauthor{\bsnm{{Schmid}}, \binits{C.}},
\bauthor{\bsnm{{Segreto}}, \binits{A.}},
\bauthor{\bsnm{{Shearer}}, \binits{A.}},
\bauthor{\bsnm{{Smith}}, \binits{A.}},
\bauthor{\bsnm{{Smith}}, \binits{P.J.}},
\bauthor{\bsnm{{Soffitta}}, \binits{P.}},
\bauthor{\bsnm{{Stergioulas}}, \binits{N.}},
\bauthor{\bsnm{{Stolarski}}, \binits{M.}},
\bauthor{\bsnm{{Stuchlik}}, \binits{Z.}},
\bauthor{\bsnm{{Tiengo}}, \binits{A.}},
\bauthor{\bsnm{{Torres}}, \binits{D.}},
\bauthor{\bsnm{{T{\"o}r{\"o}k}}, \binits{G.}},
\bauthor{\bsnm{{Turolla}}, \binits{R.}},
\bauthor{\bsnm{{Uttley}}, \binits{P.}},
\bauthor{\bsnm{{Vaughan}}, \binits{S.}},
\bauthor{\bsnm{{Vercellone}}, \binits{S.}},
\bauthor{\bsnm{{Waters}}, \binits{R.}},
\bauthor{\bsnm{{Watts}}, \binits{A.}},
\bauthor{\bsnm{{Wawrzaszek}}, \binits{R.}},
\bauthor{\bsnm{{Webb}}, \binits{N.}},
\bauthor{\bsnm{{Wilms}}, \binits{J.}}:
\batitle{{The Large Observatory for X-ray Timing (LOFT)}}.
\bjtitle{Experimental Astronomy}
\bvolume{34}(\bissue{2}),
\bfpage{415}--\blpage{444}
(\byear{2012})
\doiurl{10.1007/s10686-011-9237-2}
{\href{https://arxiv.org/abs/1107.0436}{{arXiv:1107.0436}}}
{[astro-ph.IM]}
\end{barticle}
\endbibitem

\bibitem[\protect\citeauthoryear{{Zhang} et~al.}{2025}]{zhang25}
\begin{barticle}
\bauthor{\bsnm{{Zhang}}, \binits{S.-N.}},
\bauthor{\bsnm{{Santangelo}}, \binits{A.}},
\bauthor{\bsnm{{Xu}}, \binits{Y.}},
\bauthor{\bsnm{{Feng}}, \binits{H.}},
\bauthor{\bsnm{{Lu}}, \binits{F.}},
\bauthor{\bsnm{{Chen}}, \binits{Y.}},
\bauthor{\bsnm{{Ge}}, \binits{M.}},
\bauthor{\bsnm{{Nandra}}, \binits{K.}},
\bauthor{\bsnm{{Wu}}, \binits{X.}},
\bauthor{\bsnm{{Feroci}}, \binits{M.}},
\bauthor{\bsnm{{Hernanz}}, \binits{M.}},
\bauthor{\bsnm{{Liu}}, \binits{C.}},
\bauthor{\bsnm{{He}}, \binits{H.}},
\bauthor{\bsnm{{Wang}}, \binits{Y.}},
\bauthor{\bsnm{{Jiang}}, \binits{W.}},
\bauthor{\bsnm{{Cui}}, \binits{W.}},
\bauthor{\bsnm{{Yang}}, \binits{Y.}},
\bauthor{\bsnm{{Wang}}, \binits{J.}},
\bauthor{\bsnm{{Li}}, \binits{W.}},
\bauthor{\bsnm{{Li}}, \binits{H.}},
\bauthor{\bsnm{{Du}}, \binits{Y.}},
\bauthor{\bsnm{{Liu}}, \binits{X.}},
\bauthor{\bsnm{{Meng}}, \binits{B.}},
\bauthor{\bsnm{{Wen}}, \binits{X.}},
\bauthor{\bsnm{{Zhang}}, \binits{A.}},
\bauthor{\bsnm{{Ma}}, \binits{J.}},
\bauthor{\bsnm{{Li}}, \binits{M.}},
\bauthor{\bsnm{{Li}}, \binits{G.}},
\bauthor{\bsnm{{Qi}}, \binits{L.}},
\bauthor{\bsnm{{Sun}}, \binits{J.}},
\bauthor{\bsnm{{Luo}}, \binits{T.}},
\bauthor{\bsnm{{Liu}}, \binits{H.}},
\bauthor{\bsnm{{Liu}}, \binits{X.}},
\bauthor{\bsnm{{Zhang}}, \binits{F.}},
\bauthor{\bsnm{{Luo}}, \binits{L.}},
\bauthor{\bsnm{{Zhu}}, \binits{Y.}},
\bauthor{\bsnm{{Zhao}}, \binits{Z.}},
\bauthor{\bsnm{{Sun}}, \binits{L.}},
\bauthor{\bsnm{{Yang}}, \binits{X.}},
\bauthor{\bsnm{{Wu}}, \binits{Q.}},
\bauthor{\bsnm{{Jiang}}, \binits{J.}},
\bauthor{\bsnm{{Shi}}, \binits{H.}},
\bauthor{\bsnm{{Liu}}, \binits{J.}},
\bauthor{\bsnm{{Xu}}, \binits{Y.}},
\bauthor{\bsnm{{Yang}}, \binits{S.}},
\bauthor{\bsnm{{Zhang}}, \binits{L.}},
\bauthor{\bsnm{{Han}}, \binits{D.}},
\bauthor{\bsnm{{Gao}}, \binits{N.}},
\bauthor{\bsnm{{Huo}}, \binits{J.}},
\bauthor{\bsnm{{Zhang}}, \binits{Z.}},
\bauthor{\bsnm{{Wang}}, \binits{H.}},
\bauthor{\bsnm{{Zhao}}, \binits{X.}},
\bauthor{\bsnm{{Wang}}, \binits{S.}},
\bauthor{\bsnm{{Li}}, \binits{Z.}},
\bauthor{\bsnm{{Bao}}, \binits{Z.}},
\bauthor{\bsnm{{Liu}}, \binits{Y.}},
\bauthor{\bsnm{{Wang}}, \binits{K.}},
\bauthor{\bsnm{{Wang}}, \binits{N.}},
\bauthor{\bsnm{{Wang}}, \binits{B.}},
\bauthor{\bsnm{{Wang}}, \binits{L.}},
\bauthor{\bsnm{{Wang}}, \binits{D.}},
\bauthor{\bsnm{{Ding}}, \binits{F.}},
\bauthor{\bsnm{{Sheng}}, \binits{L.}},
\bauthor{\bsnm{{Qiang}}, \binits{P.}},
\bauthor{\bsnm{{Yan}}, \binits{Y.}},
\bauthor{\bsnm{{Liu}}, \binits{Y.}},
\bauthor{\bsnm{{Wu}}, \binits{Z.}},
\bauthor{\bsnm{{Liu}}, \binits{Y.}},
\bauthor{\bsnm{{Chen}}, \binits{H.}},
\bauthor{\bsnm{{Zhang}}, \binits{Y.}},
\bauthor{\bsnm{{Liu}}, \binits{H.}},
\bauthor{\bsnm{{Altmann}}, \binits{A.}},
\bauthor{\bsnm{{Bechteler}}, \binits{T.}},
\bauthor{\bsnm{{Burwitz}}, \binits{V.}},
\bauthor{\bsnm{{Fiorini}}, \binits{C.}},
\bauthor{\bsnm{{Friedrich}}, \binits{P.}},
\bauthor{\bsnm{{Meidinger}}, \binits{N.}},
\bauthor{\bsnm{{Strecker}}, \binits{R.}},
\bauthor{\bsnm{{Baldini}}, \binits{L.}},
\bauthor{\bsnm{{Bellazzini}}, \binits{R.}},
\bauthor{\bsnm{{Bonino}}, \binits{R.}},
\bauthor{\bsnm{{Frass{\`a}}}, \binits{A.}},
\bauthor{\bsnm{{Latronico}}, \binits{L.}},
\bauthor{\bsnm{{Maldera}}, \binits{S.}},
\bauthor{\bsnm{{Manfreda}}, \binits{A.}},
\bauthor{\bsnm{{Minuti}}, \binits{M.}},
\bauthor{\bsnm{{Pesce-Rollins}}, \binits{M.}},
\bauthor{\bsnm{{Sgr{\`o}}}, \binits{C.}},
\bauthor{\bsnm{{Tugliani}}, \binits{S.}},
\bauthor{\bsnm{{Pareschi}}, \binits{G.}},
\bauthor{\bsnm{{Basso}}, \binits{S.}},
\bauthor{\bsnm{{Sironi}}, \binits{G.}},
\bauthor{\bsnm{{Spiga}}, \binits{D.}},
\bauthor{\bsnm{{Tagliaferri}}, \binits{G.}},
\bauthor{\bsnm{{Tykhonov}}, \binits{A.}},
\bauthor{\bsnm{{Paltani}}, \binits{S.}},
\bauthor{\bsnm{{Bozzo}}, \binits{E.}},
\bauthor{\bsnm{{Tenzer}}, \binits{C.}},
\bauthor{\bsnm{{Bayer}}, \binits{J.}},
\bauthor{\bsnm{{Tuo}}, \binits{Y.}},
\bauthor{\bsnm{{Liu}}, \binits{H.}},
\bauthor{\bsnm{{Zhang}}, \binits{Y.}},
\bauthor{\bsnm{{Cai}}, \binits{Z.}},
\bauthor{\bsnm{{Liu}}, \binits{H.}},
\bauthor{\bsnm{{Chen}}, \binits{W.}},
\bauthor{\bsnm{{Wang}}, \binits{C.}},
\bauthor{\bsnm{{He}}, \binits{T.}},
\bauthor{\bsnm{{Chen}}, \binits{Y.}},
\bauthor{\bsnm{{Qiu}}, \binits{C.}},
\bauthor{\bsnm{{Zhang}}, \binits{Y.}},
\bauthor{\bsnm{{Feng}}, \binits{J.}},
\bauthor{\bsnm{{Zhu}}, \binits{X.}},
\bauthor{\bsnm{{Zhou}}, \binits{H.}},
\bauthor{\bsnm{{Zheng}}, \binits{S.}},
\bauthor{\bsnm{{Song}}, \binits{L.}},
\bauthor{\bsnm{{Wang}}, \binits{J.}},
\bauthor{\bsnm{{Jia}}, \binits{S.}},
\bauthor{\bsnm{{Jiang}}, \binits{Z.}},
\bauthor{\bsnm{{Li}}, \binits{X.}},
\bauthor{\bsnm{{Zhao}}, \binits{H.}},
\bauthor{\bsnm{{Guan}}, \binits{J.}},
\bauthor{\bsnm{{Zhang}}, \binits{J.}},
\bauthor{\bsnm{{Li}}, \binits{C.}},
\bauthor{\bsnm{{Huang}}, \binits{Y.}},
\bauthor{\bsnm{{Liao}}, \binits{J.}},
\bauthor{\bsnm{{You}}, \binits{Y.}},
\bauthor{\bsnm{{Zhang}}, \binits{H.}},
\bauthor{\bsnm{{Wang}}, \binits{W.}},
\bauthor{\bsnm{{Wang}}, \binits{S.}},
\bauthor{\bsnm{{Ou}}, \binits{G.}},
\bauthor{\bsnm{{Hu}}, \binits{H.}},
\bauthor{\bsnm{{Shi}}, \binits{J.}},
\bauthor{\bsnm{{Cui}}, \binits{T.}},
\bauthor{\bsnm{{Jiang}}, \binits{X.}},
\bauthor{\bsnm{{Cheng}}, \binits{Y.}},
\bauthor{\bsnm{{Li}}, \binits{H.}},
\bauthor{\bsnm{{Xu}}, \binits{Y.}},
\bauthor{\bsnm{{Zane}}, \binits{S.}},
\bauthor{\bsnm{{Bambi}}, \binits{C.}},
\bauthor{\bsnm{{Bu}}, \binits{Q.}},
\bauthor{\bsnm{{Dall'Osso}}, \binits{S.}},
\bauthor{\bsnm{{Rosa}}, \binits{A.D.}},
\bauthor{\bsnm{{Gou}}, \binits{L.}},
\bauthor{\bsnm{{Guillot}}, \binits{S.}},
\bauthor{\bsnm{{Ji}}, \binits{L.}},
\bauthor{\bsnm{{Li}}, \binits{A.}},
\bauthor{\bsnm{{Mao}}, \binits{J.}},
\bauthor{\bsnm{{Patruno}}, \binits{A.}},
\bauthor{\bsnm{{Stratta}}, \binits{G.}},
\bauthor{\bsnm{{Taverna}}, \binits{R.}},
\bauthor{\bsnm{{Tsygankov}}, \binits{S.}},
\bauthor{\bsnm{{Uttley}}, \binits{P.}},
\bauthor{\bsnm{{Watts}}, \binits{A.L.}},
\bauthor{\bsnm{{Wu}}, \binits{X.}},
\bauthor{\bsnm{{Xu}}, \binits{R.}},
\bauthor{\bsnm{{Yi}}, \binits{S.}},
\bauthor{\bsnm{{Zhang}}, \binits{G.}},
\bauthor{\bsnm{{Zhang}}, \binits{L.}},
\bauthor{\bsnm{{Zhao}}, \binits{W.}},
\bauthor{\bsnm{{Zhou}}, \binits{P.}}:
\batitle{{The enhanced X-ray Timing and Polarimetry mission{\textemdash}eXTP
  for launch in 2030}}.
\bjtitle{Science China Physics, Mechanics, and Astronomy}
\bvolume{68}(\bissue{11}),
\bfpage{119502}
(\byear{2025})
\doiurl{10.1007/s11433-025-2786-6}
{\href{https://arxiv.org/abs/2506.08101}{{arXiv:2506.08101}}}
{[astro-ph.HE]}
\end{barticle}
\endbibitem

\bibitem[\protect\citeauthoryear{{Campana} et~al.}{2013}]{campana13}
\begin{barticle}
\bauthor{\bsnm{{Campana}}, \binits{R.}},
\bauthor{\bsnm{{Feroci}}, \binits{M.}},
\bauthor{\bsnm{{Del Monte}}, \binits{E.}},
\bauthor{\bsnm{{Mineo}}, \binits{T.}},
\bauthor{\bsnm{{Lund}}, \binits{N.}},
\bauthor{\bsnm{{Fraser}}, \binits{G.W.}}:
\batitle{{Background simulations for the Large Area Detector onboard LOFT}}.
\bjtitle{Experimental Astronomy}
\bvolume{36}(\bissue{3}),
\bfpage{451}--\blpage{477}
(\byear{2013})
\doiurl{10.1007/s10686-013-9341-6}
{\href{https://arxiv.org/abs/1305.3789}{{arXiv:1305.3789}}}
{[astro-ph.IM]}
\end{barticle}
\endbibitem

\bibitem[\protect\citeauthoryear{Bateman}{1910}]{bateman1910}
\begin{barticle}
\bauthor{\bsnm{Bateman}, \binits{H.}}:
\batitle{The solution of a system of differential equations occurring in the
  theory of radioactive transformations}.
\bjtitle{Proceedings of the Cambridge Philosophical Society}
\bvolume{15}(\bissue{pt V}),
\bfpage{423}--\blpage{427}
(\byear{1910})
\end{barticle}
\endbibitem

\bibitem[\protect\citeauthoryear{{Pressyanov}}{2002}]{pressyanov2002}
\begin{barticle}
\bauthor{\bsnm{{Pressyanov}}, \binits{D.S.}}:
\batitle{{Short solution of the radioactive decay chain equations}}.
\bjtitle{American Journal of Physics}
\bvolume{70}(\bissue{4}),
\bfpage{444}--\blpage{445}
(\byear{2002})
\doiurl{10.1119/1.1427084}
\end{barticle}
\endbibitem

\bibitem[\protect\citeauthoryear{{Onega}}{1969}]{onega1969}
\begin{barticle}
\bauthor{\bsnm{{Onega}}, \binits{R.J.}}:
\batitle{{Radioactivity Calculations}}.
\bjtitle{American Journal of Physics}
\bvolume{37}(\bissue{10}),
\bfpage{1019}--\blpage{1022}
(\byear{1969})
\doiurl{10.1119/1.1975187}
\end{barticle}
\endbibitem

\bibitem[\protect\citeauthoryear{{Moral} and
  {Pacheco}}{2003}]{moralpacheco2003}
\begin{barticle}
\bauthor{\bsnm{{Moral}}, \binits{L.}},
\bauthor{\bsnm{{Pacheco}}, \binits{A.F.}}:
\batitle{{Algebraic approach to the radioactive decay equations}}.
\bjtitle{American Journal of Physics}
\bvolume{71}(\bissue{7}),
\bfpage{684}--\blpage{686}
(\byear{2003})
\doiurl{10.1119/1.1571834}
\end{barticle}
\endbibitem

\bibitem[\protect\citeauthoryear{{Amaku} et~al.}{2010}]{amaku2010}
\begin{barticle}
\bauthor{\bsnm{{Amaku}}, \binits{M.}},
\bauthor{\bsnm{{Pascholati}}, \binits{P.R.}},
\bauthor{\bsnm{{Vanin}}, \binits{V.R.}}:
\batitle{{Decay chain differential equations: Solution through matrix
  algebra}}.
\bjtitle{Computer Physics Communications}
\bvolume{181}(\bissue{1}),
\bfpage{21}--\blpage{23}
(\byear{2010})
\doiurl{10.1016/j.cpc.2009.08.011}
\end{barticle}
\endbibitem

\bibitem[\protect\citeauthoryear{{Hauf} et~al.}{2013}]{hauf13}
\begin{barticle}
\bauthor{\bsnm{{Hauf}}, \binits{S.}},
\bauthor{\bsnm{{Kuster}}, \binits{M.}},
\bauthor{\bsnm{{Bati{\v{c}}}}, \binits{M.}},
\bauthor{\bsnm{{Bell}}, \binits{Z.W.}},
\bauthor{\bsnm{{Hoffmann}}, \binits{D.H.H.}},
\bauthor{\bsnm{{Lang}}, \binits{P.M.}},
\bauthor{\bsnm{{Neff}}, \binits{S.}},
\bauthor{\bsnm{{Pia}}, \binits{M.G.}},
\bauthor{\bsnm{{Weidenspointner}}, \binits{G.}},
\bauthor{\bsnm{{Zoglauer}}, \binits{A.}}:
\batitle{{Radioactive Decays in Geant4}}.
\bjtitle{IEEE Transactions on Nuclear Science}
\bvolume{60}(\bissue{4}),
\bfpage{2966}--\blpage{2983}
(\byear{2013})
\doiurl{10.1109/TNS.2013.2270894}
{\href{https://arxiv.org/abs/1307.0996}{{arXiv:1307.0996}}}
{[physics.comp-ph]}
\end{barticle}
\endbibitem

\end{thebibliography}

\end{document}